\documentclass[preprint,12pt]{elsarticle}

\usepackage{amssymb}
\usepackage{amsmath}
\usepackage{lineno}
\journal{Nuclear Physics B}

\begin{document}

\begin{frontmatter}



\title{\boldmath Under-coverage in high-statistics counting experiments with finite MC samples} %% Article title

\author[a,c]{C.~A.~Alexe}
\author[d]{J.~Bendavid}
\author[b,c]{L.~Bianchini}
\author[a,c]{D.~Bruschini}

\affiliation[a]{organization={Scuola Normale Superiore},
             addressline={P.zza dei Cavalieri, 7},
             city={Pisa},
             postcode={56126},
             country={Italy}}
\affiliation[b]{organization={Department of Physics E. Fermi, University of Pisa},
             addressline={Largo B. Pontecorvo, 3},
             city={Pisa},
             postcode={56127},
             country={Italy}}
\affiliation[c]{organization={Sezione di Pisa, INFN},
             addressline={Largo B. Pontecorvo, 3},
             city={Pisa},
             postcode={56127},
             country={Italy}}
\affiliation[d]{organization={European Organization for Nuclear Research (CERN)}, city={Geneva}, country={Switzerland}}

%% Abstract
\begin{abstract}
We consider the problem of setting confidence intervals on a parameter of interest from the maximum-likelihood fit of a physics model to a binned data set with a large number of bins, large event-counts per bin, and in the presence of systematic uncertainties modeled as nuisance parameters. We use the profile-likelihood ratio for statistical inference and focus on the case in which the model is determined from Monte Carlo simulated samples of finite size. We start by presenting a toy model in which the properties of widely used approximations of the profile-likelihood ratio in the asymptotic limit, which are commonly expected to hold in the high-statistics regime, are manifestly broken even if the numbers of events per bin in both the data and simulated samples are seemingly large enough to warrant their validity. We then move to the general setting to show how statistical uncertainties in the Monte Carlo predictions can affect the coverage of confidence intervals constructed in the asymptotic approximation always in the same direction, namely they lead to systematic under-coverage.
\end{abstract}

%% Keywords

\end{frontmatter}

\section{Introduction}\label{sec:intro}

Experiments in particle physics are often concerned with the problem of estimating a parameter of interest ($\mu$) from the maximum-likelihood fit of a physics model to sampling distributions constructed from a large number of identically repeated events, such as decays or collisions of particles. When the number of events is large enough to populate the full space of experimental outcomes, sampling distributions can be more conveniently approximated by histograms with a finite number of bins, resulting in a so-called {\it binned analysis}. Best-fit values for the parameter of interest (POI) are generally of limited use unless presented together with a {\it confidence interval} (CI), that is the range of allowed values in which the {\it true} value $\mu_{\rm t}$ of the POI is likely to be found within some conventional level of confidence.

Let us then assume that the experiment under consideration aims at measuring the number of occurrences of some particle physics reaction in the $n$ bins of a histogram, and that the expected number of events in each bin is related to $\mu$ according to a given physics model. Under fairly general assumptions, the event counts ${\bf y}=(y_1, \hdots,y_n)$ can be modeled as a vector of $n$ independent and Poisson-distributed random variables whose expectation value $\langle {\bf y} \rangle={\bf f}(\mu)$ is thus sufficient to fully define the joint probability density function of the data. The model ${\bf f}(\mu)$ might be affected by other sources of systematic uncertainties, related for example to imperfect knowledge of the detector response, the presence of backgrounds, or additional unknown parameters. We will assume that these uncertainties can be accounted for by extending the parameter space to include additional variables $\boldsymbol{\theta}$, commonly known as {\it nuisance parameters}. %, so that ${\bf f}$ will eventually be a function of both $\mu$ and $\boldsymbol{\theta}$. %The latter might be constrained by auxiliary measurements or be otherwise unknown.

Due to the complexity of modeling particle physics processes from generator up to detector-level, the function ${\bf f}(\mu, \boldsymbol{\theta})$ can rarely be computed analytically, leaving Monte Carlo (MC) techniques as the only viable way for its determination. Given that MC samples necessarily comprise a finite number of events, a MC-simulated model prediction will never be {\it exact}, thus introducing an additional source of systematic uncertainty. A general rule is that the size of the samples should exceed that of the data by at least a few times. However, such a request can turn out computationally expensive or even unfeasible for analyses performed on high-statistics samples of events, especially when CPU-intensive MC simulations are needed to achieve the desired level of theoretical accuracy. In such cases, one has to live with samples of MC-simulated events of size comparable or smaller than the data. The situation can become even more problematic when the simulated events are weighted~\cite{Glusenkamp}. %, which makes the treatment of the related uncertainty even more convoluted. 

The correct frequentist approach to deal with the finite size of MC samples was first introduced by Barlow and Beeston in ref.~\cite{BARLOW1993219}.  In their seminal paper, the authors proposed the idea of treating the uncertain MC predictions as additional nuisance parameters of which the available MC sample represents an ancillary measurement. Since then, several improvements in the treatment of MC uncertainties have been proposed in the literature, with particular emphasis on the case of low-statistic samples and the proper handling of event weights~\cite{BARLOW1993219,chirkin2013likelihooddescriptioncomparingdata,Glusenkamp,Arguelles}.

In this paper, we focus on the problem of setting confidence intervals on a parameter of interest in the presence of nuisance parameters and in the case of high-statistics counting experiments, for which asymptotic formulas from the theory of maximum likelihood estimators~\cite{Cowan,Kendall,Asimov} are expected to hold. We notice that the high-statistics case has received less attention in the literature compared to the low-statistics limit, where asymptotic formulas are more likely invalidated by low event counts. Furthermore, the high-statistics case is presently of great interest for precision measurements performed on large-scale data samples, such as those collected at the LHC, see e.g. refs.~\cite{ATLAS:2024wla,CMS:2024lrd} for a recent review on this subject.

A relevant example of a precision measurement performed on a large-scale data sample, and with a large number of nuisance parameters, is represented by the measurement of the $W$ boson mass by the CMS Collaboration~\cite{CMS:2024lrd}: the CMS measurement analyzed a sample of LHC collision data consisting of about $10^8$ single-muon events, binned in the kinematics and charge of the muons, resulting in approximately $3 \times 10^3$ bins, which are used to simultaneously measure the mass of the $W$ boson together with a few thousands of nuisance parameters via a maximum-likelihood fit. The model is constructed by using samples of MC-simulated events whose overall statistical power exceeds that of the data by a factor of about $4$. Alternate MC predictions corresponding to $\pm1\sigma$ variations of each nuisance parameter are constructed by reweighting the nominal MC simulation. The effect of statistical fluctuations in the alternate templates used to construct the systematic variations was observed to lead to a moderate under-coverage of Hessian uncertainties, which required an inflation of the statistical uncertainty of the MC templates to restore the correct coverage~\cite{CMS:2024lrd}. The main scope of this paper is to study the origin of such an effect from a more general perspective.

This paper is organized as follows. We start by presenting a hypothetical experiment which can be considered paradigmatic of our case of interest: the data set consists in a histogram with a large number of bins; the event counts in each bin are large; the physics model used to describe the data includes nuisance parameters and is determined from MC-simulated samples of size comparable to the data. We compute numerically the coverage of intervals constructed from CI-setting methods that rely on the asymptotic properties of maximum-likelihood estimators, such as those based on the Hessian matrix or the profile-likelihood ratio, and find that such intervals cover systematically less than their claimed confidence level. Then, we try alternative CI-setting methods that do not rely on asymptotic properties, such as those inspired by the unified approach by Feldman and Cousins~\cite{FC} based on Neyman's construction~\cite{1937RSPTA.236..333N} and again find similar issues. Inspired by our findings, we propose a heuristic definition of a confidence interval that better approaches the {\it correct} Feldman-Cousins interval. In the second part of the paper, we study the same problem from a more general perspective to ascertain in which circumstances statistical fluctuations of the MC model can lead to a systematic under-coverage of asymptotic confidence intervals. We conclude by suggesting numerical tests to quantify the impact of MC statistical fluctuations for a generic problem, remarking that the validity of the asymptotic approximation implied by Wilks' theorem should be carefully verified in the context of binned profile-likelihood analyses even when the number of measured events per bin is large and the statistical power of the simulated samples comparable to the data.
%assumption of being in the asymptotic regime may not be simply guaranteed by the large size of the data and simulated samples alone.   

\section{A paradigmatic toy model}\label{sec:toy}

%\subsection{The model}

Consider a hypothetical experiment measuring a set of event counts $y_i$, with $i=1,\hdots,n$, collected in the $n$ bins of a histogram. The measured events can be produced by either a signal or a background process, labeled by the indexes $j=1$ and $2$, respectively. The two processes are assumed to generate independent Poisson-distributed event counts $y_{ji}$ with expectation values in the $i^{\rm th}$ bin given by
\begin{equation}
\langle y_{1i} \rangle \equiv \nu_{1i} =  K,  \;\;\;
\langle y_{2i} \rangle \equiv \nu_{2i} =  
\begin{cases}
 (1+\epsilon)K & \;\; {\rm for}\; i<  \frac{n}{2} \\% \left(\frac{1+\alpha}{1-\alpha}\right)  
 K  & \;\; {\rm for}\; i\geq \frac{n}{2}
\end{cases}
%\frac{N}{n} \left( \frac{1\pm\alpha}{2-\alpha} \right),
\label{eq:truemodel}
\end{equation}
where $|\epsilon| \ll 1$ and $K\gg1$ are constant. For simplicity, we take $n$ as an even integer. In a real experiment, the constant $K$ would be related to e.g. beam flux, detector efficiency, cross sections, branching ratios, etc. In our study, we find it more convenient to express $K$ in terms of the total number of expected events ($N$) in the data sample, which can be easily determined in terms of $n$, $K$, and $\epsilon$ from eq.~\eqref{eq:truemodel}.

Let the purpose of this hypothetical experiment be a measurement of the unknown total number of signal events, henceforth denoted as {\it parameter of interest} (POI), with the unknown normalization of the background process playing the role of a {\it nuisance parameter} (NP). For this purpose, we will make use of a {physics model} which can predict the probability for a signal or background event to fall in a given bin. If this model could be derived directly from eq.~\eqref{eq:truemodel}, the data histogram could be re-binned in just two bins with no loss of information on either the POI or NP. However, the crucial assumption made here is that the physics model is not known {\it analytically} but can only be predicted from finite-size MC samples consisting of ${t}_{ji}$ simulated events in the $i^{\rm th}$ bin and for the $j^{\rm th}$ process. Since $t_{ji}$ are also random variables, they will differ from their {\it true values} by statistical fluctuations introduced by the MC generation, which could, in fact, justify preserving the full granularity of the data.

Given a set of MC predictions $t_{ji}$, the signal normalization can be extracted by parameterizing the unknown expectation value $\langle y_i \rangle$ in the $i^{\rm th}$ bin as:
\begin{equation}
f_i(\mu_1,\mu_2 \, ; \, t_{1i}, t_{2i} ) = \sum_{j=1}^{2} \mu_j \frac{t_{ji}}{k_j} %\mu_1  \frac{t_{1i}}{k_1}  +  \mu_2    \frac{t_{2i}}{k_2}
\label{eq:nui}
\end{equation} 
where {\it strength modifiers} $\mu_1$ and $\mu_2$ have been conveniently introduced for the two processes in place of their respective total number of events, and $k_j$ are constants that rescale the normalization of the MC samples to some reference value, which we are free to choose. In the following, we conveniently set $k_j$ equal to the ratio between the expectation value of $t_{ji}$ and its true value $\nu_{ji}$, so that by construction $f_i$ is an unbiased estimator of $\langle y_i\rangle$ for $\mu_1\equiv \mu_{1,{\rm t}}=1$ and $\mu_2\equiv \mu_{{2,\rm t}}=1$. With this choice, the constants $k_j$ represent the statistical power of the simulated samples relative to the data. For example, $k_1=k_2=1$ means that the total number of simulated events equals on average the number of data events, while values greater (smaller) than one indicate that more (less) events have been generated by the MC simulation than expected in data. Consistently with this assumption, numerical values for $y_{ji}$ and $t_{ji}$ in pseudo-experiments can be generated by drawing random integer numbers from the Poisson density with mean values $\nu_{ji}$ and $k_j\nu_{ji}$, respectively, that is:
\begin{equation}\label{eq:ditrt}
{\rm Prob}\left[y_{ij} \right] = \frac{e^{-\nu_{ji}} \left(\nu_{ji} \right)^{y_{ji}}}{y_{ij}!}\;\;\; {\rm and}\;\;\; {\rm Prob}\left[t_{ij} \right] = \frac{e^{-k_j \nu_{ji}} \left(k_j \nu_{ji} \right)^{t_{ji}}}{t_{ij}!}. 
\end{equation}

\begin{figure}
    \centering
    \includegraphics[width=0.45\linewidth]{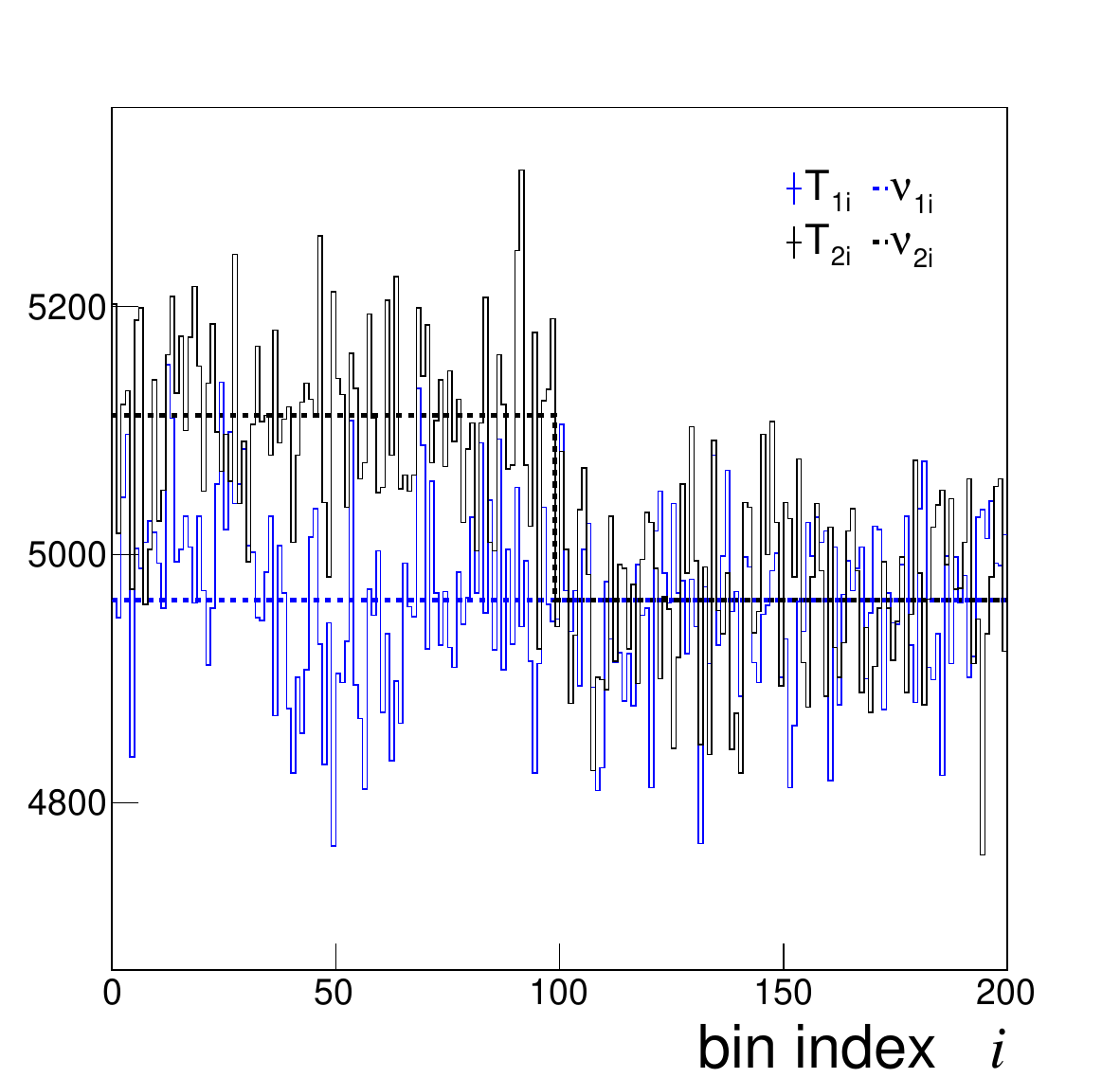}
    \includegraphics[width=0.45\linewidth]{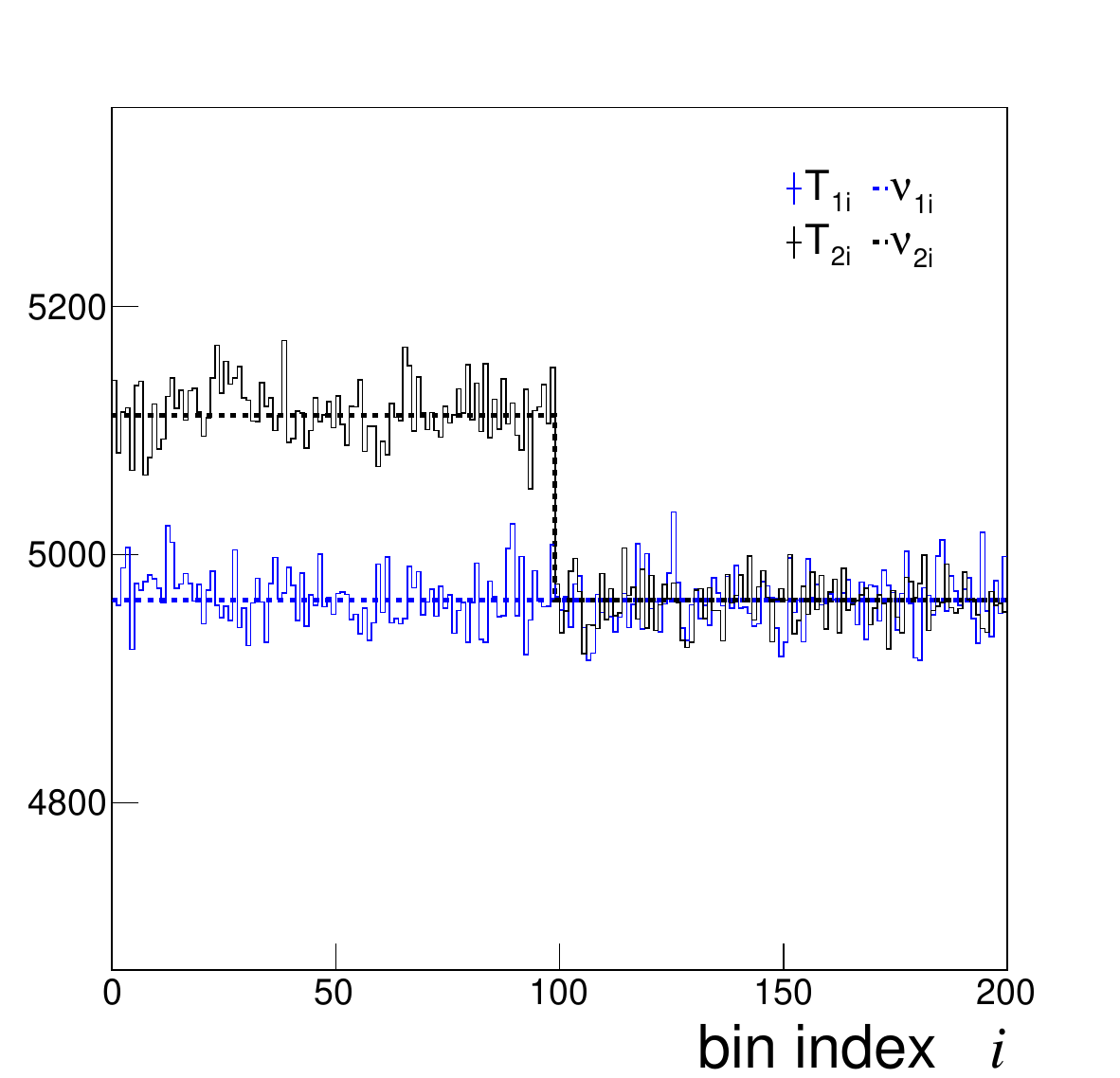}
    \caption{Examples of templates $T_{ji}$ generated as described in the text for $N=2\times 10^6$, $\epsilon=0.03$, $n=200$, and for two values of $k_1=k_2=k$, namely $k=1$ (left) and $k=10$ (right). The horizontal dashed lines in each plot provide the expected number of events per bin for the signal and background processes.}
    \label{fig:model}
\end{figure}

We notice that the rescaled MC predictions $T_{1i}=t_{1i}/k_1$ and $T_{2i}=t_{2i}/k_2$, commonly referred to as {\it templates}, also define the Jacobian matrix of the physics model ${\bf f}=(f_1, \hdots,f_n)$, with respect to its parameters $(\mu_1,\mu_2)$, that is
\begin{equation}\label{eq:jacdef}
\frac{\partial (f_1, \hdots, f_n)}{\partial(\mu_1,\mu_2)} = 
\begin{pmatrix}
T_{11} & T_{21} \\
\vdots & \vdots \\
T_{1n} & T_{2n} \\
\end{pmatrix}.
\end{equation}
The concept of Jacobian will be pivotal to section~\ref{sec:general}. Examples of templates $T_{ji}$ are shown in figure~\ref{fig:model} for representative values of $N$, $\epsilon$, $n$, and $k_j$. 

In the following, we will use the notation $\mu=\mu_1$ and $\theta=\mu_2$ to highlight the different role played by each parameter. In addition to a point estimator for the POI ($\hat{\mu}$), we also want to calculate confidence intervals $I$ at confidence level $\alpha$ with the aim that ${\rm Prob}\left[ \mu_{\rm t} \in I \right]= \alpha$, regardless of the true value $\mu_{\rm t}$. In particular,
we decide to report two-sided central confidence intervals at the conventional confidence level $\alpha =68.3\%$. We take the half-width $\hat{\sigma}_I$ of the intervals as an effective ``1$\sigma$'' standard deviation, and determine the mean and median of its sampling distribution from a large number of identically repeated pseudo-experiments. The coverage of $I$ is defined as the fraction of pseudo-experiments for which $\mu_{\rm t} \in I$.

In order to verify that the coverage of $I$ does not depend on $\mu_{\rm t}$, two additional independent data sets ${\bf y}_\pm$ are also considered. The latter are generated similarly to the left of eq.~\eqref{eq:ditrt} with the replacement
\begin{equation}
\nu_{1i} \to \nu^{\pm}_{1i} = \nu_{1i}\cdot \left(1 \pm 5\bar{\sigma}_{\rm H} \right),
\label{eq:altmodelcorr}
\end{equation}
where $\bar{\sigma}_{{\rm H}}$ is defined as the standard deviation of $\hat{\mu}$ when the MC templates $T_{ji}$ are replaced by their {\it true} values $\nu_{ji}$, or, equivalently, in the limit of infinite-size MC samples, that is $k_j\to\infty$. Clearly, $\mu_{\rm t}=1 \pm 5\bar{\sigma}_{{\rm H}}$ for these alternate data sets.

\subsection{Likelihood functions for the toy model}\label{sec:likelihoods}

The total event counts per bin $y_i$ can be treated as $n$ independent random variables drawn from the Poisson density $\mathcal{P}$, leading to the definition of the likelihood as:
\begin{equation}\label{eq:P}
L_{\rm P}(\mu, \theta) = \prod_{i=1}^{n}\mathcal{P}(y_i \, | \, f_i(\mu, \theta)) .
\end{equation}
The maximum-likelihood estimator (MLE) for the likelihood parameters is the point $(\hat{\mu},\hat{{\theta}})$ that minimizes the negative log-likelihood function (NLL), which,
apart from an irrelevant constant, can be written as
\begin{equation}\label{eq:logP}
-\ln L_{\rm P}(\mu, \theta) = \sum_{i=1}^{n} \left[ f_i(\mu, \theta \, ; \, t_{1i}, t_{2i}) - y_i\ln f_i(\mu, \theta \, ; \, t_{1i}, t_{2i}) \right].
\end{equation}
Equation~\eqref{eq:logP} neglects the fact that $t_{ji}$ are also random variables. A straightforward way to account for their stochastic nature is the Barlow-Beeston approach~\cite{BARLOW1993219}, which consists in treating the vectors ${\bf t}_{j}=(t_{j1},\hdots,t_{jn})$ as independent Poisson-distributed random variables with unknown mean values $\boldsymbol{\theta}_j$, therefore extending the likelihood to include an extra contribution:
\begin{align}
 -\ln L_{\rm P+BB}(\mu, \theta, \boldsymbol{\theta}_1, \boldsymbol{\theta}_2) 
 & =  \sum_{i=1}^{n} \left[ f_i(\mu, \theta \, ; \, \theta_{1i}, \theta_{2i}) - y_i\ln f_i(\mu, \theta \, ; \, \theta_{1i}, \theta_{2i}) \right]  \nonumber \\
& + \sum_{i=1}^{n}\sum_{j=1}^{2} \left(\theta_{ji} - t_{ji}\ln \theta_{ji} \right).
\label{eq:logPBB}
\end{align}
A major drawback of eq.~\eqref{eq:logPBB} is the growth in dimensionality: the number of likelihood parameters increases from $p=2$ to $p=2(n+1)$. However, as discussed in ref.~\cite{BARLOW1993219}, numerical minimization of eq.~\eqref{eq:logPBB} can be still recast as a two-dimensional optimization problem, at the cost of solving a non-linear system of equations~\cite{BARLOW1993219} at each step of the minimization, which can be challenging numerically~\cite{BBlite}. For this reason, a simplification to eq.~\eqref{eq:logPBB} has been proposed by Conway~\cite{BBlite}, leading to the so-called Barlow-Beeston lite likelihood:  
\begin{align}\label{eq:logPBBlite}
& -\ln L_{\rm P+BBlite}(\mu, \theta, \boldsymbol{\beta})  = \nonumber \\
& \sum_{i=1}^{n} \left[ f_i(\mu, \theta \, ; \, \beta_i t_{1i}, \beta_i  t_{2i}) - y_i\ln f_i(\mu, \theta \, ; \, \beta_i t_{1i}, \beta_i t_{2i}) \right] \nonumber \\
& + \sum_{i=1}^{n} m_i\left[\beta_{i} -  \ln \beta_i\right],
\end{align}
where the last term accounts for the total statistical uncertainty of the nominal MC simulation in the $i^{\rm th}$ bin as represented by the {\it effective} number of MC events $m_i$, defined as\footnote{The effective number of MC events can be defined such that its {\it relative} uncertainty equals the expectation for $m_i$ unweighted events:
\begin{equation}
\frac{1}{\sqrt{m_i}} = \frac{\sqrt{\rm Var\left[ T_{1i}+T_{21} \right]} }{ T_{1i} + T_{2i} } = \frac{\sqrt{\frac{t_{1i}}{k_1^2} + \frac{t_{21}}{k_2^2}}} { \frac{t_{1i}}{k_1} + \frac{t_{21}}{k_2} }
\end{equation}
For a MC sample consisting of $M$ weighted events, each one having weight $w_k$, Eq.~\eqref{eq:effective} can be generalized to $m_i={\left( \sum_{k=1}^{M} w_k\right)^2}/\left({\sum_{k=1}^{M} w_k^2}\right)$.
}
\begin{equation}\label{eq:effective}
    m_i = \left( \frac{t_{1i}}{k_1} + \frac{t_{2i}}{k_2} \right)^2 \left( \frac{t_{1i}}{k_1^2} + \frac{t_{2i}}{k_2^2}  \right)^{-1}.
\end{equation}
Equation~\eqref{eq:logPBBlite} still implies an increase in dimensionality, but this time the additional parameters $\boldsymbol{\beta}$ can be profiled analytically with a negligible computational overhead. This feature, together with its capability of catching the dominant effect of MC uncertainties, has made the Barlow-Beeston lite approach a very popular and successful solution in HEP analyses.

A summary of the bin-dependent variables introduced in the text is provided in Table~\ref{tab:glossary}. 

\begin{table}
\small
    \centering
\begin{tabular}{cll}
\hline
 Variable  & Definition  &  Role    \\
\hline
\hline
 $\nu_{ji}^{(\pm)}$ & True num. of events from the $j^{\rm th}$ process & exact \\
 $y_i$  & Num. of data events & random \\
 $t_{ji}$ & Num. of MC events from the $j^{\rm th}$ process & random \\
 $T_{ji}$ & Template value for the $j^{\rm th}$ process ($=t_{ji}/k_j$) & random \\
 $m_i$  & Effective num. of MC events & random  \\
 $\theta_{ji}$ & Barlow-Beeston NP for the $j^{\rm th}$ process & parameter \\
$\beta_i$  & Barlow-Beeston {\it lite} NP & parameter  \\
 $f_i$  & Parametric model prediction for $\langle y_i \rangle$ & exact/random \\
 \hline
\end{tabular}
\caption{Glossary of the variables defined for the $i^{\rm th}$ bin, as discussed in the text. In the last line, the model prediction $f_i$ should be considered as a fully parametric model (i.e. ``exact'') for the Barlow-Beeston likelihood and ``random'' for the Barlow-Beeston lite.}
    \label{tab:glossary}
\end{table}

\subsection{Alternate likelihood parametrization}

A data set distributed according to eq.~\eqref{eq:truemodel} is clearly not sensitive to the signal normalization when $\epsilon=0$, which translates into a degeneracy of the likelihood function with respect to its parameters. For $\epsilon>0$, the problem is mathematically well defined, but the separation between the two processes becomes increasingly more difficult as $\epsilon$ approaches zero due to the large anti-correlation between $\mu_1$ and $\mu_2$. For example, for values of the $\epsilon$ parameter of ${\cal O}(1\%)$, the linear correlation $\rho$ between the two parameters is almost complete, e.g. $\rho=-0.99989$ for $\epsilon=0.03$. In fact, one might argue that the individual measurement of $\mu_1$ and $\mu_2$ is perhaps a poorly defined problem, given that the data is much more sensitive to their sum than to the difference.

The role played by the choice of parameters is investigated by considering an alternate parametrization of the likelihood defined by the linear transformation
\begin{equation}
\mu_{1,2}^\prime=\frac{1}{2}\left(\mu_1 \mp \mu_2\right),
 \label{eq:transf}
\end{equation}
with $\mu^\prime_{1}$ ($\mu^\prime_{2}$) taken as POI (NP). Indeed, after applying the transformation in eq.~\eqref{eq:transf}, the linear correlation coefficient\footnote{We remark that the condition number of the covariance matrix is invariant under a linear transformation such as eq.~\eqref{eq:transf}, which indeed offers a better metric to characterize the problem.} between the transformed parameters changes to $0.71$ for the same value of $\epsilon$. However, we remark that the reduced correlation between parameters is not for free, given that the effective MC statistical power of the transformed templates
\begin{equation}
T_{ji}^\prime=T_{1i}+(-1)^{j+1}T_{2i}
\end{equation}
has also changed. The obvious solution would be to directly generate templates $T_{ji}^\prime$ for the transformed parameters, although this might not always be feasible for generic problems, for example because the MC generator is not flexible enough to allow for it.  

When considering the alternate parametrization of eq.~\eqref{eq:transf}, we thus set $\mu=\mu_1^\prime$ and $\theta=\mu_2^\prime$, %, since the former is more directly related to $\mu_1$ (e.g. they have almost the same variance),
and change the definition of the alternate data sets accordingly, that is:
\begin{align}
\nu^{\pm}_{1i} = \nu_{1i}\cdot \left(1 \pm 5\bar{\sigma}_{{\rm H}}^\prime \right) \;\;\; {\rm and} \;\;\;\nu^{\pm}_{2i} = \nu_{2i}\cdot \left(1 \mp 5\bar{\sigma}_{{\rm H}}^\prime \right),
 \label{eq:transf2}
\end{align}
where $\bar{\sigma}_{{\rm H}}^\prime$ is defined as the standard deviation of $\hat{\mu}_1^\prime$ for the nominal model in the limit of infinite-size MC samples. Given that $\mu_1$ and $\mu_2$ are highly correlated, and that their variances are also similar for $\epsilon\ll 1$, it is easy to see that $\bar{\sigma}_{{\rm H}}^\prime \approx \bar{\sigma}_{{\rm H}}$, so for the sake of simplicity we will use $\bar{\sigma}_{{\rm H}}^\prime = \bar{\sigma}_{{\rm H}}$ in eq.~\eqref{eq:transf2}. For this other parametrization, we have $\mu^\prime_{\rm t}=0$ for the nominal data and $\mu^\prime_{\rm t}=\pm 5\bar{\sigma}_{\rm H}$ for the alternate. 

The experiment can now be interpreted as measuring simultaneously the total number of events together with the asymmetry ${\cal A}$ between the two halves of the histogram. Indeed, by denoting the number of expected events in the first (second) half of the bins as $N_{<}$ ($N_{>}$), it is easy to prove that for the true model
\begin{equation}\label{eq:asymm}
{\cal A} = \frac{N_{<}-N_{>}}{N_{<}+N_{>}} = \frac{\epsilon}{4}\left( 1 - \frac{\mu_1^\prime}{\mu_2^\prime} + {\cal O}(\epsilon)\right).
\end{equation}
The statistical uncertainty on $\mu_2^\prime$ is of ${\cal O}(N^{-1/2})$, so it will be highly constrained when $N$ is large. Hence, the uncertainty on ${\cal A}$, which is directly proportional to the uncertainty on the ratio $\mu^\prime_1/\mu_2^\prime$ as for eq.~\eqref{eq:asymm}, is mostly driven by the limited precision on $\mu_1^\prime$. Measurements of asymmetries between two portions of a distribution occur frequently in Particle Physics. The determination of the $W$ boson mass at hadron colliders from the distribution of transverse momenta of the decaying leptons, already mentioned in section~\ref{sec:intro}, is a relevant example of this kind. Indeed, this measurement can be interpreted, at least to a first approximation, as the determination of the asymmetry of the transverse momentum distribution around the Jacobian peak~\cite{Rottoli:2023xdc}. 

\subsection{Likelihood functions in the Gaussian approximation}\label{sec:gaus}

In the limit $y_i\gg1$, numerical approximations of the negative log-likelihood function can be made. Indeed, the latter consists in a sum of terms of the same form which can be expanded around $f_i = y_i$ yielding:
\begin{equation}\label{eq:approx}
f_i - y_i\ln f_i = \frac{\left(f_i - y_i\right)^2}{y_i}  + \frac{2}{3} \frac{\left(f_i - y_i\right)^3}{y_i^2}  + \frac{1}{2}\frac{ \left( f_i - y_i\right)^4}{y_i^3} + \hdots,
\end{equation} 
where the dots indicate terms that either do not depend on $f_i$ or are of higher-order in the expansion. Neglecting terms beyond the quadratic order amounts to approximate the joint Poisson density as a Gaussian multivariate with covariance matrix ${\bf V}={\rm diag}({\bf y})$. For the toy model, it has been verified that a joint Poisson density and its Gaussian approximation result in consistent point estimators and confidence intervals for all values of $N$ and $n$ considered in this study. For example, for $N=2 \times 10^6$ and $n=200$, the number of events per bin is $y_i \sim10^4$, and the second term in the expansion represents a correction of order $(y_i - \nu_i)/y_i \sim \mathcal{O}(\sqrt{y_i})/y_i \sim 1\%$. The complete list of likelihood functions in the Gaussian approximation used in the study is presented in~\ref{app:likelihoods}.

Another practical advantage of the Barlow-Beeston lite approach discussed in section~\ref{sec:likelihoods} becomes evident in the Gaussian approximation. To see this, we first notice that the right-hand side of eq.~\eqref{eq:logP}, which would be the correct NLL function in the case of infinite-size MC samples, converges to a quadratic function of $(\theta,\mu)$ in the limit $y_i\gg1$, with error matrix given by ${\bf V}$. Profiling the $\boldsymbol{\beta}$ parameters in eq.~\eqref{eq:logPBBlite} introduces a nonlinearity in the parameters, which now appear also in the squared errors at the denominator of the profile likelihood function, see eq.~\eqref{eq:profiledGBBlite} in the appendix.
Formally, this function would be the same quadratic function as in the infinite-size MC case, provided that the error matrix is replaced by ${\bf V}+{\bf V}_{\rm MC}$, where ${\bf V}_{\rm MC}$ is a diagonal matrix with $i^{\rm th}$ diagonal element equal to $f^2_{i}/m_i$. This additional contribution to the error matrix depends on $(\mu,\theta)$, thus formally breaking the quadratic dependence. However, we notice that, for values of $(\mu,\theta)$ around the minimum, $f_i$ is constrained to take values of order $y_i$ by the {penalty} term $(y_i-f_i)^2$ appearing at the numerator of the profile likelihood function, so that $f^2_{i}/m_i\approx y_i/k$ to a very good approximation, where $k\sim m_i/y_i$ represents the total statistical power of the MC simulation in the $i^{\rm th}$ bin relative to the data.

This observation allows us to conclude that the profiled likelihood for the Barlow-Beeston lite case, can be approximated, at least numerically, as a quadratic function of $\mu$ and $\theta$ with error matrix given by ${\bf V}+{\bf V}_{\rm MC} \approx {\bf V}\cdot (1+k^{-1})$. In the following, we will denote the latter function as ``Gaus. + MC stat.''. As expected, numerical differences between minimizing the Barlow-Beeston lite likelihood or its Gaus. + MC stat. approximation are found to be negligible in the context of the toy study.

\subsection{Confidence intervals in the asymptotic limit} \label{sec:CIasy}

Based on the properties of the maximum-likelihood estimators and of the profile-likelihood ratio in the asymptotic limit~\cite{Kendall}, we consider the following methods to set confidence intervals on $\mu$.  
\begin{itemize}
    \item {\bf Hessian uncertainty}\\
The Hessian uncertainty $\hat{\sigma}_{\rm H}$ is defined as the square root of the last diagonal element of the inverse Hessian matrix of the NLL function at the minimum:
\begin{equation}
    {\bf C} = \left( - \frac{\partial^2 \ln L}{ \partial^2 (\boldsymbol{\theta},\mu) } \right)^{-1},
    \label{eq:hessian}
\end{equation}
where $\boldsymbol{\theta}$ is a shorthand for the complete set of nuisance parameters.
In the asymptotic limit, the maximum-likelihood estimators are known to be Gaussian-distributed around their true values~\cite{Cowan}, with covariance matrix given by the expectation value of ${\bf C}$, so that
\begin{equation}
\left[ \hat{\mu} - \hat{\sigma}_{\rm H}, \; \hat{\mu} + \hat{\sigma}_{\rm H} \right],
\end{equation}\label{eq:Ihessian}
is (asymptotically) a 68.3\% CL interval for $\mu$.
\item {\bf Profile-Likelihood ratio} \\
The profile-likelihood ratio (PLR) is a function of $\mu$ defined as~\cite{Kendall}
\begin{equation}
t_{\mu} = -2 \ln \frac{ L(\mu, {\hat{\boldsymbol{ \theta}}}_\mu) }{L( \hat{\mu}, {\hat{\boldsymbol{ \theta}}})},
\label{eq:plr} 
\end{equation} 
where $\hat{\mu}$ and $\hat{{\boldsymbol\theta}}$ are the maximum-likelihood estimators for $\mu$ and $\boldsymbol{\theta}$, respectively, while ${\hat{\boldsymbol{ \theta}}}_\mu$ is the NP point that minimizes the NLL at a fixed value of $\mu$. In the asymptotic limit, the distribution of $t_{\mu}$ tends to a chi-square with one degree of freedom as for Wilks' theorem~\cite{Wilks}, with deviations expected at the ${\cal O}(1/\sqrt{N})$ level, with $N$ related to the size of the data. If this high-order correction can be neglected, the interval
%$\left[ \mu_{-}, \mu_{+}\right]$, defined such that $t_{\mu_{-}}=t_{\mu_{+}}=1$, is again a 68.3\% CL interval, that is
%If the asymptotic limit is not valid, a confidence interval can be still defined as the set of points
\begin{equation}
\{ \mu \, : \, t_\mu\leq 1 \}.
\label{eq:Iplr}  
\end{equation}
is again a 68.3\% CL interval for $\mu$.
\end{itemize}
Given that $\hat{\mu}$ is also asymptotically Gaussian-distributed, it follows that the confidence intervals defined by eq.~\eqref{eq:hessian} and~\eqref{eq:plr} coincide in the asymptotic limit.

\subsection{Numerical results}

The minimization of the negative log-likelihood function is performed numerically using the {\tt MIGRAD} minimizer implemented in {\tt Minuit2}~\cite{JAMES1975343}, with PLR scans performed by {\tt MINOS} and Hessian-based uncertainties computed by {\tt HESSE}. Linear algebra operations are performed using the {\tt Eigen} library~\cite{eigenweb} in double precision. The nominal configuration parameters for the toy model are reported in table~\ref{tab:nominal_parameters}.
The effect of changing their values is studied in section~\ref{sec:recovery}. For simplicity, $k_1=k_2=k$ has been assumed\footnote{For example, this would be appropriate when the signal and background can be defined as sub-processes of a same MC sample. The generalization to arbitrary $k_j$ is also straightforward and is not a critical aspect for this study.}. Coverage, mean, and median of the $1\sigma$ confidence interval for $\mu=\mu_1$ are obtained from an ensemble of $10^4$ identically repeated pseudo-experiments, and reported in table~\ref{tab:nominal_corr}, while the results for the other choice $\mu=\mu^\prime_1$ are shown in table~\ref{tab:nominal}. 

The expected Hessian uncertainty $\bar{\sigma}_{\rm H}$ for the nominal model and in the limit of infinite-size MC samples, that is when $T_{ji}=\nu_{ji}$, can be computed analytically. Using well-known results, which will be further elaborated in section~\ref{sec:general}, it is easy to prove that
\begin{equation}
    \bar{\sigma}_{\rm H} = \frac{4}{|\epsilon|\sqrt{N}}\left( 1 + {\cal O}(\epsilon)\right)\approx 0.094,
\end{equation}
in very good agreement with the numerical result shown in the first row of tables~\ref{tab:nominal_corr} and~\ref{tab:nominal}. Also, one can easily verify that the linear correlation between $\mu_1$ and $\mu_2$ differs from $-1$ by terms of ${\cal O}(\epsilon^2)$. Differences between the Hessian uncertainty for the nominal and alternate models in table~\ref{tab:nominal_corr} are mostly due to the different number of expected events when $\mu_{\rm t}$ is changed.

\begin{table}
\small
    \centering
\begin{tabular}{cccc}
\hline
   $N \, \left[\times 10^6\right]$ & $\epsilon \, \left[\times 0.01\right]$ & $n$ & $k_1=k_2=k$       \\
\hline
\hline
 ${\bf 2}$, $10$, $100$ & ${\bf 3}$, $6$, $12$ & $20$, $100$, ${\bf 200}$ & ${\bf 1}$, $10$, $40$ \\
  \hline
\end{tabular}
\caption{Nominal (in bold) and alternate configuration parameters for the toy model.}
    \label{tab:nominal_parameters}
\end{table}

\begin{table}
\scriptsize
    \centering
\begin{tabular}{lccccc}
Likelihood     & Minimim. & CI method      & $\mu_{\rm t}-1$ & Coverage    &  $\hat{\sigma}$ (mean, median)          \\
\hline
\hline
1) Gauss (asympt.)  & Analytic & Hessian        & 0                     &  $0.683$    &  $0.095, \; 0.095$ \\
                 &          &                &$+5\bar{\sigma}_{{\rm H} }$   &  $0.683$    &  $0.106, \; 0.106$ \\
                 &          &                &$-5\bar{\sigma}_{{\rm H} }$   &  $0.683$    &  $0.083, \; 0.083$ \\
\hline
2) Poisson          & Numeric  & Hessian        & 0                     &  $0.521(5)$ &  $0.058, \; 0.057$ \\
                 &          &                &$+5\bar{\sigma}_{{\rm H} }$   &  $0.160(4)$ &  $0.064, \; 0.064$ \\
                 &          &                &$-5\bar{\sigma}_{{\rm H} }$   &  $0.064(2)$ &  $0.050, \; 0.050$ \\
\hline
3) Gauss            & Numeric  & Hessian        & 0                     &  $0.520(5)$ &  $0.057, \; 0.057$ \\
                 &          &                &$+5\bar{\sigma}_{{\rm H} }$   &  $0.161(4)$ &  $0.064, \; 0.064$ \\
                 &          &                &$-5\bar{\sigma}_{{\rm H} }$   &  $0.063(2)$ &  $0.050, \; 0.050$ \\
\hline
4) Gauss            & Analytic & Hessian        & 0                     &  $0.520(5)$ &  $0.057, \; 0.057$ \\
                 &          &                &$+5\bar{\sigma}_{{\rm H} }$   &  $0.161(4)$ &  $0.064, \; 0.064$ \\
                 &          &                &$-5\bar{\sigma}_{{\rm H} }$   &  $0.063(2)$ &  $0.050, \; 0.050$ \\
\hline
5) Gauss + MC stat. & Analytic & Hessian        & 0                     &  ${\bf 0.682}(5)$ &  $0.081, \; 0.081$ \\
                 &          &                &$+5\bar{\sigma}_{{\rm H} }$   &  $0.230(4)$ &  $0.086, \; 0.085$ \\
                 &          &                &$-5\bar{\sigma}_{{\rm H} }$   &  $0.124(3)$ &  $0.076, \; 0.076$ \\
\hline
6) Gauss + BB-lite  & Numeric  & Hessian        & 0                     &  ${\bf 0.679}(5)$ &  $0.080, \; 0.080$ \\
                 &          &                &$+5\bar{\sigma}_{{\rm H} }$   &  $0.259(4)$ &  $0.094, \; 0.094$ \\
                 &          &                &$-5\bar{\sigma}_{{\rm H} }$   &  $0.097(3)$ &  $0.066, \; 0.066$ \\
\hline
7) Gauss + BB       & Numeric  & Hessian        & 0                     &  $0.439(5)$ &  $0.137, \; 0.132$ \\
                 &          &                &$+5\bar{\sigma}_{{\rm H} }$   &  $0.449(5)$ &  $0.163, \; 0.157$ \\
                 &          &                &$-5\bar{\sigma}_{{\rm H} }$   &  $0.454(5)$ &  $0.116, \; 0.111$ \\
\hline
8) Gauss + BB       & Numeric  & PLR       & 0                     &  $0.454(5)$ &  $0.141, \; 0.135$ \\
                 &          &                &$+5\bar{\sigma}_{{\rm H} }$   &  $0.461(5)$ &  $0.170, \; 0.162$ \\
                 &          &                &$-5\bar{\sigma}_{{\rm H} }$   &  $0.469(5)$ &  $0.116, \; 0.111$ \\
\hline
9) Gauss + BB       & Numeric  & sPFC& 0                     &  $0.471(5)$ &  $0.160, \; 0.146$ \\
                 &          &                &$+5\bar{\sigma}_{{\rm H} }$   &  $0.464(5)$ &  $0.184, \; 0.169$ \\
                 &          &                &$-5\bar{\sigma}_{{\rm H} }$   &  $0.614(5)$ &  $0.166, \; 0.155$ \\
\hline
10) Gauss + BB       & Numeric  & PFC & 0                     &  $0.484(5)$ &  $0.172, \; 0.157$ \\
                 &          &                &$+5\bar{\sigma}_{{\rm H} }$   &  $0.546(5)$ &  $0.211, \; 0.188$ \\
                 &          &                &$-5\bar{\sigma}_{{\rm H} }$   &  ${\bf 0.656}(5)$ &  $0.176, \; 0.160$ \\
\hline
11) Gauss + BB       & Numeric  & Barlett    & 0                     &  $0.480(5)$ &  $0.168, \; 0.155$ \\
                 &          &                &$+5\bar{\sigma}_{{\rm H} }$   &  $0.513(5)$ &  $0.205, \; 0.184$ \\
                 &          &                &$-5\bar{\sigma}_{{\rm H} }$   &  $0.615(5)$ &  $0.166, \; 0.152$ \\
\hline
12) Gauss + BB       & Numeric  & CH     & 0                   &  ${\bf 0.681}(5)$ &  $0.236, \; 0.221$ \\
                 &          &                           &$+5\bar{\sigma}_{{\rm H} }$ &  $0.776(4)$ &  $0.296, \; 0.283$ \\
                 &          &                           &$-5\bar{\sigma}_{{\rm H} }$ &  $0.847(4)$ &  $0.281, \; 0.276$ \\
\hline
13) Gauss + BB       & Numeric  & FC Cheat       & 0                     &  ${\bf 0.664}(5)$ &  $0.229, \; 0.207$ \\
                 &          &                &$+5\bar{\sigma}_{{\rm H} }$   &  ${\bf 0.673}(5)$ &  $0.258, \; 0.236$ \\
                 &          &                &$-5\bar{\sigma}_{{\rm H} }$   &  ${\bf 0.680}(5)$ &  $0.190, \; 0.171$ \\
\hline
14) Gauss + BB       & Numeric  & Heuristic   & 0                   &  ${\bf 0.678}(5)$ &  $0.228, \; 0.221$ \\
                 &          &                &$+5\bar{\sigma}_{{\rm H} }$ &  ${\bf 0.660}(5)$ &  $0.256, \; 0.248$ \\
                 &          &                &$-5\bar{\sigma}_{{\rm H} }$ &  $0.720(4)$ &  $0.204, \; 0.197$ \\
\hline
\hline
\end{tabular}
\caption{ Coverage, mean, and median of the $1\sigma$ confidence interval for $\mu=\mu_{1}$ for the toy model with $N=2\times 10^6$, $n=200$, $k=1$, and $\epsilon=0.03$, obtained from an ensemble of $10^4$ identically repeated pseudo-experiments. Cases where the observed coverage agrees with the expectation of $68.3\%$ within a relative $\pm5\%$ tolerance are highlighted with a bold font. When present, the number within parenthesis refers to the statistical error on the last digit.}
    \label{tab:nominal_corr}
\end{table}

\begin{table}
\scriptsize
    \centering
\begin{tabular}{lccccc}
Likelihood     & Minimim. & CI method      & $\mu_{\rm t}$ & Coverage    &  $\hat{\sigma}$ (mean, median)          \\
\hline
\hline
1) Gauss (asympt.)  & Analytic & Hessian        & 0                     &  $0.683$    &  $0.094, \; 0.094$ \\
                 &          &                &$+5\bar{\sigma}_{{\rm H} }$   &  $0.683$    &  $0.094, \; 0.094$ \\
                 &          &                &$-5\bar{\sigma}_{{\rm H} }$   &  $0.683$    &  $0.094, \; 0.094$ \\
\hline
2) Poisson          & Numeric  & Hessian        & 0                     &  $0.516(5)$ &  $0.057, \; 0.057$ \\
                 &          &                &$+5\bar{\sigma}_{{\rm H} }$   &  $0.001(1)$ &  $0.057, \; 0.057$ \\
                 &          &                &$-5\bar{\sigma}_{{\rm H} }$   &  $0.002(1)$ &  $0.057, \; 0.057$ \\
\hline
3) Gauss            & Numeric  & Hessian        & 0                     &  $0.516(5)$ &  $0.057, \; 0.057$ \\
                 &          &                &$+5\bar{\sigma}_{{\rm H} }$   &  $0.001(1)$ &  $0.057, \; 0.057$ \\
                 &          &                &$-5\bar{\sigma}_{{\rm H} }$   &  $0.002(1)$ &  $0.057, \; 0.057$ \\
\hline
4) Gauss            & Analytic & Hessian        & 0                     &  $0.516(5)$ &  $0.057, \; 0.057$ \\
                 &          &                &$+5\bar{\sigma}_{{\rm H} }$   &  $0.001(1)$ &  $0.057, \; 0.057$ \\
                 &          &                &$-5\bar{\sigma}_{{\rm H} }$   &  $0.002(1)$ &  $0.057, \; 0.057$ \\
\hline
5) Gauss + MC stat. & Analytic & Hessian        & 0                     &  ${\bf 0.675}(5)$ &  $0.080, \; 0.080$ \\
                 &          &                &$+5\bar{\sigma}_{{\rm H} }$   &  $0.004(1)$ &  $0.080, \; 0.080$ \\
                 &          &                &$-5\bar{\sigma}_{{\rm H} }$   &  $0.004(1)$ &  $0.080, \; 0.080$ \\
\hline
6) Gauss + BB-lite  & Numeric  & Hessian        & 0                     &  ${\bf 0.674}(5)$ &  $0.080, \; 0.080$ \\
                 &          &                &$+5\bar{\sigma}_{{\rm H} }$   &  $0.004(1)$ &  $0.080, \; 0.080$ \\
                 &          &                &$-5\bar{\sigma}_{{\rm H} }$   &  $0.004(1)$ &  $0.081, \; 0.080$ \\
\hline
7) Gauss + BB       & Numeric  & Hessian        & 0                     &  $0.443(5)$ &  $0.139, \; 0.133$ \\
                 &          &                &$+5\bar{\sigma}_{{\rm H} }$   &  $0.468(5)$ &  $0.147, \; 0.140$ \\
                 &          &                &$-5\bar{\sigma}_{{\rm H} }$   &  $0.465(5)$ &  $0.147, \; 0.141$ \\
\hline
8) Gauss + BB       & Numeric  & PLR       & 0                     &  $0.448(5)$ &  $0.140, \; 0.134$ \\
                 &          &                &$+5\bar{\sigma}_{{\rm H} }$   &  $0.474(5)$ &  $0.152, \; 0.144$ \\
                 &          &                &$-5\bar{\sigma}_{{\rm H} }$   &  $0.470(5)$ &  $0.143, \; 0.137$ \\
\hline
9) Gauss + BB       & Numeric  & sPFC& 0                     &  $0.474(5)$ &  $0.158, \; 0.144$ \\
                 &          &                &$+5\bar{\sigma}_{{\rm H} }$   &  $0.597(5)$ &  $0.211, \; 0.200$ \\
                 &          &                &$-5\bar{\sigma}_{{\rm H} }$   &  $0.590(5)$ &  $0.191, \; 0.184$ \\
\hline
10) Gauss + BB       & Numeric  & PFC & 0                     &  $0.471(5)$ &  $0.172, \; 0.156$ \\
                 &          &                &$+5\bar{\sigma}_{{\rm H} }$   &  ${\bf 0.671}(5)$ &  $0.222, \; 0.196$ \\
                 &          &                &$-5\bar{\sigma}_{{\rm H} }$   &  ${\bf 0.675}(5)$ &  $0.212, \; 0.192$ \\
\hline
11) Gauss + BB       & Numeric  & Barlett    & 0                     &  $0.466(5)$ &  $0.167, \; 0.154$ \\
                 &          &                &$+5\bar{\sigma}_{{\rm H} }$   &  $0.633(5)$ &  $0.213, \; 0.188$ \\
                 &          &                &$-5\bar{\sigma}_{{\rm H} }$   &  $0.635(5)$ &  $0.201, \; 0.181$ \\
\hline
12) Gauss + BB       & Numeric  & CH     & 0                   &  ${\bf 0.675}(5)$ &  $0.235, \; 0.220$ \\
                 &          &                           &$+5\bar{\sigma}_{{\rm H} }$ &  $0.884(3)$ &  $0.345, \; 0.352$ \\
                 &          &                           &$-5\bar{\sigma}_{{\rm H} }$ &  $0.884(3)$ &  $0.347, \; 0.354$ \\
\hline
13) Gauss + BB       & Numeric  & FC Cheat       & 0                     &  ${\bf 0.677}(5)$ &  $0.228, \; 0.206$ \\
                 &          &                &$+5\bar{\sigma}_{{\rm H} }$   &  ${\bf 0.682}(5)$ &  $0.232, \; 0.210$ \\
                 &          &                &$-5\bar{\sigma}_{{\rm H} }$   &  ${\bf 0.672}(5)$ &  $0.221, \; 0.199$ \\
\hline
14) Gauss + BB       & Numeric  & Heuristic   & 0                   &  ${\bf 0.675}(5)$ &  $0.229, \; 0.222$ \\
                 &          &                &$+5\bar{\sigma}_{{\rm H} }$ &  ${\bf 0.703}(5)$ &  $0.241, \; 0.232$ \\
                 &          &                &$-5\bar{\sigma}_{{\rm H} }$ &  ${\bf 0.700}(5)$ &  $0.242, \; 0.233$ \\
\hline
\hline
\end{tabular}
\caption{ Coverage, mean, and median of the $1\sigma$ confidence interval for $\mu=\mu^{\prime}_{1}$ for the toy model with $N=2\times 10^6$, $n=200$, $k=1$, and $\epsilon=0.03$, obtained from an ensemble of $10^4$ identically repeated pseudo-experiments. Cases where the observed coverage agrees with the expectation of $68.3\%$ within a relative $\pm5\%$ tolerance are highlighted with a bold font. When present, the number within parenthesis refers to the statistical error on the last digit.}
    \label{tab:nominal}
\end{table}

The outcome of the numerical study can be summarized as follows.

\begin{figure}
    \centering
    \includegraphics[width=0.7\linewidth]{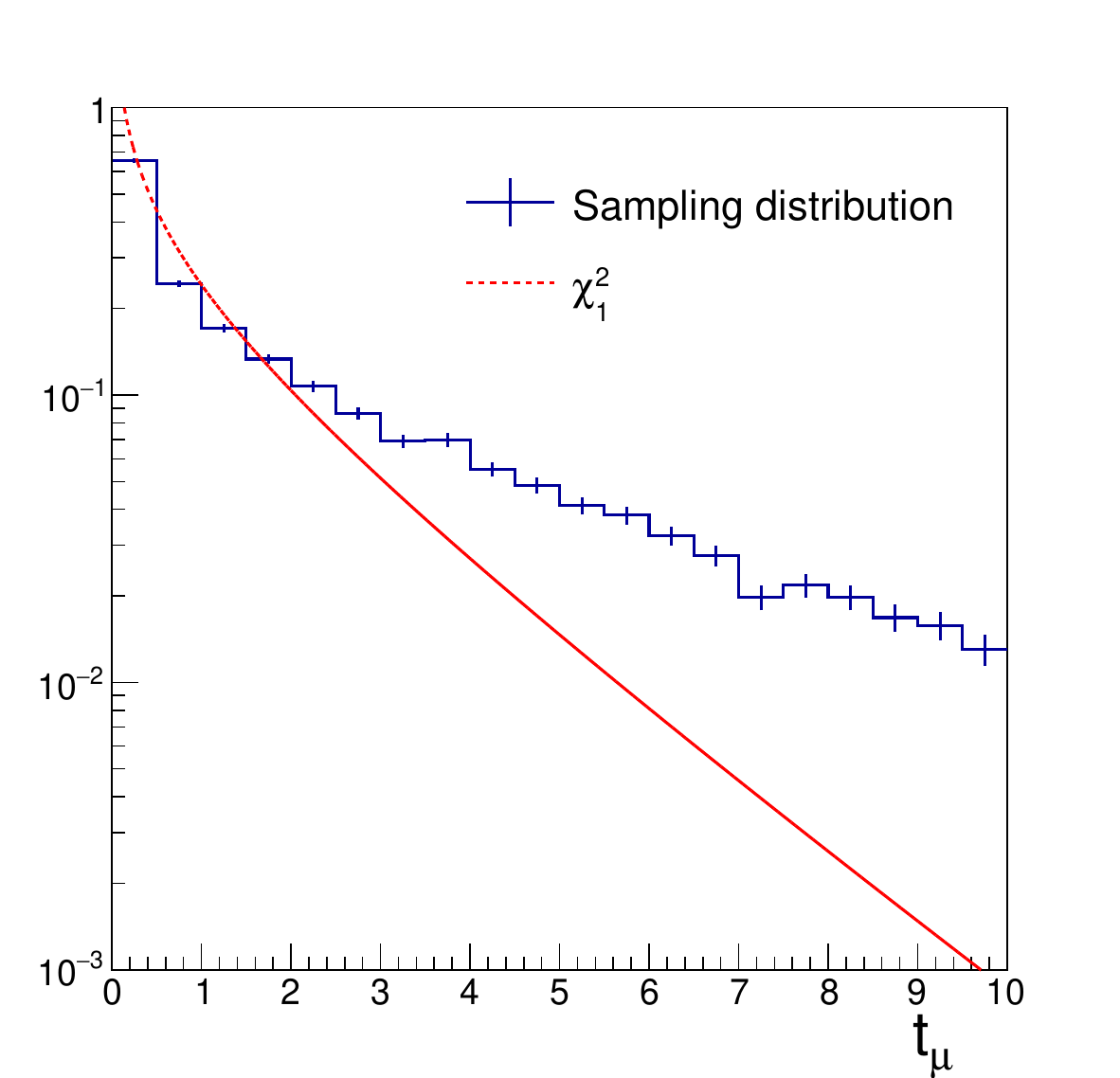}
    \caption{Sampling distribution of $t_\mu$, the profile-likelihood function in the Gaussian approximation, evaluated at $\mu \equiv \mu^\prime_1=0$ for data distributed according to the nominal model (i.e. with $\mu^\prime_{\rm t}=0$) with $N=2\times 10^6$, $\epsilon=0.03$, $n=200$, and $k=1$, obtained from an ensemble of $10^4$ identically repeated pseudo-experiments.}
    \label{fig:tstat}
\end{figure}

\begin{figure}
    \centering
    \includegraphics[width=0.45\linewidth]{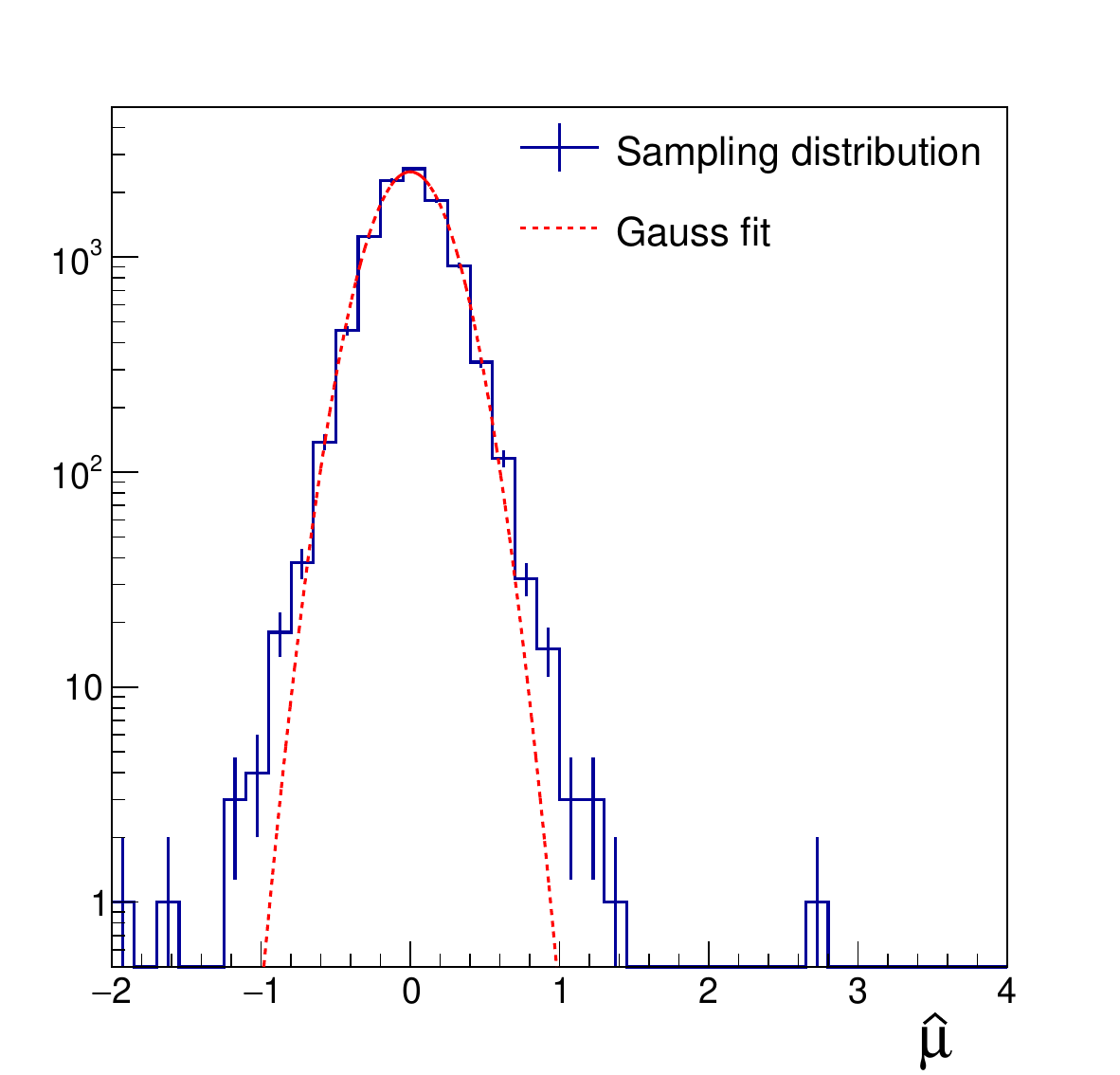}
    \includegraphics[width=0.45\linewidth]{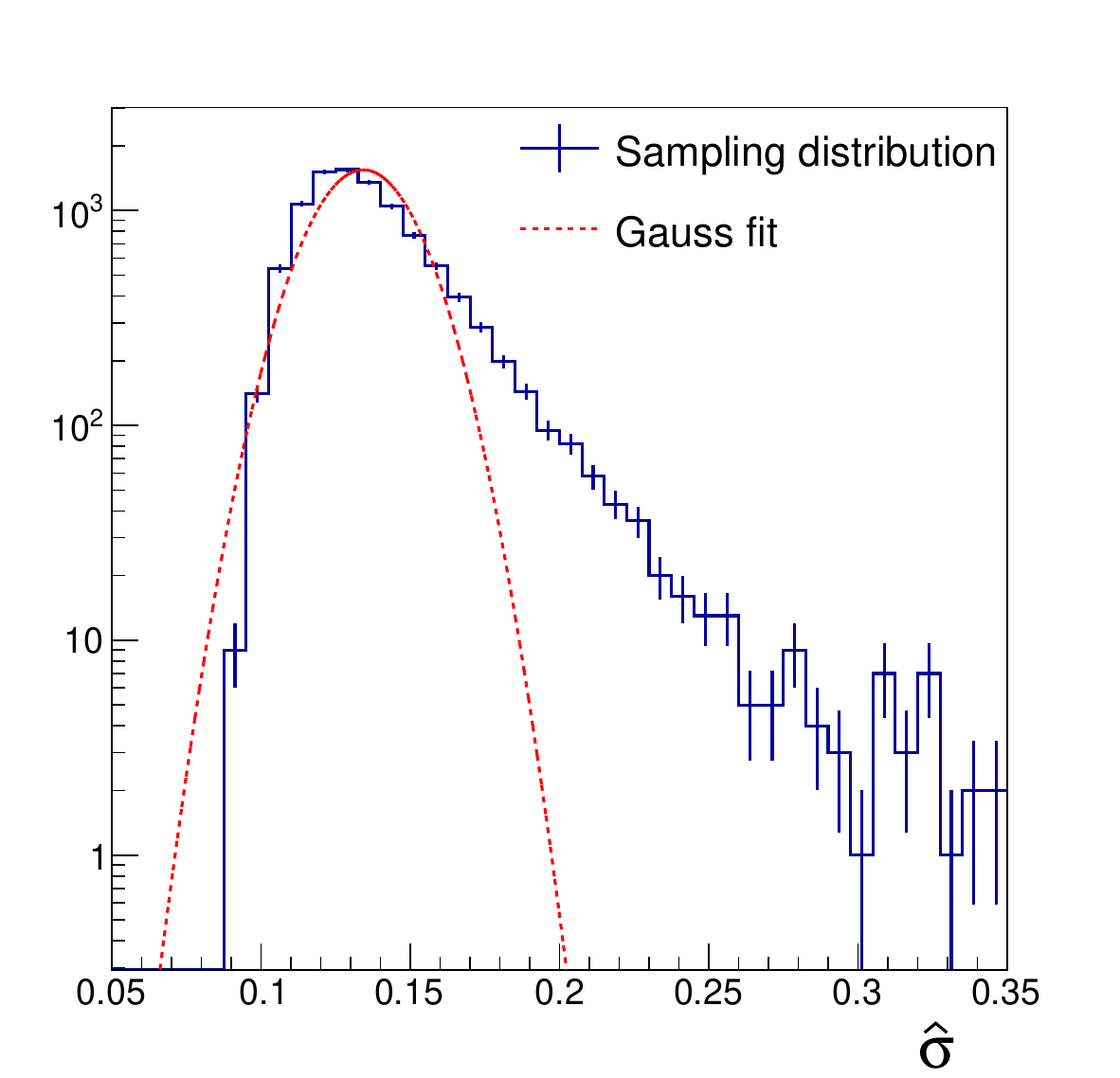}
    \caption{Sampling distributions of the point-estimator $\hat{\mu} \equiv \hat{\mu}^\prime_1$ and of its Hessian uncertainty $\hat{\sigma}_{\rm H}$ obtained from the Gauss + Barlow-Beeston likelihood applied to data distributed according to the nominal  model (i.e. with $\mu_{\rm t}=0$) with $N=2\times 10^6$, $\epsilon=0.03$, $n=200$, and $k=1$, obtained from an ensemble of $10^4$ identically repeated pseudo-experiments. For both distributions, a Gaussian fit is overlaid for reference.}
    \label{fig:hat}
\end{figure}

\begin{enumerate}
    \item The Poisson ($2^{\rm nd}$ row) and Gaussian ($3^{\rm rd}$ row) likelihood provide consistent results in terms of coverage and $1\sigma$ uncertainty when using the definition of confidence intervals based on the Hessian matrix. In the Gaussian approximation, numerical minimization ($3^{\rm th}$ row) or an analytical solution ($4^{\rm th}$ row) provide identical results. As expected, by neglecting the stochastic nature of the MC templates, these likelihoods bring to severe under-coverage: the uncertainties are about twice smaller than $\bar{\sigma}_{\rm H}$ ($1^{\rm st}$ row).
   
    \item The Hessian uncertainty for the Gauss + MC stat. likelihood ($5^{\rm th}$ row) gives the correct coverage for the nominal data set, but under-covers for the alternate data sets. The $1\sigma$ uncertainty is still smaller than $\bar{\sigma}_{\rm H}$, further suggesting that these confidence intervals will generally under-cover. Results obtained by using either the Gauss + MC stat. ($5^{\rm th}$ row) or Gauss + BBlite likelihood ($6^{\rm th}$ row) are quantitatively similar. 
    
    \item The Gauss + BB likelihood with Hessian uncertainty ($7^{\rm th}$ row) provides uncertainties that are finally larger than in the asymptomatic limit, but intervals still under-cover for both the nominal and alternate models. The level of under-coverage is similar for the three cases, suggesting that this method can at least provide confidence intervals with uniform coverage.
    
    It has been verified for several pseudo-data sets that a numerical minimization of the Poisson or Gaussian + BB likelihoods leads to identical numerical results. %, similarly to what already observed for the infinite MC statistic case.
    Motivated by the practical equivalence between the two, in the following we will concentrate on the latter since it can be minimized more efficiently. % and focus on the Gauss + Barlow-Beeston likelihood.
    
    \item Uncertainties based on either the Hessian matrix ($7^{\rm th}$ row) or the profile-likelihood ratio ($8^{\rm th}$ row) in the Gaussian approximation are similar, showing that for each pseudo-data the log-likelihood function is reasonably parabolic around the minimum. However, the sampling distribution of $t_{\mu}$ is remarkably different from a chi-square with one degree of freedom, as illustrated by figure~\ref{fig:tstat} for $\mu\equiv\mu_1^\prime=0$ on pseudo-data generated at the true value $\mu_{\rm t}=0$. The sampling distributions of the MLE estimator $\hat{\mu}$ and of the estimated Hessian uncertainty $\hat{\sigma}_{\rm H}$ are shown in figure~\ref{fig:hat}. They both deviate visibly from a Gaussian distribution, with the latter featuring a long tail towards the right. Similar results are obtained when considering the other parameterization.
  
    \item Fitting $(\mu_1,\mu_2)$ or $(\mu_1^\prime,\mu_2^\prime)$ leads to qualitatively very similar results in terms of coverage. Numerical differences for $\mu_{\rm t}=1\pm 5\bar{\sigma}_{\rm H}$ are mostly due to the different definition of alternate data, see e.g. eq.~\eqref{eq:altmodelcorr} and eq.~\eqref{eq:transf2}. Hence, we conclude that under-coverage occurs independently from the parametrization being used.
\end{enumerate}

The main conclusion of this study is that CI-setting methods based on the asymptotic properties of the MLE or PLR are clearly invalidated even though the sample sizes are large in an absolute sense for both data and simulation. This motivates us to go beyond the asymptotic approximation.

\subsection{Beyond the asymptotic approximation}

We have considered the following CI-setting methods inspired by the classical Feldman and Cousins (FC) unified approach~\cite{FC}. The main feature common to all methods is their use of the profile-likelihood ratio as an ordering principle with no {\it priori} assumption on its statistical distribution. The latter is indeed determined {\it a posteriori} from a collection of randomly generated pseudo-experiments. At variance with the original Feldman-Cousins formulation, the main challenge here is represented by the presence of nuisance parameters.      

\begin{itemize}
\item {\bf Profiled Feldman-Cousins} \\
If the asymptotic approximation is not valid, the profile-likelihood ratio $t_{\mu}$ in eq.~\eqref{eq:plr} is not necessarily chi-square distributed and the coverage properties of eq.~\eqref{eq:Iplr} are not guaranteed. In the so-called profiled Feldman-Cousins approach~\cite{acero2024}, a best guess of the nuisance parameters $\hat{\boldsymbol{\theta}}_\mu$ is first obtained by minimizing the NLL function at a fixed value of $\mu$. Then, for each point $\mu$ sampling the range of values that we wish to include/exclude from the confidence interval, an ensemble of so-called Feldman-Cousins pseudo-experiments~\cite{FC} are generated by drawing event counts $y_i$ around $f_i(\mu, \hat{\boldsymbol{\theta}}_\mu)$. The sampling distribution $P_{\rm FC}$ of $t_\mu$ is then used to determine a critical value $c_\mu(\alpha)$ at significance level $\alpha$, such that
\begin{equation}\label{eq:FC}
\int_{0}^{c_\mu(\alpha)} dt_\mu \, P_{\rm FC}(t_\mu) = \alpha. 
\end{equation}  
The confidence interval $I_{\rm FC}$ is defined as the set of points for which $t_\mu$ is less than the critical value $c_\mu(\alpha)$. 
Here, we are interested in understanding whether an interval based on eq.~\eqref{eq:FC} would contain $\mu_{\rm t}$ with the right probability for at least the nominal and alternate toy models. For this purpose, we determine the critical values $c_\mu(\alpha)$ for $\mu_k\in\{1, 1\pm5\bar{\sigma}_{{\rm H}}\}$ and evaluate the coverage for pseudo-data distributed according to $\mu_{\rm t}=\mu_k$. % and compute the probability for $t_\mu$ to be smaller than the corresponding critical values.
Since we mostly care about coverage, we don't perform the full Feldman-Cousins procedure to determine an exact confidence interval for each pseudo-experiment, which would be very time demanding, but follow instead a simplified procedure, namely, for the true and alternate models we define the confidence interval as
\begin{equation}
\{ \mu \, : \, t_\mu \leq c_{\mu_k} \}, \;\;\; {\rm with} \;\;\; \mu_k \in \{ 1, 1\pm5\bar{\sigma}_{{\rm H}}\}.
\label{eq:IFC}
\end{equation} 
This construction leads to what we denote by {\it profiled FC} intervals. Notice that eq.~\eqref{eq:IFC} is not equivalent to the full Feldman-Cousins construction because the critical values $c_\mu$ are not computed for all possible values of $\mu$. The two definitions would eventually lead to the same intervals if the sampling distribution $P_{\rm FC}$ were independent from $\mu$. % offers a useful definition to estimate e.g the interval size.

\item {\bf Simplified Feldman-Cousins} \\
We considered an alternative construction of the FC interval where the best-fit estimators $(\hat{\mu},\hat{\boldsymbol{\theta}})$ are first determined for all parameters, including the POI; then, an ensemble of FC pseudo-experiments are generated by drawing Poisson distributed event counts around $f_i(\hat{\mu}, \hat{\boldsymbol{\theta}})$; finally, the sampling distribution of $t_\mu$ and critical value $c_\mu$ are computed for $\mu=\hat{\mu}$. The resulting interval, which will be denoted as {\it simplified FC}, is then defined similarly to eq.~\eqref{eq:IFC} with $\mu_k=\hat{\mu}$.

\item {\bf Profile-likelihood-ratio with Barlett's correction}\\
In case the asymptotic approximation is not valid, it has been suggested by Barlett~\cite{Barlett,CowanUnc} to modify the profile-likelihood ratio $t_\mu$ such that its expectation value equals its asymptotic value of one (for one degree of freedom), that is $t_\mu \to t^\prime_\mu = t_\mu/\langle t_\mu \rangle$. Thus, starting from the sampling distributions obtained from the FC pseudo-experiments, $\langle t_\mu \rangle$ is determined for $\mu\in\{1, 1\pm 5\bar{\sigma}_{{\rm H}}\}$, and the corresponding confidence interval defined as
\begin{equation}
 \{ \mu \, : \, t^\prime_\mu\leq 1 \}.
\label{eq:IFCBarlett} 
\end{equation} 

\item {\bf Cousins-Highlands}\\
In the profiled FC approach, all nuisance parameters, including those related to the MC statistical uncertainties, are fixed to their best-fit to the data, but their post-fit uncertainty is neglected when the FC pseudo-experiments are generated. This can be partly accounted for by following an hybrid Bayesian-frequentist approach first proposed by Cousins and Highland~\cite{COUSINS1992331}. In the Cousins-Highland (CH) approach, the bin contents $y_i$ in each FC pseudo-experiment are drawn from different central values $f_i( \mu, \tilde{\boldsymbol{\theta}}_\mu)$, where $\tilde{\boldsymbol{\theta}}_\mu$ is sampled from a multivariate normal distribution with mean $\hat{\boldsymbol{\theta}}_\mu$ and covariance matrix given by the {\it a posteriori} covariance matrix of $\hat{\boldsymbol{\theta}}_\mu$. The resulting sampling distribution $P_{\rm HC}$ differs from $P_{\rm FC}$, generally leading to more conservative intervals.
%We define the confidence interval accordingly as:
%\begin{equation}
%I_{\rm FH} = \{ \mu \, : \, t_\mu\leq c_{\hat{\mu}} \}.
%\end{equation}\label{eq:IFH}  
\item {\bf Feldman-Cousins ``cheat''}\\
Finally, we considered an ideal, albeit unrealistic, construction of confidence intervals which is still based on the PLR, but with all parameters fixed to their {\it true} values for pseudo-experiments generation. In this case, the sampling distribution of $t_\mu$ from the FC pseudo-experiments matches the sampling distribution of the pseudo-data so that the coverage should come out right by construction. We will use this definition as a reference for the true $1\sigma$ uncertainty. 
\end{itemize}

%\subsubsection{Numerical results}

Using the numerical setup described in section~\ref{sec:CIasy}, we studied the performances of the newly introduced CI-setting methods. All results are reported in tables~\ref{tab:nominal_corr} and~\ref{tab:nominal}. The main observations can be summarized as follows. 
\begin{enumerate}
    \item The profiled Feldman-Cousins approaches ($9^{\rm th}$ and $10^{\rm th}$ rows) provide intervals that under-cover in general, although less than from using the PLR. However, the coverage is found to depend on the value of $\mu_{\rm t}$. As expected, the profiled method ($10^{\rm th}$ row) performs better than the simplified one ($9^{\rm th}$ row), suggesting that the distribution $P_{\rm FC}(t_\mu)$ depends on $\mu$.
    
    \item The Barlett correction to the PLR method ($11^{\rm th}$ row) provides results similar to the profiled FC construction ($10^{\rm th}$ row), showing that the sampling distribution $P_{\rm FC}$ from the FC pseudo-experiments is reasonably close to a chi-square with one degree of freedom. This is also in agreement with the previous observation that intervals based on either the Hessian matrix or the PLR function are almost identical.
 
    \item The Cousins-Highlands method ($12^{\rm th}$ row) leads to the largest uncertainty, but the intervals now over-cover, with a coverage level that depends on $\mu_{\rm t}$.
   
    \item The Feldman-Cousins ``cheat'' method provides the right coverage by construction: the right CI size for $\mu$ would be about twice as large as the one obtained from the Hessian or PLR method on the full Barlow-Beeston likelihood, i.e. what one would have obtained based on the asymptotic properties of maximum-likelihood estimators.
\end{enumerate}

The main conclusion of the study is that none of the realistic CI-setting methods considered here can provide the right coverage for the toy model: the asymptotic properties of the PLR are clearly invalidated by non-negligible corrections implied by Wilks' theorem, which might have been unexpected given the large size of the data sample or the number of bins; on the other hand, methods that do not rely on asymptotic properties suffer from the problem of handling nuisance parameters, most critically, those related to statistical fluctuations of the MC templates.

However, one could reasonably expect that both issues will be solved in the limit of negligible MC statistical uncertainties, and the right coverage will eventually be restored. This can be achieved in multiple ways, the most obvious being by increasing the size of the simulated samples relative to the data or, if possible, by using larger samples for both. Reducing the number of bins might also help. Alternatively, we might conclude that the problem is related to some intrinsic feature of the model. We test these hypotheses by studying alternate model configurations in the next subsection.

\subsection{Recovering the correct coverage}\label{sec:recovery}

The results obtained by using alternate values for $N$, $\epsilon$, $n$, and $k$ are collected in appendix~\ref{app:alt}. The main observations from this study can be summarized as follows.

\begin{enumerate}
    \item {\bf Increasing the size of the MC samples}. Tables~\ref{tab:nominal_lumi10} and~\ref{tab:nominal_lumi40} collect the results obtained by increasing the statistical power of the MC simulation relative to the data ($k$) by factors of $10$ and $40$, respectively. As one would have naively expected, the impact of the limited MC statistics becomes increasingly less relevant as the MC sample size grows compared to the data. For $k=10$, the correct behavior is still not fully achieved, while for $k=40$ all likelihood functions and CI setting methods provide roughly the same intervals and the right coverage is recovered for all cases. This shows that the asymptotic properties of the MLE can be ultimately recovered once the MC template fluctuations become sufficiently small.
    
    \item {\bf Improving the signal-to-background separation}. The results reported in tables~\ref{tab:nominal_asy0p03} and~\ref{tab:nominal_asy0p06} have been obtained by choosing $\epsilon=0.06$ and $\epsilon=0.12$, respectively. As the value of $\epsilon$ increases, the model becomes better behaved, as also reflected by a reduction in the condition number of the Hessian matrix. The coverage of the Barlow-Beeston likelihood with Hessian or PLR uncertainties improves, but the correct coverage cannot be recovered even for $\epsilon\sim {\cal O}(10\%)$. The level of under-coverage from the FC methods remains stable. %Thus, it seems that methods based on the post-fit distribution of $t_\mu$ remain sensitive to bin-by-bin fluctuations in the post-fit templates, regardless of conditioning.
    
    \item {\bf Reducing the number of bins}. The results reported in tables~\ref{tab:nominal_bins100} and~\ref{tab:nominal_bins20} have been obtained by rebinning the histograms by factors of $2$ and $10$, respectively. As $n$ gets smaller, the coverage of the Hessian- or PLR-method improves steadily, but the intervals slightly under-cover even for $n=20$. As expected, fluctuations of MC templates across bins tend to average out with a coarser binning. Again, the level of under-coverage from the FC method does not change significantly.
    
    \item {\bf Increasing the size of both data and simulation}. The results in tables~\ref{tab:nominal_x10} and~\ref{tab:nominal_x100} have been obtained by increasing the number of events in data by factors of $10$ and $100$, respectively, while keeping the same proportion between data and simulation. For sufficiently large values of $N$, the asymptotic limit is ultimately reached when using the full Barlow-Beeston likelihood function~\eqref{eq:logPBB}, so that CI-setting methods based on the Hessian uncertainty or the PLR can finally provide the right coverage. The FC-methods still suffer from systematic under-coverage.
\end{enumerate}

\subsection{Discussion}

The main results of the toy study are summarized in Table~\ref{tab:summary_toy}. The correct likelihood function for the problem under consideration is given by eq.~\eqref{eq:logPBB}. Neglecting the stochastic nature of the MC templates $t_{ji}$, as done, for example, in eq.~\eqref{eq:logP}, has a sizable impact in terms of both the coverage and the {\it size} of confidence intervals already with an equal amount of data and simulated events. The ``Barlow-Beeston lite'' or ``MC stat.'' approaches treat the uncertain MC predictions as an additional source of statistical noise to the data. This accounts in part for the random fluctuations of MC templates, but is not sufficient to restore the right coverage or interval size, as demonstrated by the difference between using the lite vs. full Barlow-Beeston methods. When using the correct likelihood function in the Gaussian approximation, the asymptotic properties of maximum-likelihood estimators and of the profile-likelihood ratio are manifestly broken even if the number of events per bin is large enough to justify, for example, the approximation of the NLL as a chi-square function. We found that the asymptotic properties can eventually be recovered by applying a sufficiently large scaling of the data and/or simulation sizes, but we could not find a first-principle indication of which scale can be considered sufficiently ``large''.
%This suggests that statistical fluctuations of the templates can artificially augment the sensitivity to the POI, leading to both biased estimators and Hessian uncertainties that are smaller than the actual spread of the MLE.

Approaches based on the post-fit distribution of the profile-likelihood ratio in the Guassian approximation suffer from a related problem: the post-fit nuisance parameters, in particular those related with the fluctuations of MC templates, are never exactly set to their true value. The footprint of the fluctuated MC templates survives in the post-fit distribution of $t_\mu$, leading to an artificially augmented sensitivity to $\mu$. We remark that the problem of under-coverage of confidence intervals in the presence of nuisance parameters is well-known already~\cite{conrad2002,acero2024,Punzi,Cranmer}.
It can be ultimately ascribed to the fact that the standard Neyman's construction~\cite{1937RSPTA.236..333N} for confidence intervals on one parameter of interest is not uniquely defined when the likelihood depends on multiple nuisance parameters, since some prescription on how to project a larger multi-dimensional space onto a one-dimensional closed interval must be supplied~\cite{Punzi,Cranmer}. Approaches based on the profile-likelihood function are just instances of the infinitely many ways to achieve such a dimensional reduction. While this matter has been mostly addressed in low-statistics counting experiments and for relatively simple problems, we have shown that the problem can also arise in high-statistics measurements.

\subsubsection{An heuristic confidence interval}

Motivated by these considerations, we propose an {\it heuristic} confidence interval which combines Hessian uncertainties in the limit of infinite-size MC samples with those obtained from two variants of the Barlow-Beeston method. More precisely, we define
\begin{equation}
I_{\rm heur} = \left[ \hat{\mu} - \hat{\sigma}_{\rm heur}, \; \hat{\mu} + \hat{\sigma}_{\rm heur} \right],
\label{eq:Iheur} 
\end{equation}
where $\hat{\mu}$ is the maximum-likelihood estimator obtained by minimizing the full Barlow-Beeston likelihood from eq.~\eqref{eq:logPBB}, while $\hat{\sigma}_{\rm heur}$ is defined as
\begin{equation}
\hat{\sigma}_{\rm heur} = \bar{\sigma}_{{\rm H}} \cdot \sqrt{1 + \frac{1}{k} } \cdot \left( \frac{ \hat{\sigma}_{\rm H}^{\rm BBfull} }{\hat{\sigma}_{\rm H}^{\rm BBlite}} \right),
\label{eq:Iheur2}
\end{equation}
where $\bar{\sigma}_{{\rm H}}$ is the Hessian uncertainty in the limit of infinite MC statistics, $k$ is the statistical power of the MC simulation relative to the data, and $\hat{\sigma}_{\rm H}^{\rm BBfull}$ ($\hat{\sigma}_{\rm H}^{\rm BBlite} $) is the Hessian uncertainty derived from the full (lite) Barlow-Beeston likelihoods. We remark the importance of using the maximum-likelihood estimator $\hat{\mu}$ based on the {\it full} Barlow-Beeston likelihood rather than its lite approximation, since the latter does not provide an unbiased estimator of $\mu$.

The rationale behind eq.~\eqref{eq:Iheur2} is the following. The Hessian uncertainty derived from the full Barlow-Beeston likelihood underestimates the real statistical spread of $\hat{\mu}$ due to the presence of spurious constraints which break the asymptotic approximation. Conversely, the Hessian uncertainty $\bar{\sigma}_{{\rm H}}$ would match the standard deviation of $\hat{\mu}$, {\it if the MC templates were exact}. We take it as our starting point. Our intuition is that statistical uncertainties on the MC templates should at least add in quadrature to the Poisson uncertainty of the data, which corresponds to the square root in eq.~\eqref{eq:Iheur2}. However, assigning an overall uncertainty to the total MC prediction per bin is not fully correct as it ignores the interplay between the $\mu$ parameter and MC statistical fluctuations pertaining to different templates. In fact, the existence of such an interplay is made evident by the breaking of the Barlow-Beeston lite approximation. The ratio between Barlow-Beeston {\it full} and {\it lite} Hessian uncertainties, corresponding to the last term on the right-hand side of eq.~\eqref{eq:Iheur2}, provides an effective correction factor to account for the increased statistical uncertainty when the MC templates $\theta_{ji}$ are allowed to float in the fit independently from each other.

This heuristic definition ($14^{\rm th}$ row of tables~\ref{tab:nominal_corr} and~\ref{tab:nominal}) provides good coverage properties for all values of $\mu_{\rm t}$, as well as for different choices of model configurations. This suggests that $\hat{\sigma}_{\rm heur}$ could be at least used as a proxy for the real statistical spread of the $\hat{\mu}$ estimator in the presence of fluctuations of both the data and the simulated samples, and so it provides a practical test to gauge the goodness of asymptotic confidence intervals. This observation is intriguing and perhaps worth being further investigated, although a rigorous proof of its validity for a generic problem is beyond the scope of this work. We should remark, however, that $\hat{\sigma}_{\rm heur}$ is defined in terms of $\bar{\sigma}_{\rm H}$, which for a realistic problem might not be known {\it a priori}. In sec.~\ref{sec:scaling} we will address the problem of estimating $\bar{\sigma}_{\rm H}$ when only finite-size MC samples are available.

Two remarks are worth making here. First, we note that eq.~\eqref{eq:Iheur2} might prove difficult to compute for realistic models that are more complex than our simple toy model, especially when the likelihood function has to handle correlations between different templates. In these cases, more sophisticated solutions should be found. Second, we have restricted our discussion to confidence intervals in one dimension, namely for a single parameter of interest, although in some cases one might be interested in setting confidence intervals in more dimensions. How to assess (and possibly correct) under-coverage in more dimensions is beyond the scope of this work.\\

In the remainder of the paper, we will study the problem of under-coverage of asymptotic confidence intervals from a more general perspective with the goal of identifying the origin of the observed effect. 

\begin{table}
\footnotesize
    \centering
\begin{tabular}{llccccc}
\hline
 MC stat. & CI & Nominal & $k=40$ & $\epsilon=0.12$ &  $n=20$ & $N=2\times 10^8$ \\
\hline
\hline
none & Hessian & $\times$ & $\checkmark$ & $\times$ & $\times$ & $\times$\\
MC stat. & Hessian  & $\times$ & $\checkmark$ & $\times$ & $\times$ & $\checkmark$\\
BB-lite & Hessian  & $\times$ & $\checkmark$ & $\times$ & $\times$ & $\checkmark$\\
BB &  Hessian/PLR & $\times$ & $\checkmark$ & $\checkmark$ & $\times$ & $\checkmark$\\
BB & PFC/Barlett & $\times$ & $\checkmark$ & $\times$ & $\times$ & $\times$\\
BB & CH & $\times$ & $\checkmark$ & $\checkmark$ & $\times$ & $\checkmark$\\
BB & FC Cheat & $\checkmark$ & $\checkmark$ & $\checkmark$ & $\checkmark$ & $\checkmark$\\
BB & Heuristic & $\checkmark$ & $\checkmark$ & $\checkmark$ & $\checkmark$ & $\checkmark$\\
  \hline
\end{tabular}
\caption{Summary of the coverage properties for different model configurations. A thick (cross) indicates that the right coverage, within a relative $\pm5\%$ tolerance, is (not) attained for both the nominal and alternate data sets. In this table, ``BB'' is a shorthand for Barlow-Beeston.}
    \label{tab:summary_toy}
\end{table}

\section{The general case}\label{sec:general}

The main scope of this section is to characterize in more generality the role played by statistical fluctuations of MC templates in the determination of confidence intervals constructed in the asymptotic approximation. For this purpose, the following approach is followed.

We start by assuming that the MC templates are {\it exact}. This approximation allows us to derive a set of well-known analytic results for point estimators and confidence intervals based on the asymptotic properties of maximum-likelihood estimators~\cite{Behnke}. The same formalism can also accommodate extensions of the likelihood function that account for finite-size effects in an approximate way, such as the Gauss + MC stat. or Barlow-Beeston {\it lite} method as discussed in section~\ref{sec:toy}, which are in practice equivalent to an overall rescaling of the statistical power of the data. Albeit approximate, these methods are still relevant in light of their diffusion in HEP analyses. We will then treat statistical fluctuations of MC templates as perturbations on top of the true unknown templates and collect their contributions to asymptotic formulas as an overall bias term, whose sign and magnitude will then be the main subject of study.

We remark that our treatment is not directly applicable to the {\it full} Barlow-Beeston method, because the introduction of additional nuisance parameters (associated with the individual MC predictions) breaks the quadratic dependence of the NLL on the parameters, which is a crucial assumption in our derivation. For example, in the context of the toy study discussed in section~\ref{sec:toy}, the implementation of the full Barlow-Beeston method has been found to affect the coverage of asymptotic confidence intervals in a significant way, although still insufficient to restore the right coverage. We will come back to the validity of the Barlow-Beeston lite approximation at the end of this section. We can argue, however, that the main results derived in this section are expected to be relevant even for the full Barlow-Beeston method. Indeed, the main source of under-coverage that will be identified by our study is related to the occurrence of {\it low-statistics} regimes along particular linear combinations of the data, represented in this context by a finite number of event counts, despite the latter being large in absolute: in such situations, asymptotic formulas for maximum-likelihood estimators are likely to be just not applicable even if the exact likelihood functions were considered.

\subsection{Methodology}\label{sec:method}

Consider a binned data set represented by independent Poisson-distributed event counts ${\bf y}$ with expectation value parametrized by the functional model ${\bf f}(\mu, \boldsymbol{\theta})$, where $\mu$ is the parameter of interest and $\boldsymbol{\theta}=(\theta_1, \hdots, \theta_p)$ is a vector of nuisance parameters, which are assumed to be not constrained by auxiliary measurements. The generalization to the case where both constrained and unconstrained parameters are present is discussed in~\ref{app:additional_constr}, where an interesting connection between the present work and the main subject of ref.~\cite{Cowan:2018lhq} is pointed out.

As discussed previously, in the limit of large event counts, that is, for $y_i\gg1$ for $i=1,\hdots,n$, the negative log-likelihood function can be approximated as
\begin{align}\label{eq:chi2}
-\ln { L}(\mu, {\boldsymbol{\theta}}) = \frac12 \left({\bf y} - {\bf f}(\mu,\boldsymbol{\theta}) \right)^T {\bf V}^{-1} \left({\bf y} - {\bf f}(\mu,\boldsymbol{\theta}) \right) + \hdots %\equiv \chi^2( \mu,{\boldsymbol{\theta}} )
\end{align}
where ${\bf V}={\rm diag}({\bf y})$, and the dots stand for an overall constant plus terms that are of higher-order in $(y_i-f_i)$.
Aside from a factor of $1/2$, the right-hand side of eq.~\eqref{eq:chi2} coincides the well-known Neyman's $\chi^2$ test-statistic~\cite{Neyman}. We take ${\bf f}$ to be a differentiable function of both $\mu$ and $\boldsymbol{\theta}$ and expand it around a suitable initial point $(\mu_0, \boldsymbol{\theta}_0)$:
\begin{equation}\label{eq:lin}
{\bf f}(\mu, \boldsymbol{\theta}) = 
{\bf f}_{0} + {\bf j}(\mu -\mu_0) + {\bf J} (\boldsymbol{\theta} - \boldsymbol{\theta}_{0}) + \cdots,
\end{equation}
where ${\bf j}$ and ${\bf J}$ are the $n\times1$ and $n\times p$ blocks of the Jacobian matrix of ${\bf f}$ with respect to $\mu$ and ${\boldsymbol{\theta}}$, respectively, and the dots stand for higher-order terms in the Taylor expansion.
While this approximation is exact for linear functions, such as the toy model of section~\ref{sec:toy}, for generic functions it should be iterated around the intermediate minima of eq.~\eqref{eq:chi2}, until convergence\footnote{The validity of the linear approximation is model dependent and should be assessed case by case. In particular, functional models that change in a very asymmetric way around the true value of the parameters (e.g. due to non-negligible contributions from the second-order derivatives) should be handled with care.}. Hence, we will assume that ${\bf j}$ and ${\bf J}$ are evaluated at a point in the parameter space which is sufficiently close to the global minimum of the NLL function. Without loss of generality, we take ${\bf J}$ as a full-rank matrix, otherwise the model should be redefined to remove the degeneracy. 

\subsubsection{The profile-likelihood ratio}\label{sec:llsq}

Thanks to eq.~\eqref{eq:lin}, the function on the right-hand side of eq.~\eqref{eq:chi2} can be minimized by solving a linear system of equations, leading to the well-known {\it linear least-square estimator}~\cite{Cowan}. To this aim, we can first put the right-hand side of eq.~\eqref{eq:chi2} in the more concise form:
 \begin{align}\label{eq:rearrange}
-2 \ln L(\mu, {\boldsymbol{\theta}})  = \lVert {\bf d} - {\bf b}(\mu - \mu_0) - {\bf A} (\boldsymbol{\theta} - \boldsymbol{\theta}_{0}) \rVert^2,
\end{align}
where $\lVert \cdot \rVert$ indicates the Euclidean norm, while the newly introduced quantities are defined as
 \begin{align}\label{eq:rearrange2}
 %{\bf r}_{\mu} & = {\bf d} - {\bf b}(\mu - \mu_0) \\ 
 {\bf d} & = {\bf V}^{-\frac12}({\bf y} - {\bf f}_0)  \\
 {\bf b} & =  {\bf V}^{-\frac12}{\bf j} \\
 {\bf A} & = {\bf V}^{-\frac12} {\bf J}
\end{align}
%For convenience of notation, the vectors ${\bf b}$ and ${\bf d}$ have been introduced.
In the following, we will equivalently refer to ${\bf b}$ and ${\bf A}$ as parts of the {\it Jacobian matrix}, although they formally differ from ${\bf j}$ and ${\bf J}$ for a left multiplication by ${\bf V}^{-\frac12}$. As we shall see, the effect of finite-size MC samples on asymptotic confidence intervals can be fully characterized in terms of ${\bf b}$ and ${\bf A}$.

The NP point that minimizes eq.~\eqref{eq:rearrange} at a fixed value of $\mu$ is provided by the {\it profile likelihood} estimator $\hat{ \boldsymbol{\theta} }_{\mu}$. By inserting the expression for $\hat{ \boldsymbol{\theta} }_{\mu}$ into eq.~\eqref{eq:rearrange}, the log-likelihood function becomes a quadratic function of $\mu$, which can be again minimized yielding the maximum-likelihood estimator $\hat{\mu}$. Explicit formulas for $\hat{ \boldsymbol{\theta} }_{\mu}$ and $\hat{\mu}$ are provided in~\ref{app:additional_hessian}.
After some straightforward calculations, we can finally write the profile log-likelihood function\footnote{We remark that Eq.~\eqref{eq:chi2min} would be the {\it exact} profile-likelihood ratio only in the Gaussian approximation.} as
\begin{equation}\label{eq:chi2min}
 t_\mu  = -2 \ln \frac{ L ( \mu, \hat{\boldsymbol{\theta}}_\mu ) }{ L(\hat{\mu}, \hat{\boldsymbol{\theta}} )}=  {\bf b}^T {\bf U} {\bf b}\cdot (\mu-\hat{\mu})^2,
\end{equation}
where the matrix ${\bf U}$ is defined as
\begin{equation}
{\bf U}=  {\bf 1} - {\bf A} ( {\bf A}^T {\bf A} )^{-1}  {\bf A}^T.
\label{eq:defU}
\end{equation}
According to Wilks' theorem~\cite{Wilks}, $t_\mu$ is asymptotically distributed as a chi-square with one degree-of-freedom, so that the quadratic form
\begin{equation}\label{eq:sigmaH}
S = {\bf b}^T {\bf U} {\bf b} = {\bf b}^T {\bf b} - {\bf b}^T  {\bf A} ( {\bf A}^T {\bf A} )^{-1}  {\bf A}^T{\bf b} = \sigma_{\rm H}^{-2}
\end{equation}
must be equal to the inverse variance of $\hat{\mu}$. We notice that $\sigma_{\rm H}$ can be equivalently obtained from the diagonal elements of the covariance matrix ${\bf C}$, defined as the inverse-Hessian matrix of the NLL function, see eq.~\eqref{eq:hessian}. A proof of this statement, together with an explicit expression of ${\bf C}$ in terms of ${\bf b}$ and ${\bf A}$, can be found in~\ref{app:additional_hessian}. 

The matrix ${\bf U}$ introduced in eq.~\eqref{eq:defU} is a $n\times n$ symmetric, positive semi-definite, and {\it idempotent} matrix\footnote{A symmetric square matrix ${\bf U}$ is said to be idempotent if ${\bf U}^T{\bf U}={\bf U}={\bf U}^T$.}. The same properties pertain to the matrix ${\bf 1} - {\bf U}$, where ${\bf 1}$ denotes the $n\times n$ identity matrix. We exploit the fact that an idempotent matrix can be decomposed as the sum of the outer product of its normalized eigenvectors, and rewrite eq.~\eqref{eq:defU} in the equivalent form
\begin{equation}\label{eq:U}
{\bf U} = {\bf 1} - \sum_{j=1}^{p} {\bf u}_j {\bf u}_j^T,
\end{equation}
where ${\bf u}_j$ are linearly independent vectors of unit norm associated with the $p$ eigenvalues of the square matrix ${\bf A}({\bf A}^T{\bf A})^{-1}{\bf A}^T$, which are all equal to one. It can be easily proved that the ${\bf u}_j$ eigenvectors provide an orthonormal basis to the linear space spanned by the $p$ linearly-independent columns $ {\bf a}_1, \hdots, {\bf a}_p$ of ${\bf A}$. Being associated with degenerate eigenvalues, the vectors ${\bf u}_j$ obtained from the eigen-decomposition of ${\bf 1}-{\bf U}$ are not necessarily orthogonal, but can always be made such by using, for example, the Gram-Schmidt process, so they can be taken as mutually orthogonal with no loss of generality. %This also means that the choice of these vectors is arbitrary. 

The formalism developed so far shows that the Hessian uncertainty $\sigma_{\rm H}$ is simply related to the norm of the projection of the vector ${\bf b}$ onto the space {\it orthogonal} to the linear space spanned by the columns of ${\bf A}$. For brevity, we will denote the latter by ${\cal V}_{\bf A}$. Given that the ${\bf u}_j$ vectors provide a complete basis of ${\cal V}_{\bf A}$, the ${\bf b}$ vector can be always decomposed as
\begin{equation}\label{eq:decompb}
{\bf b} = \sum_{j=1}^p \left( {\bf b}^T{\bf u}_j \right) {\bf u}_j + {\bf b}_{\perp},    
\end{equation}
where ${\bf b}_{\perp}$ is vector of dimension $n$ orthogonal to ${\cal V}_{\bf A}$, i.e. ${\bf b}^T_{\perp}{\bf u}_j=0$ for $j=1,
\hdots,p$. It is convenient to consider two limiting cases, namely:
\begin{itemize}
\item the ${\bf b}$ vector has a {\it large projection} onto ${\cal V}_{\bf A}$, that is $\lVert {\bf b}_{\perp} \rVert \ll \lVert {\bf b}\rVert $;
\item the ${\bf b}$ vector is {\it almost orthogonal} to ${\cal V}_{\bf A}$, so that $\lVert {\bf b}_{\perp} \rVert \approx \lVert {\bf b}\rVert $.
\end{itemize}
The extent to which a given problem can be classified according to either of the two cases can be quantified by the positive-definite metric $\rho_\mu$ defined as\footnote{It can be proved that $\rho_\mu^2$ is equal to the so-called {\it global correlation} coefficient, which is used, for example, in Minuit~\cite{JAMES1975343} for fit diagnostics. See~\ref{app:additional_hessian} for further details.}
\begin{align}\label{eq:globalrho1}
\rho_\mu = \sqrt{\frac{   {\bf b}^T{\bf A} ( {\bf A}^T {\bf A} )^{-1}  {\bf A}^T {\bf b}  }{{\bf b}^T{\bf b}}},
\end{align}
which is always less or equal than one, so that the two cases would correspond to either $\rho_\mu\to 1 \; {\rm or}\; 0$, respectively. 

\subsection{Impact of MC fluctuations on $\sigma_{\rm H}$ }\label{sec:hessuncMCfinite}

\begin{figure}
    \centering
    \includegraphics[width=0.7\linewidth]{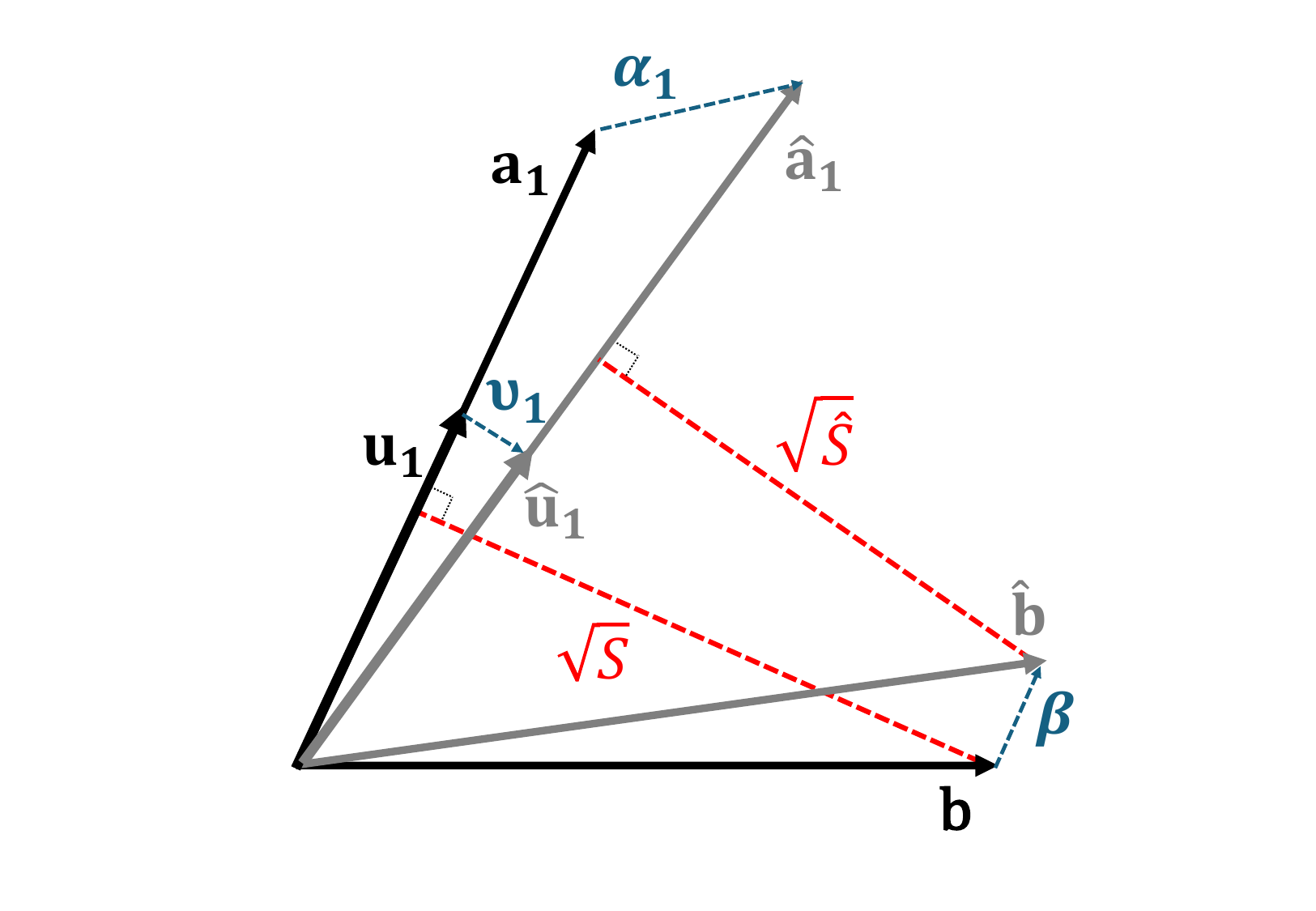}
    \caption{A sketch showing the effect of statistical fluctuations on the relevant quantities introduced in the text. The simple case $n=2$ and $p=1$ has been considered.}
    \label{fig:simple1}
\end{figure}

The results presented in section~\ref{sec:llsq} assume that the model ${\bf f}$ is known exactly. Suppose instead that ${\bf f}$ is determined from MC samples consisting of finite numbers of events, thus resulting in perturbed estimators $\hat{\bf b}={\bf b}+\boldsymbol{\beta}$ and $\hat{\bf A}$, with the latter giving rise to a perturbed matrix $\hat{\bf U}$, defined as in eq.~\eqref{eq:defU} with the replacement ${\bf A}\to\hat{\bf A}$. Since $\hat{\bf U}$ must also be idempotent, it admits an expansion in terms of a new basis of eigenvectors $\hat{\bf u}_j = {\bf u}_j + \boldsymbol{\nu}_j$. By expanding the {\it perturbed} value $\hat{S}$ around its true value $S$, and taking the expectation value in the {\it space of MC predictions}, that is the mean value of $\hat{S}$ if infinitely many identical replicas of the MC samples could be generated, we obtain:
\begin{align}\label{eq:U3terAvg}
\langle \hat{S} \rangle & = {S} + 2\langle \boldsymbol{\beta}^T \rangle {\bf U}{\bf b} +{\rm Tr} \left[ {\bf U} \langle \boldsymbol{\beta}\boldsymbol{\beta}^T \rangle \right]    \nonumber \\
& - 2\sum_{j=1}^p \left({\bf b}^T {\bf u}_j \right) \left( {\bf b}^T \langle \boldsymbol{\nu}_j\rangle \right) - {\bf b}^T\left( \sum_{j=1}^p \langle \boldsymbol{\nu}_j\boldsymbol{\nu}_j^T \rangle\right) {\bf b} \nonumber \\
&  - 2 \sum_{j=1}^p \left( {\rm Tr} \left[ \langle \boldsymbol{\nu}_j \boldsymbol{\beta}^T\rangle \, {\bf u}_j{\bf b}^T\right] +  {\rm Tr} \left[ \langle \boldsymbol{\nu}_j \boldsymbol{\beta}^T\rangle\right]   {\rm Tr} \left[ {\bf u}_j{\bf b}^T \right] \right) + \hdots.
\end{align}
The derivation of eq.~\eqref{eq:U3terAvg} from a perturbative expansion of $\hat{S}$ is presented in~\ref{app:additional_Avg}. Figure~\ref{fig:simple1} helps visualize the geometrical interpretation of ${\hat S}$, and the effect of statistical fluctuations on ${\bf A}$ and ${\bf b}$ for a representative problem with $n=2$.

The correlation terms on the last line of eq.~\eqref{eq:U3terAvg} are model dependent, so additional assumptions are needed to proceed. We notice that in many cases the columns of the Jacobian of ${\bf f}$ with respect to its $p+1$ parameters can be computed independently from each other. For example, this applies when $\mu$ and $\boldsymbol{\theta}$ are strength modifiers relative to templates obtained from independently simulated processes, which is, for example, the case considered by Barlow and Beeston in their original work~\cite{BARLOW1993219}. Conversely, there is at least another relevant case in which the assumption of uncorrelated templates might seem inappropriate, that is when nuisance parameters are associated with systematic uncertainties estimated from event-by-event re-weightings of the ``nominal'' MC sample. In this case, correlations between the columns of ${\bf A}$ cannot be ignored {\it a priori} because of correlated Poisson fluctuations of the MC sample. However, if the dominant source of statistical uncertainty on the MC prediction originates from the variance of the event weights rather than from the purely Poisson uncertainty, different re-weightings of the same MC sample could be still treated as approximately independent. To proceed further, we make the assumption that the covariance matrix $\langle \boldsymbol{\nu}_j \boldsymbol{\beta}^T\rangle$ is zero and consider the remaining terms in eq.~\eqref{eq:U3terAvg}.

\subsection{Contribution from the ${\bf A}$ matrix}

Let us start by analyzing the bias terms in eq.~\eqref{eq:U3terAvg} related to fluctuations of the ${\bf A}$ matrix.
The expectation values $\langle \boldsymbol{\nu}_j \rangle$ are generally different from zero. Indeed, from the normalization condition $\lVert \hat{\bf u}_j\rVert =1$ it follows that
\begin{equation}\label{eq:nujnorm}
2{\bf u}_j^T  \langle\boldsymbol{\nu}_j \rangle = -\langle\boldsymbol{\nu}_j^2\rangle \neq 0.
\end{equation}
We notice that the effect of statistical fluctuations on the elements of $\hat{\bf A}$ is to stretch and rotate its column vectors. However, $S$ cannot depend on their normalization, since a global re-scaling of the columns can always be absorbed by a redefinition of their associated nuisance parameter\footnote{This can be also proved by noticing that the matrix ${\bf U}$ is invariant under a right multiplication of ${\bf A}$ by an invertible matrix $\boldsymbol{\Lambda}$, that is ${\bf A}\to {\bf A}\boldsymbol{\Lambda}$, of which a column-wise re-scaling is a special case.}. Without loss of generality, we can then assume $\lVert{\bf a}_j\rVert=1$ and fix their orientation by requiring e.g. ${\bf a}_j^T{\bf b}\geq0$. Random fluctuations on the normalized vectors ${\bf a}_j$ generate perturbations $\boldsymbol{\alpha}_j$ corresponding to small rotations in a $n$-dimensional space, which must however average to zero when projected onto the sub-space orthogonal to ${\bf a}_j$, so that:
\begin{align}\label{eq:nujsmall}
\langle  \boldsymbol{\alpha}_j \rangle = -\frac{\langle \boldsymbol{\alpha}_j^2 \rangle}{2} {\bf a}_j,
\end{align}
where ${\langle \boldsymbol{\alpha}_j^2 \rangle }$ is the variance of the angle by which the vector ${\bf a}_j$ is rotated. It is easy to see that ${\langle \boldsymbol{\alpha}_j^2 \rangle }$ must be of order $n\cdot\Delta_j^2$, with $\Delta_j$ being the average relative MC uncertainty on the elements of ${\bf a}_j$. Under the same assumptions, one can easily prove that $\langle \boldsymbol{\alpha}_j\boldsymbol{\alpha}_j^T\rangle$ must be a matrix that projects onto the subspace orthogonal to ${\bf a}_j$ with equal probability along each of its $n-1$ directions, that is
\begin{align}\label{eq:nujnujTsmall}
\langle  \boldsymbol{\alpha}_j \boldsymbol{\alpha}_j^T \rangle =  \frac{\langle \boldsymbol{\alpha}^2_j \rangle}{n-1} \left({\bf 1} - {\bf a}_j{\bf a}_j^T\right).
\end{align}
By means of eq.~\eqref{eq:nujsmall} and~\eqref{eq:nujnujTsmall}, the right-hand side of eq.~\eqref{eq:U3terAvg} can be expanded further.

In the following, we will restrict to the case ${\bf a}_j^T{\bf a}_k=\delta_{jk}$, that is the column vectors of the matrix ${\bf A}$ are orthogonal to each other.
%\footnote{This also implies a diagonal matrix ${\bf W}$.}.
This hypothesis makes the analytic treatment simpler and its interpretation more intuitive, without precluding the general validity, as discussed in~\ref{app:additional_general}. We also notice that this special case has some physical interest: for example, it is appropriate when nuisance parameters parametrize systematic effects that behave in an almost ``orthogonal'' way in the space of observables ${\bf y}$, for example when they are determined from a PCA analysis~\cite{Pearson01111901,Hotelling}. 

Under the assumption of orthonormal column vectors, we can exploit the arbitrariness in the definition of the ${\bf u}_j$ vectors by choosing ${\bf u}_j={\bf a}_j$.
The second line on eq.~\eqref{eq:U3terAvg} can be expanded by means of eq.~\eqref{eq:nujsmall} and~\eqref{eq:nujnujTsmall}, obtaining after some straightforward calculation:
\begin{equation}\label{eq:extrachi5}
 \frac{n}{n-1}\sum_{j=1}^p
\left( {\bf b}^T {\bf a}_j \right)^2\langle \boldsymbol{\alpha}_i^2 \rangle - \frac{{\bf b}^T{\bf b}}{n-1}\left( \sum_{j=1}^p \langle \boldsymbol{\alpha}_i^2 \rangle \right). 
\end{equation}

Let us first consider the case of a large value of $\rho_\mu$. Aside from numerical factors, the two pieces appearing in eq.~\eqref{eq:extrachi5} are of the same order of magnitude, but of opposite sign. However, the former is scaled by a factor of $n$, thus giving a dominant {\it positive} contribution when the number of bins is large, which is our case of interest. This can be intuitively understood by noticing that when $n\gg p$, statistical fluctuations on ${\bf A}$ have a larger probability of shuffling around the basis vectors by introducing new directions along which ${\bf b}$ can have a finite projection, rather than rotate the vectors inside the same space, which would leave $S$ unchanged. These extra contributions add quadratically to $\lVert{\bf b}_{\perp}\rVert^2$ and can be numerically relevant if the latter is small.
As discussed in~\ref{app:additional_cheb}, the same conclusion can be proved for generic $n$ under some more restrictive assumptions.

We now consider the opposite limiting case, that is $\rho_\mu \to 0$. The first term on the second line of eq.~\eqref{eq:extrachi5} can be now neglected compared to the second term, which is instead always negative. Thus, we see that in this case the effect of random fluctuations of the MC templates is to reduce the value of $\langle \hat{S} \rangle$ relative to ${S}$ by a factor of 
\begin{align}\label{eq:extrachi6}
1- \frac{1}{n-1}\left( \sum_{j=1}^p \langle \boldsymbol{\alpha}_i^2 \rangle \right).
\end{align}
This result is also in line with our intuition: if ${\bf b}$ is already orthogonal to the linear space spanned by the columns of ${\bf A}$, fluctuations on the latter can only generate spurious components of ${\bf b}$ inside the linear space ${\cal V}_{\bf A}$, thus reducing the norm of ${\bf b}_\perp$. However, we notice that the relative decrease implied by eq.~\eqref{eq:extrachi6} is only of order of the squared relative uncertainty on the elements of $\tilde{\bf A}$, and is not scaled by $n$, so its overall impact might well be numerically less relevant than the effect discussed in the previous case.

For intermediate values of $\rho_\mu$, competition between positive and negative bias terms to $\langle \hat{S} \rangle$ does not allow for a conclusive statement on the overall sign, and the bias term should be determined case by case. This can be well illustrated, for example, in a two-dimensional setting, as discussed in detail in~\ref{app:additional_2D}. In this simplified setup, the ratio $\langle{\hat S}\rangle/S$ is found to depend on $\rho_\mu$ as expected from eqs.~\eqref{eq:extrachi5} and~\eqref{eq:extrachi6}.

\subsection{Contribution from the ${\bf b}$ vector}

Finally, we consider the role of statistical fluctuations on the ${\bf b}$ vector. Since the MC prediction for $\hat{\bf b}$ is randomly distributed around the true value ${\bf b}$, the expectation value of $\boldsymbol{\beta}$ is zero, so that terms linear in $\langle \boldsymbol{\beta} \rangle$ do not contribute to the bias term in eq.~\eqref{eq:U3terAvg}.
The quadratic term on the first line of eq.~\eqref{eq:extrachi5} is positive-definite\footnote{This can be also seen as consequence of Jensen's inequality applied to $\hat{S}$, which is a convex function of ${\bf b}$.}. If we make the reasonable assumption $\langle \beta_i\beta_j\rangle = \langle \beta^2_i\rangle \delta_{ij}$, that is the MC predictions for different bins are statistically uncorrelated, the quadratic term can be further expanded as
\begin{equation} \label{eq:extrab}
{\rm Tr}\left[ {\bf U} \, \langle \boldsymbol{\beta} \boldsymbol{\beta}^T\rangle \right] =n \left(\sum_{i=1}^n\frac{ 1}{n} \langle \beta_i^2\rangle \right) - \sum_{j=1}^p \left( \sum_{i=1}^n u_{j,i}^2 \langle \beta_i^2\rangle\right),
\end{equation}
where the two terms in parentheses correspond to different ways of averaging the bin-by-bin variances $\langle \beta_i^2\rangle$. If the distribution of the latter is narrowly concentrated around an average value $\overline{\langle \beta^2\rangle}$, the right-hand side of eq.~\eqref{eq:extrab} is of order $\sim (n-p)\overline{\langle \beta^2\rangle}$, so it grows linearly with the number of bins. Thus, we see that statistical fluctuations on ${\bf b}$ can induce sizable positive biases to $\langle \hat{S} \rangle$ for any value of $\rho_\mu$, provided that $(n-p)\overline{\langle \beta^2\rangle}$ is comparable to $S$. %Again, large values of $\rho_\mu$ can make the impact of eq.~\eqref{eq:extrab} relatively more significant.\\

\subsection{Discussion and connection with the toy model}

We now summarize the assumptions and the main implications of eq.~\eqref{eq:U3terAvg}. We have assumed that the MC estimators for ${\bf b}$ and ${\bf A}$ are independent from each other and that statistical fluctuations of their elements can be treated as Gaussian-distributed noise around their true values. We have considered separately the impact of MC fluctuations on the blocks ${\bf A}$ and ${\bf b}$ of the Jacobian matrix in the form of a bias term to the quadratic form $S$, defined by eq.~\eqref{eq:sigmaH}. If ${\bf b}$ has a large overlap with the linear space spanned by the columns of ${\bf A}$, or, equivalently, when $\rho_\mu$ is close to one, then the bias terms associated with fluctuations of the ${\bf A}$ matrix are positive-definite and increase with the number of bins $n$. Fluctuations of the ${\bf b}$ vector also grow with $n$ and are always positive-definite. In both cases, $ \hat{S}$ would differ from the true value $S$ by a positive offset, implying that the Hessian uncertainty, estimated as $\hat{\sigma}_{\rm H}=\hat{S} ^{-1/2}$, would be underestimated\footnote{More correctly, a positive bias on $\hat{S}$ would also imply a negative bias on $\hat{\sigma}_{\rm H}$ if the variance of $\hat{S}$ is small, so that $\langle \hat{S}^{-1/2} \rangle \approx \langle \hat{S} \rangle^{-1/2}$. In particular, by Taylor expansion, we find that $\frac{3}{8}\frac{ \langle \hat{S}^2 \rangle - \langle \hat{S} \rangle^2 }{\langle\hat{S}\rangle^2} \ll1 $ should hold, a condition that is more likely to be satisfied when the number of bins is large, as for the law of large numbers.}. 

These findings can finally explain the observations made in the context of the toy model of sec.~\ref{sec:toy}: when the original parametrization is considered (i.e. $\mu=\mu_1$), the linear correlation coefficient $\rho=\rho_\mu$ is indeed close to unity, implying a large bias term from fluctuations of the matrix ${\bf A}=(T_{21}/\sqrt{y_1},\hdots,T_{2n}/\sqrt{y_n})^T$; for $\mu=\mu_1^\prime$, instead, the smaller correlation reduces the impact of statistical fluctuations on ${\bf A}$, and the bulk of finite-size MC effects are now propagated to $\hat{\sigma}_{\rm H}$ through a noisy vector ${\bf b} = (T_{11}^\prime/\sqrt{y_1},\hdots,T_{1n}^\prime/\sqrt{y_n})^T$.     

Remarkably, none of the bias terms appearing in eq.~\eqref{eq:extrachi5} is directly proportional to the true value $S$ by a naive factor of ${\cal O}(1)\cdot k^{-1}$, with $k$ related to the effective statistical power of the simulation relative to the data. The lack of proportionality also means that the bias and true value can stand in a ratio which is {\it a priori} arbitrary. This is one of the main observations of this work. Incidentally, this also explains how the Barlow-Beeston lite approach~\cite{BBlite} can be insufficient to account for MC uncertainties. As discussed in sec.~\ref{eq:approx}, the lite version of the Barlow-Beeston likelihood is practically equivalent to a rescaling of the Poisson variance of the data by a factor of $(1+k^{-1})$, implying an {\it increase} of the Hessian uncertainty $\hat{\sigma}_{\rm H}$ by the square root of the same factor. However, a bias in $\hat{S}$, which is both positive and comparable to $S$, might {\it reduce} $\hat{\sigma}_{\rm H}$ by a potentially larger factor, ultimately leading to the non-physical situation of confidence intervals that are smaller than they would be in the limit of infinite-size MC samples.

As a last step in our discussion, we address the question of providing quantitative estimates for the bias terms appearing in eq.~\eqref{eq:U3terAvg}.

\subsection{Asymptotic uncertainties from finite-size samples}\label{sec:scaling}

The bias term generated by statistical fluctuations of the Jacobian matrix could be estimated in principle from eq.~\eqref{eq:U3terAvg} by computing the covariance matrices of ${\bf b}$ and of the column vectors of ${{\bf A}}$, for example by bootstrap resampling, and using their nominal MC predictions $\hat{\bf b}$ and $\hat{\bf A}$ in place of their true, unknown values. The asymptotic Hessian uncertainty $\bar{\sigma}_{\rm H}={S}^{-1/2} $, i.e. the Hessian uncertainty in the limit of infinite MC statistics, could be determined from eq.~\eqref{eq:U3terAvg} after subtracting the bias contribution.

A possibly simpler way to estimate $\bar{\sigma}_{\rm H}$ consists in analyzing the scaling of the Hessian uncertainty $\hat{\sigma}_{\rm H}$ as a function of the MC-to-data ratio $k$ at a fixed size $N$ of the data. Indeed, the following scaling laws can be easily derived:
\begin{align}\label{eq:scaling}
& {\bf b} \sim \sqrt{N}, \;\;\; {\bf U} \sim 1,\;\;\; \langle \boldsymbol{\alpha}_j^2 \rangle \sim \frac{1}{kN}, \;\;\; \langle \boldsymbol{\beta}\boldsymbol{\beta}^T \rangle \sim  \frac{1}{k}, %, \;\;\;  {\bf C}_{ \boldsymbol{\nu}_j  {\boldsymbol{\beta}} } \sim  \frac{\sqrt{N}}{k}.
\end{align}
from which one gets:
\begin{align}\label{eq:scaling2}
\bar{\sigma}_{\rm H} = \hat{\sigma}_{{\rm H}} \cdot  \sqrt{ 1 + \frac{\Delta_N}{k} }, 
\end{align}
where $\Delta_N$ is a constant independent from $k$, and $\hat{\sigma}_{{\rm H}}$ is the Hessian uncertainty obtained from a MC sample with $k$-times the statistical power of the data. %, and $\delta = b/a$.
The right-hand side of eq.~\eqref{eq:scaling2} contains two unknowns: $\bar{\sigma}_{\rm H}$ and $\Delta_N$. If $N$ is kept fixed, but a fraction $1/\xi$, with $\xi>1$ of the MC events is randomly selected and used to construct a new model prediction $\hat{\bf f}$, then the new Hessian uncertainty will be a fraction $r_{\xi}<1$ of the uncertainty from the full sample. Then, one can readily solve for $\bar{\sigma}_{\rm H}$, obtaining:
\begin{align}\label{eq:scaling4}
\bar{\sigma}_{\rm H} = \left( \frac{r_\xi \sqrt{\xi-1}}{\sqrt{\xi r^2_\xi - 1}} \right) \hat{\sigma}_{{\rm H}}.
\end{align}
This result could be improved by taking the expectation values of $\hat{\sigma}_{{\rm H}}$ over several bootstrap re-samplings of the MC sample.

We can test the validity of eq.~\eqref{eq:scaling4} for the toy model of section~\ref{sec:toy}. Here, the mean Hessian uncertainty for $N=2\times 10^6$, $n=200$, and $\alpha=0.03$ is found to be $\hat{\sigma}_{ {\rm H}}=0.0572$ ($0.0447$) when $k=1$ ($k=0.5$), from which the asymptotic uncertainty can be predicted to be $0.0956$ from eq.~\eqref{eq:scaling4}, in good agreement with the true value of $0.0950$. Extensions of eq~\eqref{eq:scaling4} to the Barlow-Beeston lite likelihood are discussed in~\ref{app:additional_scaling}.

Although $\bar{\sigma}_{\rm H}$ does not provide a reliable estimate of the actual statistical spread of $\hat{\mu}$ induced by fluctuations of both the data and the MC samples, it can provide useful information nonetheless. For example, values of $\bar{\sigma}_{\rm H}$ significantly larger than $\hat{\sigma}_{\rm H}$ would signal the presence of spurious constraints on the POI which are likely to invalidate the asymptotic properties of maximum-likelihood estimators even in the presence of a proper account of MC statistical uncertainties, as done for example in the full Barlow-Beeston method. Moreover, $\bar{\sigma}_{\rm H}$ is one of the inputs needed to construct the heuristic confidence interval of eq.~\eqref{eq:Iheur2}.

\subsection{Testing the Barlow-Beeston lite approximation}\label{sec:litevsfull}

Suppose that $\bar{\sigma}_{\rm H}$ calculated from eq.~\eqref{eq:scaling2} is in good agreement with $\hat{\sigma}_{\rm H}$, or, equivalently, that the Hessian uncertainty is stable when the size of the MC sample is reduced. One might still want to ensure that the lite version of the Barlow-Beeston method represents an adequate approximation for a given problem. For this purpose, we can use the formalism developed in the previous section. We start by noticing that the MLE estimators $\hat{\boldsymbol{\theta}}$ and $\hat{\mu}$, besides being functions of ${\bf y}$, also depend on ${\bf j}$, ${\bf J}$ and ${\bf f}_0$ through the matrix ${\bf B}$, defined in~\ref{app:additional_hessian}, and the vector ${\bf d}$ defined by eq.~\eqref{eq:rearrange2}, which are all random variables of known variance.
If the latter can be treated as independent and Gaussian-distributed variables with standard deviations $\Delta B_{ik}$ and $\Delta d_{i}$, the contribution to the variance of the MLE estimator $\hat{\mu}$ induced by fluctuations of the MC templates can be determined from standard error propagation as
\begin{equation}\label{eq:jacvar4}
\hat{\sigma}^2_{\rm MC} = \sum_{i=1}^{n}   \left[\sum_{k=1}^{p}\left( \frac{\partial \hat{\mu} }{\partial B_{ik}} \right)^2  (\Delta B_{ik})^2 + \left( \frac{\partial \hat{\mu}}{\partial d_{i}} \right)^2 (\Delta d_{i})^2 \right]. 
\end{equation}
Analytical expressions for the two partial derivatives appearing in eq.~\eqref{eq:jacvar4} have been derived in~\ref{app:additional_dermu}.
Equation~\eqref{eq:jacvar4} can be easily extended to generic covariances for $\hat{\bf B}$ and $\hat{\bf d}$. However, we remark that the right-hand side of eq.~\eqref{eq:jacvar4} can only be evaluated using the MC prediction $\hat{\bf B}$ and $\hat{\bf d}$ derived from finite-size MC samples, so eq.~\eqref{eq:jacvar4} might be a biased estimator of the real variance of $\hat{\mu}$ under fluctuations of the MC templates. With this caveat in mind, we could still estimate the total variance of the maximum-likelihood estimator $\hat{\mu}$ under fluctuations of both the data and the MC samples as
\begin{equation}\label{eq:jacvar5}
\hat{\sigma}_{\rm H}^{\rm BBfull} \approx \sqrt{\hat{\sigma}_{\rm H}^2 + \hat{\sigma}_{\rm MC}^2}.
\end{equation}
Values of $\hat{\sigma}_{\rm MC}$ larger than the Barlow-Beeston {\it lite} expectation of $\hat{\sigma}_{\rm H}/\sqrt{k}$, or, equivalently values of $\hat{\sigma}_{\rm H}^{\rm BBfull}$ in excess of the naive expectation $\hat{\sigma}_{\rm H}\sqrt{1+k^{-1}}$, should be taken as an indication that the lite approach is insufficient to provide a proper account of MC statistical uncertainties, suggesting that the full approach should be adopted instead. For example, for the toy study presented in section~\ref{sec:toy}, we find that $\hat{\sigma}_{\rm MC}$ estimated from eq.~\eqref{eq:jacvar4} is a factor of $\sim1.5$ larger than $\hat{\sigma}_{\rm H}/\sqrt{k}$ for the nominal model configuration. Interestingly, this ratio decreases monotonically towards $\sim1$ when $N$ or $\epsilon$ are increased, or when $n$ is reduced, corroborating the observation that the Barlow-Beeston lite approach becomes more efficient in these limits. Conversely, we find that the ratio approaches a plateau of $\sim 1.6$ for $k\gg1$, when $n$, $N$ and $\epsilon$ are fixed to their default values, implying that the Barlow-Beeston lite version will never be equivalent to its full implementation for at least the nominal model configuration\footnote{In practice this difference is irrelevant when $k$ is large because the contribution from MC statistical uncertainties to the total variance will also be negligible in this limit.}. A summary of numerical results is reported in table~\ref{tab:nominal_jac}.

Finally, we remark that $\hat{\sigma}_{\rm MC}$ could also be estimated by considering the variance of $\hat{\mu}$ over multiple bootstrap re-samplings of the MC samples. However, eq.~\eqref{eq:jacvar4} offers the advantage that it is fully analytical and that, differently from resampling techniques, the estimators for ${\bf B}$ and ${\bf d}$ can be computed using the {\it full} statistical power of the simulated samples.
%Finally, we remark that eq.~\eqref{eq:jacvar5} could be also used to estimate the ratio $\hat{\sigma}_{\rm H}^{\rm BBfull}/\hat{\sigma}_{\rm H}^{\rm BBlite}$, which was the last piece needed to construct the heuristic confidence interval of eq.~\eqref{eq:Iheur2}.

\begin{table}
\small
    \centering
\begin{tabular}{cccc}
   $k \times (1,10,40)$ & $n \times (\frac{1}{2},\frac{1}{10},\frac{1}{20})$ &  $\epsilon\times (2,4,8)$ & $N\times (10,10^2,10^3)$       \\
\hline
\hline
 ${\bf 1.51}, 1.63,\, 1.65$ & $1.40, \, 1.14, \, 1.07$ & $1.27, \, 1.10, \, 1.03$ & $1.14, 1.02, 1.00$ \\
  \hline
\end{tabular}
\caption{Ratio between $\hat{\sigma}_{\rm MC}$ and $\hat{\sigma}_{\rm H}/ \sqrt{k}$ for the toy model of section~\ref{sec:toy} by varying one configuration parameter at the time. The value for the nominal model is given by the number in bold.}
    \label{tab:nominal_jac}
\end{table}

\section{Conclusions}\label{sec:end}

In this work, we have studied the problem of setting confidence intervals on a parameter of interest measured from the binned analysis of high-statistics and high-granularity data sets, focusing on the case where the likelihood function depends on nuisance parameters and is constructed from finite-size MC samples.
We have set ourselves in the limit of large event counts in all bins for both the data and the MC-simulated samples, such that asymptotic formulas from the theory of maximum-likelihood estimators would be naively expected to hold. In particular, by assuming Wilks' theorem, central confidence intervals can be defined from either the profile-likelihood ratio or the inverse Hessian matrix of the likelihood function in the Gaussian approximation.
%The natural way to accommodate the uncertain MC predictions is to extend the likelihood function to include additional nuisance parameters, as originally proposed by Barlow and Beeston~\cite{BARLOW1993219}.

We have designed a simple toy model consisting of a signal and background process where the unknown signal (background) normalization is treated as the parameter of interest (nuisance parameter). We have further assumed that the likelihood function is constructed using Monte Carlo simulations of finite size. This toy model depends on configuration parameters that we are free to choose: the total number of expected events, the number of bins, a model parameter that increases the signal-to-background separation, and the statistical power of the MC samples relative to the data. By inspecting a large number of pseudo-experiments, we could establish numerically that asymptotic formulas are broken even when the statistical power of the simulated samples is a few times larger than that of the data. In particular, confidence intervals based on Wilks' theorem suffer from systematic under-coverage. Consequently, we have tried alternative methods to set confidence intervals based on the post-fit distribution of the profile-likelihood ratio, such as the profiled Feldman-Cousins approach or variants thereof~\cite{acero2024,Barlett,COUSINS1992331}, but none of them succeeded in recovering the correct coverage.

Methods based on the asymptotic properties of the maximum-likelihood estimator ultimately yield the right coverage, but only in the limit of a very large scaling of the data and/or simulated sample size, though we could not find a first-principle indication of how large such a scaling should be. Finally, by combining the Hessian uncertainties obtained from different likelihood functions, we have worked out a heuristic prescription for a confidence interval, defined in eq.~\eqref{eq:Iheur}, which gets closer to the correct Feldman-Cousins interval for this problem, thus providing a practical test to quantify the level of under-coverage when other standard confidence intervals are used. This prescription is model-independent and could in principle be extended to a generic analysis, provided that the point-estimator and Hessian uncertainty from the full Barlow-Beeston method can be both computed.  

Inspired by the toy model, we have studied the problem from a more general perspective, with the goal of understanding in which circumstances statistical fluctuations in the MC prediction can reduce the uncertainty predicted for the parameter of interest, ultimately leading to under-coverage of confidence intervals constructed in the asymptotic approximation. We found that this artificially augmented sensitivity can be expressed in terms of a bias term to the quadratic form $S$ defined in eq.~\eqref{eq:sigmaH} which, in the limit of infinite size MC samples, is equal to the inverse variance of the maximum-likelihood estimator of the POI. We have identified and characterized two main sources of bias which then leads to under-coverage: fluctuations on the Jacobian matrix of the model function with respect to the POI, and fluctuations on the Jacobian matrix with respect to the nuisance parameters. In the latter case, the effect can be greatly enhanced in the presence of a large global correlation of the parameter of interest, as defined by eq.~\eqref{eq:globalrho1}. Both effects can explain the results of the toy study.
Interestingly, the bias term is not proportional to $S$ by a naive factor of $k^{-1}$, suggesting that extensions of the likelihood functions that include the effect of finite MC samples in an effective way, such as the popular Barlow-Beeston lite method~\cite{BBlite}, can be insufficient to provide a full account of these effects.

Based on general scaling laws, we propose a practical test, described in section~\ref{sec:scaling}, to estimate the standard deviation that the MLE for the parameter of interest would have in the limit of infinite-size MC samples. As the latter should always be smaller than the standard deviation for finite-size MC samples, it could be used as a reference to gauge how far an analysis is from the asymptotic regime. This test only requires the possibility of repeating the maximum-likelihood fit using a subset of the MC simulated samples. We recommend it to be performed in data analyses to ensure the validity of asymptotic formulas, especially in case the Barlow-Beeston lite approximation is used.
Although the full Barlow-Beeston method should always be preferred, as it represents the correct likelihood function for finite-size MC samples, its implementation is often challenging in practice. In section~\ref{sec:litevsfull} we suggest a test, based on analytical formulas, to verify that the Barlow-Beeston lite approach provides an adequate approximation of the exact likelihood function. %We remark, however, that even using the full Barlow-Beeston likelihood, asymptotic properties cannot be given for granted. 

While Wilks' theorem for the profile-likelihood ratio guarantees that the desired asymptotic behavior will eventually be attained, the assumption of being in the asymptotic regime could be wrong even when the event counts are large in absolute terms for both data and simulation. This suggests that the MC statistical uncertainty that needs to be ``small'' to be safely in the asymptotic regime is not necessarily provided by total number of observed/simulated events bin, nor by the total number of bins. Care should be therefore taken when reporting the uncertainty on parameters of interest based on asymptotic formulas when the data model is derived from finite-size MC samples and additional nuisance parameters are profiled. This seems to be the case also for high-statistics experiments, and even when the statistical power of the simulated samples is comparable to that of the data.

\newpage

\appendix

\section{Approximate likelihood functions for the toy model} \label{app:likelihoods}

The likelihood functions described below encode different levels of approximating the exact log-likelihood function eq.~\eqref{eq:logPBB}. For each definition, it should be understood that $\mu_1$ and $\mu_2$ are the appropriate functions of $\mu$ and $\theta$, depending on whether the original or transformed parameters are being considered.
\begin{itemize}
\item \textbf{Poisson without MC stat.}
\begin{equation}\label{eq:logPexpl}
-\ln L_{\rm P}(\mu, \theta) = \sum_{i=1}^{n} \left[ \left(\mu_1  \frac{t_{1i}}{k_1} + \mu_2 \frac{t_{2i}}{k_2}\right) - y_i\ln \left( \mu_1 \frac{t_{1i}}{k_1} + \mu_2 \frac{t_{2i}}{k_2} \right) \right]
\end{equation}
This would be the correct likelihood function if the MC predictions $t_{ji}$ were exact. For finite MC samples, it must be considered as an approximation. In the context of the toy study, it is mostly considered as a reference. It is expected to provide unbiased estimators for $\mu$ and confidence intervals with the right coverage only in the limit $k_j\to\infty$.
\item \textbf{Gauss without MC stat.}
\begin{equation}
-\ln L_{\rm G}(\mu,\theta) = \frac{1}{2} \sum_{i=1}^{n}   \frac{\left(y_i - \left(\mu_1  \frac{t_{1k}}{k_1} + \mu_2 \frac{t_{2k}}{k_2}\right) \right)^2 }{y_i}. 
\label{eq:logG}
\end{equation}
This function provides an excellent approximation of the joint Poisson likelihood of eq.~\eqref{eq:logPexpl} in the limit $y_i\gg1$. An interesting feature of eq.~\eqref{eq:logG} is that its minimum can be found analytically by solving a linear system of equations~\cite{Cowan}. Again, for finite MC samples, it represents an approximation of the correct likelihood. It is considered here mostly to justify the Gaussian approximation of the likelihood function. %when $\mu$ and $\theta$ are the only likelihood parameters.
\item \textbf{Gauss + Barlow-Beeston}
\begin{align}\label{eq:logGBB}
& -\ln L_{\rm G + BB}(\mu, \theta, \boldsymbol{\theta}_1, \boldsymbol{\theta}_2) = \\ \nonumber 
&\frac{1}{2} \sum_{i=1}^{n} \left[ \frac{\left( y_i - \left( \mu_1 \frac{\theta_{1i}}{k_1} + \mu_2 \frac{\theta_{2i}}{k_2}\right) \right)^2 }{y_i}  + \sum_{j=1}^{2} \frac{\left( \theta_{ji} - t_{ji}\right)^2}{ t_{ji} } \right].
\end{align}
This is the Barlow-Beeston likelihood of eq.~\eqref{eq:logPBB} in the Gaussian approximation. It is considered here to justify the Gaussian approximation of the likelihood function even in the presence of nuisance parameters for the uncertain MC predictions. Furthermore, it offers the advantage that the latter can be profiled analytically. Indeed, it is easy to prove that \begin{equation}\label{eq:profiledGBB}
(\hat{\mu},\hat{\theta}) = \underset{\mu,\theta}{\rm argmin} \left[ 
 \sum_{i=1}^{n} \frac{\left( y_i - \left( \mu_1 \frac{t_{1i}}{k_1} + \mu_2 \frac{t_{2i}}{k_2}\right)\right)^2 }{y_i + \mu_1^2  \frac{t_{1i}}{k_1^2} + \mu_2^2  \frac{t_{2i}}{k_2^2}}   \right]
\end{equation}
For the toy study, it has been verified that minimizing eq.~\eqref{eq:logGBB} over the full set of parameters or solving eq.~\eqref{eq:profiledGBB} leads to identical results within numerical accuracy.
\item \textbf{Gauss + Barlow-Beeston lite}
\begin{align}\label{eq:logBBlite}
& -\ln L_{\rm G + BBlite}(\mu,\theta, \boldsymbol{\beta}) = \\ \nonumber
& \frac{1}{2} \sum_{i=1}^{n} \left[  \frac{ \left( y_i - \beta_i \cdot \left( \mu_1 \frac{t_{1i}}{k_1} + \mu_2 \frac{t_{2i}}{k_2} \right) \right)^2 }{y_i}  + \left( \beta_i - 1\right)^2m_{i} \right].
\end{align}
This is a well-known and widely used approximation of the Barlow-Beeston likelihood~\cite{BBlite}. It consists in assigning a unique parameter $\beta_i$ for all MC processes in each bin constrained by the total effective relative uncertainty $1/\sqrt{m_i}$, as defined in eq.~\eqref{eq:effective}. It is studied along with eq.~\eqref{eq:logGBB} as it represents a widely adopted approximation to the full problem. Again, the $\boldsymbol{\beta}$ parameters can be profiled analytically, leading to:
\begin{equation}\label{eq:profiledGBBlite}
(\hat{\mu},\hat{\theta}) = \underset{\mu,\theta}{\rm argmin} \left[ 
 \sum_{i=1}^{n} \frac{\left( y_i - \left( \mu_1 \frac{t_{1i}}{k_1} + \mu_2 \frac{t_{2i}}{k_2}\right)\right)^2 }{y_i +  \frac{ \left(\mu_1  \frac{t_{1i}}{k_1} + \mu_2 \frac{t_{2i}}{k_2}\right)^2}{m_i} }    \right]
\end{equation}
\item \textbf{Gauss + MC stat.}
\begin{equation}
-\ln L_{\rm G + MC}(\mu,\theta) =  \frac{1}{2} \sum_{i=1}^{n}  \frac{\left(y_i - \left(\mu_1 \frac{t_{1i}}{k_1} + \mu_2 \frac{t_{2i}}{k_1}\right) \right)^2 }{y_i + \frac{t_{1i}}{k_1^2} + \frac{t_{2i}}{k_2^2}}.
\label{eq:logGMC}
\end{equation}
This is the simplest extension of eq.~\eqref{eq:logG} that accounts for statistical uncertainties on the MC templates: the total MC variance per bin is added to the Poisson variance of the data. The global minimum can still be found analytically. In the context of the toy study, it is considered to prove that the ``{Gauss + Barlow-Beeston lite}'' likelihood can be well approximated by a Neyman's $\chi^2$ test-statistic with augmented statistical uncertainties.
\end{itemize}

\newpage

\section{Alternative model configurations}\label{app:alt}

Numerical results for coverage, mean, and median of the $1\sigma$ confidence interval for $\mu=\mu^{\prime}_{1}$, obtained for alternative  configurations of the toy model of section~\ref{sec:toy}, are reported in this Appendix. 

The results reported in Table~\ref{tab:nominal_lumi10} (\ref{tab:nominal_lumi40}) have been obtained by increasing the data-to-simulation ratio $k$ by a factor of 10 (40) compared to the nominal, while Table~\ref{tab:nominal_asy0p03} (\ref{tab:nominal_asy0p06}) corresponds to increasing the asymmetry factor $\epsilon$ by a factor of 2 (4). The results reported in Table~\ref{tab:nominal_bins100} (\ref{tab:nominal_bins20}) has been obtained by reducing the number of bins $n$ by a factor of 2 (10).
Finally, the results reported in Table~\ref{tab:nominal_x10} (\ref{tab:nominal_x100}) have been obtained by increasing the total number of events $N$ by a factor of 10 (100) compared to the nominal.

\begin{table}
\scriptsize
    \centering
\begin{tabular}{cccccc}
Likelihood     & Minimim. & CI method      & $\mu_{\rm t}$ & Coverage    &  $\hat{\sigma}$ (mean, median)          \\
\hline
\hline
Gauss (asympt.)  & Analytic & Hessian        & 0                     &  $0.683$    &  $0.094, \; 0.094$ \\
                 &          &                &$+5\bar{\sigma}_{{\rm H} }$   &  $0.683$    &  $0.094, \; 0.094$ \\
                 &          &                &$-5\bar{\sigma}_{{\rm H} }$   &  $0.683$    &  $0.094, \; 0.094$ \\
\hline
Poisson          & Numeric  & Hessian        & 0                     &  ${\bf 0.661}(5)$ &  $0.087, \; 0.087$ \\
                 &          &                &$+5\bar{\sigma}_{{\rm H} }$   &  $0.529(5)$ &  $0.087, \; 0.087$ \\
                 &          &                &$-5\bar{\sigma}_{{\rm H} }$   &  $0.532(5)$ &  $0.087, \; 0.087$ \\
\hline
Gauss            & Numeric  & Hessian        & 0                     &  ${\bf 0.663}(5)$ &  $0.087, \; 0.087$ \\
                 &          &                &$+5\bar{\sigma}_{{\rm H} }$   &  $0.528(5)$ &  $0.087, \; 0.087$ \\
                 &          &                &$-5\bar{\sigma}_{{\rm H} }$   &  $0.532(5)$ &  $0.087, \; 0.087$ \\
\hline
Gauss            & Analytic & Hessian        & 0                     &  ${\bf 0.663}(5)$ &  $0.087, \; 0.087$ \\
                 &          &                &$+5\bar{\sigma}_{{\rm H} }$   &  $0.528(5)$ &  $0.087, \; 0.087$ \\
                 &          &                &$-5\bar{\sigma}_{{\rm H} }$   &  $0.532(5)$ &  $0.087, \; 0.087$ \\
\hline
Gauss + MC stat. & Analytic & Hessian        & 0                     &  ${\bf 0.683}(5)$ &  $0.091, \; 0.091$ \\
                 &          &                &$+5\bar{\sigma}_{{\rm H} }$   &  $0.552(5)$ &  $0.091, \; 0.091$ \\
                 &          &                &$-5\bar{\sigma}_{{\rm H} }$   &  $0.554(5)$ &  $0.091, \; 0.091$ \\
\hline
Gauss + BB-lite  & Numeric  & Hessian        & 0                     &  ${\bf 0.683}(5)$ &  $0.091, \; 0.091$ \\
                 &          &                &$+5\bar{\sigma}_{{\rm H} }$   &  $0.552(5)$ &  $0.091, \; 0.091$ \\
                 &          &                &$-5\bar{\sigma}_{{\rm H} }$   &  $0.554(5)$ &  $0.091, \; 0.091$ \\
\hline
Gauss + BB       & Numeric  & Hessian        & 0                     &  $0.647(5)$ &  $0.099, \; 0.099$ \\
                 &          &                &$+5\bar{\sigma}_{{\rm H} }$   &  ${\bf 0.649}(5)$ &  $0.100, \; 0.100$ \\
                 &          &                &$-5\bar{\sigma}_{{\rm H} }$   &  $0.638(5)$ &  $0.100, \; 0.100$ \\
\hline
Gauss + BB       & Numeric  & PLR       & 0                     &  $0.647(5)$ &  $0.099, \; 0.099$ \\
                 &          &                &$+5\bar{\sigma}_{{\rm H} }$   &  ${\bf 0.650}(5)$ &  $0.100, \; 0.100$ \\
                 &          &                &$-5\bar{\sigma}_{{\rm H} }$   &  $0.639(5)$ &  $0.100, \; 0.100$ \\
\hline
Gauss + BB       & Numeric  & sPFC& 0                     &  ${\bf 0.662}(5)$ &  $0.103, \; 0.103$ \\
                 &          &                &$+5\bar{\sigma}_{{\rm H} }$   &  $0.745(4)$ &  $0.124, \; 0.123$ \\
                 &          &                &$-5\bar{\sigma}_{{\rm H} }$   &  $0.746(4)$ &  $0.123, \; 0.122$ \\
\hline
Gauss + BB       & Numeric  & PFC & 0                     &  ${\bf 0.654}(5)$ &  $0.102, \; 0.102$ \\
                 &          &                &$+5\bar{\sigma}_{{\rm H} }$   &  $0.753(4)$ &  $0.124, \; 0.123$ \\
                 &          &                &$-5\bar{\sigma}_{{\rm H} }$   &  $0.747(4)$ &  $0.123, \; 0.122$ \\
\hline
Gauss + BB       & Numeric  & Barlett    & 0                     &  ${\bf 0.653}(5)$ &  $0.102, \; 0.102$ \\
                 &          &                &$+5\bar{\sigma}_{{\rm H} }$   &  $0.740(4)$ &  $0.121, \; 0.121$ \\
                 &          &                &$-5\bar{\sigma}_{{\rm H} }$   &  $0.738(4)$ &  $0.121, \; 0.120$ \\
\hline
Gauss + BB       & Numeric  & CH     & 0                   &  ${\bf 0.679}(5)$ &  $0.107, \; 0.107$ \\
                 &          &                           &$+5\bar{\sigma}_{{\rm H} }$ &  $0.788(4)$ &  $0.135, \; 0.134$ \\
                 &          &                           &$-5\bar{\sigma}_{{\rm H} }$ &  $0.785(4)$ &  $0.134, \; 0.133$ \\
\hline
Gauss + BB       & Numeric  & FC Cheat       & 0                     &  ${\bf 0.680}(5)$ &  $0.108, \; 0.106$ \\
                 &          &                &$+5\bar{\sigma}_{{\rm H} }$   &  ${\bf 0.684}(5)$ &  $0.109, \; 0.107$ \\
                 &          &                &$-5\bar{\sigma}_{{\rm H} }$   &  ${\bf 0.690}(5)$ &  $0.108, \; 0.106$ \\
\hline
Gauss + BB       & Numeric  & Heuristic   & 0                   &  ${\bf 0.683}(5)$ &  $0.107, \; 0.107$ \\
                 &          &                &$+5\bar{\sigma}_{{\rm H} }$ &  ${\bf 0.686}(5)$ &  $0.108, \; 0.108$ \\
                 &          &                &$-5\bar{\sigma}_{{\rm H} }$ &  ${\bf 0.677}(5)$ &  $0.109, \; 0.109$ \\
\hline
\hline
\end{tabular}
\caption{ Coverage, mean, and median of the $1\sigma$ confidence interval for $\mu=\mu^{\prime}_{1}$ for the toy model with $N=2\times 10^6$, $n=200$, $k=10$, and $\epsilon=0.03$, obtained from an ensemble of $10^4$ identically repeated pseudo-experiments. Cases where the observed coverage agrees with the expectation of $68.3\%$ within a relative $\pm5\%$ tolerance are highlighted with a bold font. When present, the number within parenthesis refers to the statistical error on the last digit.}
    \label{tab:nominal_lumi10}
\end{table}

\begin{table}
\scriptsize
    \centering
\begin{tabular}{cccccc}
Likelihood     & Minimim. & CI method      & $\mu_{\rm t}$ & Coverage    &  $\hat{\sigma}$ (mean, median)          \\
\hline
\hline
Gauss (asympt.)  & Analytic & Hessian        & 0                     &  $0.683$    &  $0.094, \; 0.094$ \\
                 &          &                &$+5\bar{\sigma}_{{\rm H} }$   &  $0.683$    &  $0.094, \; 0.094$ \\
                 &          &                &$-5\bar{\sigma}_{{\rm H} }$   &  $0.683$    &  $0.094, \; 0.094$ \\
\hline
Poisson          & Numeric  & Hessian        & 0                     &  ${\bf 0.669}(5)$ &  $0.092, \; 0.092$ \\
                 &          &                &$+5\bar{\sigma}_{{\rm H} }$   &  ${\bf 0.672}(5)$ &  $0.092, \; 0.092$ \\
                 &          &                &$-5\bar{\sigma}_{{\rm H} }$   &  ${\bf 0.665}(5)$ &  $0.092, \; 0.092$ \\
\hline
Gauss            & Numeric  & Hessian        & 0                     &  ${\bf 0.669}(5)$ &  $0.092, \; 0.092$ \\
                 &          &                &$+5\bar{\sigma}_{{\rm H} }$   &  ${\bf 0.672}(5)$ &  $0.092, \; 0.092$ \\
                 &          &                &$-5\bar{\sigma}_{{\rm H} }$   &  ${\bf 0.667}(5)$ &  $0.092, \; 0.092$ \\
\hline
Gauss            & Analytic & Hessian        & 0                     &  ${\bf 0.669}(5)$ &  $0.092, \; 0.092$ \\
                 &          &                &$+5\bar{\sigma}_{{\rm H} }$   &  ${\bf 0.672}(5)$ &  $0.092, \; 0.092$ \\
                 &          &                &$-5\bar{\sigma}_{{\rm H} }$   &  ${\bf 0.667}(5)$ &  $0.092, \; 0.092$ \\
\hline
Gauss + MC stat. & Analytic & Hessian        & 0                     &  ${\bf 0.675}(5)$ &  $0.093, \; 0.093$ \\
                 &          &                &$+5\bar{\sigma}_{{\rm H} }$   &  ${\bf 0.679}(5)$ &  $0.093, \; 0.093$ \\
                 &          &                &$-5\bar{\sigma}_{{\rm H} }$   &  ${\bf 0.672}(5)$ &  $0.094, \; 0.094$ \\
\hline
Gauss + BB-lite  & Numeric  & Hessian        & 0                     &  ${\bf 0.675}(5)$ &  $0.093, \; 0.093$ \\
                 &          &                &$+5\bar{\sigma}_{{\rm H} }$   &  ${\bf 0.678}(5)$ &  $0.093, \; 0.093$ \\
                 &          &                &$-5\bar{\sigma}_{{\rm H} }$   &  ${\bf 0.672}(5)$ &  $0.094, \; 0.094$ \\
\hline
Gauss + BB       & Numeric  & Hessian        & 0                     &  ${\bf 0.665}(5)$ &  $0.095, \; 0.095$ \\
                 &          &                &$+5\bar{\sigma}_{{\rm H} }$   &  ${\bf 0.674}(5)$ &  $0.096, \; 0.096$ \\
                 &          &                &$-5\bar{\sigma}_{{\rm H} }$   &  ${\bf 0.668}(5)$ &  $0.096, \; 0.096$ \\
\hline
Gauss + BB       & Numeric  & PLR       & 0                     &  ${\bf 0.665}(5)$ &  $0.095, \; 0.095$ \\
                 &          &                &$+5\bar{\sigma}_{{\rm H} }$   &  ${\bf 0.675}(5)$ &  $0.096, \; 0.096$ \\
                 &          &                &$-5\bar{\sigma}_{{\rm H} }$   &  ${\bf 0.669}(5)$ &  $0.096, \; 0.096$ \\
\hline
Gauss + BB       & Numeric  & sPFC& 0                     &  ${\bf 0.679}(5)$ &  $0.096, \; 0.096$ \\
                 &          &                &$+5\bar{\sigma}_{{\rm H} }$   &  ${\bf 0.689}(5)$ &  $0.098, \; 0.098$ \\
                 &          &                &$-5\bar{\sigma}_{{\rm H} }$   &  ${\bf 0.679}(5)$ &  $0.098, \; 0.098$ \\
\hline
Gauss + BB       & Numeric  & PFC & 0                     &  ${\bf 0.675}(5)$ &  $0.096, \; 0.096$ \\
                 &          &                &$+5\bar{\sigma}_{{\rm H} }$   &  ${\bf 0.682}(5)$ &  $0.098, \; 0.098$ \\
                 &          &                &$-5\bar{\sigma}_{{\rm H} }$   &  ${\bf 0.690}(5)$ &  $0.098, \; 0.098$ \\
\hline
Gauss + BB       & Numeric  & Barlett    & 0                     &  ${\bf 0.675}(5)$ &  $0.096, \; 0.096$ \\
                 &          &                &$+5\bar{\sigma}_{{\rm H} }$   &  ${\bf 0.681}(5)$ &  $0.098, \; 0.098$ \\
                 &          &                &$-5\bar{\sigma}_{{\rm H} }$   &  ${\bf 0.690}(5)$ &  $0.098, \; 0.098$ \\
\hline
Gauss + BB       & Numeric  & CH     & 0                   &  ${\bf 0.681}(5)$ &  $0.098, \; 0.098$ \\
                 &          &                           &$+5\bar{\sigma}_{{\rm H} }$ &  ${\bf 0.688}(5)$ &  $0.100, \; 0.100$ \\
                 &          &                           &$-5\bar{\sigma}_{{\rm H} }$ &  ${\bf 0.698}(5)$ &  $0.100, \; 0.100$ \\
\hline
Gauss + BB       & Numeric  & FC Cheat       & 0                     &  ${\bf 0.683}(5)$ &  $0.098, \; 0.097$ \\
                 &          &                &$+5\bar{\sigma}_{{\rm H} }$   &  ${\bf 0.689}(5)$ &  $0.098, \; 0.097$ \\
                 &          &                &$-5\bar{\sigma}_{{\rm H} }$   &  ${\bf 0.684}(5)$ &  $0.098, \; 0.098$ \\
\hline
Gauss + BB       & Numeric  & Heuristic   & 0                   &  ${\bf 0.675}(5)$ &  $0.098, \; 0.098$ \\
                 &          &                &$+5\bar{\sigma}_{{\rm H} }$ &  ${\bf 0.683}(5)$ &  $0.098, \; 0.098$ \\
                 &          &                &$-5\bar{\sigma}_{{\rm H} }$ &  ${\bf 0.678}(5)$ &  $0.098, \; 0.098$ \\
\hline
\hline
\end{tabular}
\caption{ Coverage, mean, and median of the $1\sigma$ confidence interval for $\mu=\mu^{\prime}_{1}$ for the toy model with $N=2\times 10^6$, $n=200$, $k=40$, and $\epsilon=0.03$, obtained from an ensemble of $10^4$ identically repeated pseudo-experiments. Cases where the observed coverage agrees with the expectation of $68.3\%$ within a relative $\pm5\%$ tolerance are highlighted with a bold font. When present, the number within parenthesis refers to the statistical error on the last digit.}
    \label{tab:nominal_lumi40}
\end{table}

\begin{table}
\scriptsize
    \centering
\begin{tabular}{cccccc}
Likelihood     & Minimim. & CI method      & $\mu_{\rm t}$ & Coverage    &  $\hat{\sigma}$ (mean, median)          \\
\hline
\hline
Gauss (asympt.)  & Analytic & Hessian        & 0                     &  $0.683$    &  $0.047, \; 0.047$ \\
                 &          &                &$+5\bar{\sigma}_{{\rm H} }$   &  $0.683$    &  $0.047, \; 0.047$ \\
                 &          &                &$-5\bar{\sigma}_{{\rm H} }$   &  $0.683$    &  $0.047, \; 0.047$ \\
\hline
Poisson          & Numeric  & Hessian        & 0                     &  $0.519(5)$ &  $0.039, \; 0.039$ \\
                 &          &                &$+5\bar{\sigma}_{{\rm H} }$   &  $0.248(4)$ &  $0.039, \; 0.039$ \\
                 &          &                &$-5\bar{\sigma}_{{\rm H} }$   &  $0.254(4)$ &  $0.039, \; 0.039$ \\
\hline
Gauss            & Numeric  & Hessian        & 0                     &  $0.520(5)$ &  $0.039, \; 0.039$ \\
                 &          &                &$+5\bar{\sigma}_{{\rm H} }$   &  $0.249(4)$ &  $0.039, \; 0.039$ \\
                 &          &                &$-5\bar{\sigma}_{{\rm H} }$   &  $0.255(4)$ &  $0.039, \; 0.039$ \\
\hline
Gauss            & Analytic & Hessian        & 0                     &  $0.520(5)$ &  $0.039, \; 0.039$ \\
                 &          &                &$+5\bar{\sigma}_{{\rm H} }$   &  $0.249(4)$ &  $0.039, \; 0.039$ \\
                 &          &                &$-5\bar{\sigma}_{{\rm H} }$   &  $0.255(4)$ &  $0.039, \; 0.039$ \\
\hline
Gauss + MC stat. & Analytic & Hessian        & 0                     &  ${\bf 0.680}(5)$ &  $0.056, \; 0.056$ \\
                 &          &                &$+5\bar{\sigma}_{{\rm H} }$   &  $0.362(5)$ &  $0.056, \; 0.056$ \\
                 &          &                &$-5\bar{\sigma}_{{\rm H} }$   &  $0.372(5)$ &  $0.056, \; 0.056$ \\
\hline
Gauss + BB-lite  & Numeric  & Hessian        & 0                     &  ${\bf 0.680}(5)$ &  $0.056, \; 0.056$ \\
                 &          &                &$+5\bar{\sigma}_{{\rm H} }$   &  $0.363(5)$ &  $0.055, \; 0.055$ \\
                 &          &                &$-5\bar{\sigma}_{{\rm H} }$   &  $0.372(5)$ &  $0.056, \; 0.056$ \\
\hline
Gauss + BB       & Numeric  & Hessian        & 0                     &  $0.592(5)$ &  $0.067, \; 0.067$ \\
                 &          &                &$+5\bar{\sigma}_{{\rm H} }$   &  $0.600(5)$ &  $0.068, \; 0.067$ \\
                 &          &                &$-5\bar{\sigma}_{{\rm H} }$   &  $0.598(5)$ &  $0.068, \; 0.068$ \\
\hline
Gauss + BB       & Numeric  & PLR       & 0                     &  $0.593(5)$ &  $0.067, \; 0.067$ \\
                 &          &                &$+5\bar{\sigma}_{{\rm H} }$   &  $0.602(5)$ &  $0.068, \; 0.068$ \\
                 &          &                &$-5\bar{\sigma}_{{\rm H} }$   &  $0.599(5)$ &  $0.068, \; 0.067$ \\
\hline
Gauss + BB       & Numeric  & sPFC& 0                     &  $0.500(5)$ &  $0.055, \; 0.054$ \\
                 &          &                &$+5\bar{\sigma}_{{\rm H} }$   &  $0.532(5)$ &  $0.060, \; 0.059$ \\
                 &          &                &$-5\bar{\sigma}_{{\rm H} }$   &  $0.528(5)$ &  $0.059, \; 0.059$ \\
\hline
Gauss + BB       & Numeric  & PFC & 0                     &  $0.560(5)$ &  $0.067, \; 0.061$ \\
                 &          &                &$+5\bar{\sigma}_{{\rm H} }$   &  $0.628(5)$ &  $0.072, \; 0.064$ \\
                 &          &                &$-5\bar{\sigma}_{{\rm H} }$   &  $0.628(5)$ &  $0.071, \; 0.063$ \\
\hline
Gauss + BB       & Numeric  & Barlett    & 0                     &  $0.553(5)$ &  $0.065, \; 0.061$ \\
                 &          &                &$+5\bar{\sigma}_{{\rm H} }$   &  $0.598(5)$ &  $0.070, \; 0.063$ \\
                 &          &                &$-5\bar{\sigma}_{{\rm H} }$   &  $0.595(5)$ &  $0.069, \; 0.062$ \\
\hline
Gauss + BB       & Numeric  & CH     & 0                   &  ${\bf 0.687}(5)$ &  $0.081, \; 0.077$ \\
                 &          &                           &$+5\bar{\sigma}_{{\rm H} }$ &  $0.794(4)$ &  $0.099, \; 0.090$ \\
                 &          &                           &$-5\bar{\sigma}_{{\rm H} }$ &  $0.789(4)$ &  $0.097, \; 0.088$ \\
\hline
Gauss + BB       & Numeric  & FC Cheat       & 0                     &  ${\bf 0.691}(5)$ &  $0.081, \; 0.071$ \\
                 &          &                &$+5\bar{\sigma}_{{\rm H} }$   &  ${\bf 0.681}(5)$ &  $0.083, \; 0.072$ \\
                 &          &                &$-5\bar{\sigma}_{{\rm H} }$   &  ${\bf 0.678}(5)$ &  $0.081, \; 0.070$ \\
\hline
Gauss + BB       & Numeric  & Heuristic   & 0                   &  ${\bf 0.682}(5)$ &  $0.080, \; 0.080$ \\
                 &          &                &$+5\bar{\sigma}_{{\rm H} }$ &  ${\bf 0.690}(5)$ &  $0.081, \; 0.081$ \\
                 &          &                &$-5\bar{\sigma}_{{\rm H} }$ &  ${\bf 0.684}(5)$ &  $0.082, \; 0.081$ \\
\hline
\hline
\end{tabular}
\caption{ Coverage, mean, and median of the $1\sigma$ confidence interval for $\mu=\mu^{\prime}_{1}$ for the toy model with $N=2\times 10^6$, $n=200$, $k=1$, and $\epsilon=0.06$, obtained from an ensemble of $10^4$ identically repeated pseudo-experiments. Cases where the observed coverage agrees with the expectation of $68.3\%$ within a relative $\pm5\%$ tolerance are highlighted with a bold font. When present, the number within parenthesis refers to the statistical error on the last digit.}
    \label{tab:nominal_asy0p03}
\end{table}

\begin{table}
\scriptsize
    \centering
\begin{tabular}{cccccc}
Likelihood     & Minimim. & CI method      & $\mu_{\rm t}$ & Coverage    &  $\hat{\sigma}$ (mean, median)          \\
\hline
\hline
Gauss (asympt.)  & Analytic & Hessian        & 0                     &  $0.683$    &  $0.024, \; 0.024$ \\
                 &          &                &$+5\bar{\sigma}_{{\rm H} }$   &  $0.683$    &  $0.024, \; 0.024$ \\
                 &          &                &$-5\bar{\sigma}_{{\rm H} }$   &  $0.683$    &  $0.024, \; 0.024$ \\
\hline
Poisson          & Numeric  & Hessian        & 0                     &  $0.523(5)$ &  $0.022, \; 0.022$ \\
                 &          &                &$+5\bar{\sigma}_{{\rm H} }$   &  $0.485(5)$ &  $0.022, \; 0.022$ \\
                 &          &                &$-5\bar{\sigma}_{{\rm H} }$   &  $0.498(5)$ &  $0.022, \; 0.022$ \\
\hline
Gauss            & Numeric  & Hessian        & 0                     &  $0.522(5)$ &  $0.022, \; 0.022$ \\
                 &          &                &$+5\bar{\sigma}_{{\rm H} }$   &  $0.486(5)$ &  $0.022, \; 0.022$ \\
                 &          &                &$-5\bar{\sigma}_{{\rm H} }$   &  $0.496(5)$ &  $0.022, \; 0.022$ \\
\hline
Gauss            & Analytic & Hessian        & 0                     &  $0.522(5)$ &  $0.022, \; 0.022$ \\
                 &          &                &$+5\bar{\sigma}_{{\rm H} }$   &  $0.486(5)$ &  $0.022, \; 0.022$ \\
                 &          &                &$-5\bar{\sigma}_{{\rm H} }$   &  $0.496(5)$ &  $0.022, \; 0.022$ \\
\hline
Gauss + MC stat. & Analytic & Hessian        & 0                     &  ${\bf 0.689}(5)$ &  $0.032, \; 0.032$ \\
                 &          &                &$+5\bar{\sigma}_{{\rm H} }$   &  $0.648(5)$ &  $0.032, \; 0.032$ \\
                 &          &                &$-5\bar{\sigma}_{{\rm H} }$   &  ${\bf 0.655}(5)$ &  $0.032, \; 0.032$ \\
\hline
Gauss + BB-lite  & Numeric  & Hessian        & 0                     &  ${\bf 0.688}(5)$ &  $0.032, \; 0.032$ \\
                 &          &                &$+5\bar{\sigma}_{{\rm H} }$   &  $0.648(5)$ &  $0.032, \; 0.032$ \\
                 &          &                &$-5\bar{\sigma}_{{\rm H} }$   &  ${\bf 0.656}(5)$ &  $0.032, \; 0.032$ \\
\hline
Gauss + BB       & Numeric  & Hessian        & 0                     &  ${\bf 0.663}(5)$ &  $0.033, \; 0.033$ \\
                 &          &                &$+5\bar{\sigma}_{{\rm H} }$   &  ${\bf 0.658}(5)$ &  $0.033, \; 0.033$ \\
                 &          &                &$-5\bar{\sigma}_{{\rm H} }$   &  ${\bf 0.661}(5)$ &  $0.034, \; 0.034$ \\
\hline
Gauss + BB       & Numeric  & PLR       & 0                     &  ${\bf 0.664}(5)$ &  $0.033, \; 0.033$ \\
                 &          &                &$+5\bar{\sigma}_{{\rm H} }$   &  ${\bf 0.658}(5)$ &  $0.034, \; 0.033$ \\
                 &          &                &$-5\bar{\sigma}_{{\rm H} }$   &  ${\bf 0.662}(5)$ &  $0.034, \; 0.034$ \\
\hline
Gauss + BB       & Numeric  & sPFC& 0                     &  $0.521(5)$ &  $0.025, \; 0.025$ \\
                 &          &                &$+5\bar{\sigma}_{{\rm H} }$   &  $0.512(5)$ &  $0.025, \; 0.025$ \\
                 &          &                &$-5\bar{\sigma}_{{\rm H} }$   &  $0.519(5)$ &  $0.025, \; 0.025$ \\
\hline
Gauss + BB       & Numeric  & PFC & 0                     &  $0.596(5)$ &  $0.030, \; 0.027$ \\
                 &          &                &$+5\bar{\sigma}_{{\rm H} }$   &  $0.593(5)$ &  $0.030, \; 0.028$ \\
                 &          &                &$-5\bar{\sigma}_{{\rm H} }$   &  $0.592(5)$ &  $0.031, \; 0.028$ \\
\hline
Gauss + BB       & Numeric  & Barlett    & 0                     &  $0.588(5)$ &  $0.029, \; 0.027$ \\
                 &          &                &$+5\bar{\sigma}_{{\rm H} }$   &  $0.580(5)$ &  $0.030, \; 0.027$ \\
                 &          &                &$-5\bar{\sigma}_{{\rm H} }$   &  $0.579(5)$ &  $0.030, \; 0.027$ \\
\hline
Gauss + BB       & Numeric  & CH     & 0                   &  ${\bf 0.700}(5)$ &  $0.035, \; 0.033$ \\
                 &          &                           &$+5\bar{\sigma}_{{\rm H} }$ &  ${\bf 0.711}(5)$ &  $0.036, \; 0.033$ \\
                 &          &                           &$-5\bar{\sigma}_{{\rm H} }$ &  ${\bf 0.706}(5)$ &  $0.036, \; 0.033$ \\
\hline
Gauss + BB       & Numeric  & FC Cheat       & 0                     &  ${\bf 0.683}(5)$ &  $0.035, \; 0.031$ \\
                 &          &                &$+5\bar{\sigma}_{{\rm H} }$   &  ${\bf 0.683}(5)$ &  $0.036, \; 0.031$ \\
                 &          &                &$-5\bar{\sigma}_{{\rm H} }$   &  ${\bf 0.684}(5)$ &  $0.036, \; 0.031$ \\
\hline
Gauss + BB       & Numeric  & Heuristic   & 0                   &  ${\bf 0.689}(5)$ &  $0.035, \; 0.035$ \\
                 &          &                &$+5\bar{\sigma}_{{\rm H} }$ &  ${\bf 0.685}(5)$ &  $0.035, \; 0.035$ \\
                 &          &                &$-5\bar{\sigma}_{{\rm H} }$ &  ${\bf 0.689}(5)$ &  $0.035, \; 0.035$ \\
\hline
\hline
\end{tabular}
\caption{ Coverage, mean, and median of the $1\sigma$ confidence interval for $\mu=\mu^{\prime}_{1}$ for the toy model with $N=2\times 10^6$, $n=200$, $k=1$, and $\epsilon=0.12$, obtained from an ensemble of $10^4$ identically repeated pseudo-experiments. Cases where the observed coverage agrees with the expectation of $68.3\%$ within a relative $\pm5\%$ tolerance are highlighted with a bold font. When present, the number within parenthesis refers to the statistical error on the last digit.}
    \label{tab:nominal_asy0p06}
\end{table}

\begin{table}
\scriptsize
    \centering
\begin{tabular}{cccccc}
Likelihood     & Minimim. & CI method      & $\mu_{\rm t}$ & Coverage    &  $\hat{\sigma}$ (mean, median)          \\
\hline
\hline
Gauss (asympt.)  & Analytic & Hessian        & 0                     &  $0.683$    &  $0.094, \; 0.094$ \\
                 &          &                &$+5\bar{\sigma}_{{\rm H} }$   &  $0.683$    &  $0.094, \; 0.094$ \\
                 &          &                &$-5\bar{\sigma}_{{\rm H} }$   &  $0.683$    &  $0.094, \; 0.094$ \\
\hline
Poisson          & Numeric  & Hessian        & 0                     &  $0.530(5)$ &  $0.069, \; 0.069$ \\
                 &          &                &$+5\bar{\sigma}_{{\rm H} }$   &  $0.060(2)$ &  $0.069, \; 0.069$ \\
                 &          &                &$-5\bar{\sigma}_{{\rm H} }$   &  $0.067(2)$ &  $0.069, \; 0.069$ \\
\hline
Gauss            & Numeric  & Hessian        & 0                     &  $0.531(5)$ &  $0.069, \; 0.069$ \\
                 &          &                &$+5\bar{\sigma}_{{\rm H} }$   &  $0.060(2)$ &  $0.069, \; 0.069$ \\
                 &          &                &$-5\bar{\sigma}_{{\rm H} }$   &  $0.067(3)$ &  $0.069, \; 0.069$ \\
\hline
Gauss            & Analytic & Hessian        & 0                     &  $0.531(5)$ &  $0.069, \; 0.069$ \\
                 &          &                &$+5\bar{\sigma}_{{\rm H} }$   &  $0.060(2)$ &  $0.069, \; 0.069$ \\
                 &          &                &$-5\bar{\sigma}_{{\rm H} }$   &  $0.067(3)$ &  $0.069, \; 0.069$ \\
\hline
Gauss + MC stat. & Analytic & Hessian        & 0                     &  ${\bf 0.691}(5)$ &  $0.098, \; 0.097$ \\
                 &          &                &$+5\bar{\sigma}_{{\rm H} }$   &  $0.107(3)$ &  $0.098, \; 0.097$ \\
                 &          &                &$-5\bar{\sigma}_{{\rm H} }$   &  $0.110(3)$ &  $0.098, \; 0.098$ \\
\hline
Gauss + BB-lite  & Numeric  & Hessian        & 0                     &  ${\bf 0.692}(5)$ &  $0.098, \; 0.097$ \\
                 &          &                &$+5\bar{\sigma}_{{\rm H} }$   &  $0.107(3)$ &  $0.097, \; 0.097$ \\
                 &          &                &$-5\bar{\sigma}_{{\rm H} }$   &  $0.110(3)$ &  $0.098, \; 0.098$ \\
\hline
Gauss + BB       & Numeric  & Hessian        & 0                     &  $0.545(5)$ &  $0.137, \; 0.134$ \\
                 &          &                &$+5\bar{\sigma}_{{\rm H} }$   &  $0.549(5)$ &  $0.144, \; 0.140$ \\
                 &          &                &$-5\bar{\sigma}_{{\rm H} }$   &  $0.548(5)$ &  $0.145, \; 0.141$ \\
\hline
Gauss + BB       & Numeric  & PLR       & 0                     &  $0.552(5)$ &  $0.138, \; 0.134$ \\
                 &          &                &$+5\bar{\sigma}_{{\rm H} }$   &  $0.557(5)$ &  $0.150, \; 0.145$ \\
                 &          &                &$-5\bar{\sigma}_{{\rm H} }$   &  $0.556(5)$ &  $0.141, \; 0.137$ \\
\hline
Gauss + BB       & Numeric  & sPFC& 0                     &  $0.481(5)$ &  $0.126, \; 0.121$ \\
                 &          &                &$+5\bar{\sigma}_{{\rm H} }$   &  $0.537(5)$ &  $0.149, \; 0.143$ \\
                 &          &                &$-5\bar{\sigma}_{{\rm H} }$   &  $0.543(5)$ &  $0.138, \; 0.134$ \\
\hline
Gauss + BB       & Numeric  & PFC & 0                     &  $0.529(5)$ &  $0.151, \; 0.138$ \\
                 &          &                &$+5\bar{\sigma}_{{\rm H} }$   &  $0.637(5)$ &  $0.181, \; 0.153$ \\
                 &          &                &$-5\bar{\sigma}_{{\rm H} }$   &  $0.635(5)$ &  $0.169, \; 0.145$ \\
\hline
Gauss + BB       & Numeric  & Barlett    & 0                     &  $0.523(5)$ &  $0.147, \; 0.136$ \\
                 &          &                &$+5\bar{\sigma}_{{\rm H} }$   &  $0.599(5)$ &  $0.172, \; 0.149$ \\
                 &          &                &$-5\bar{\sigma}_{{\rm H} }$   &  $0.594(5)$ &  $0.161, \; 0.141$ \\
\hline
Gauss + BB       & Numeric  & CH     & 0                   &  ${\bf 0.692}(5)$ &  $0.193, \; 0.180$ \\
                 &          &                           &$+5\bar{\sigma}_{{\rm H} }$ &  $0.815(4)$ &  $0.266, \; 0.249$ \\
                 &          &                           &$-5\bar{\sigma}_{{\rm H} }$ &  $0.821(4)$ &  $0.253, \; 0.237$ \\
\hline
Gauss + BB       & Numeric  & FC Cheat       & 0                     &  ${\bf 0.688}(5)$ &  $0.191, \; 0.170$ \\
                 &          &                &$+5\bar{\sigma}_{{\rm H} }$   &  ${\bf 0.683}(5)$ &  $0.199, \; 0.174$ \\
                 &          &                &$-5\bar{\sigma}_{{\rm H} }$   &  ${\bf 0.683}(5)$ &  $0.186, \; 0.164$ \\
\hline
Gauss + BB       & Numeric  & Heuristic   & 0                   &  ${\bf 0.691}(5)$ &  $0.186, \; 0.183$ \\
                 &          &                &$+5\bar{\sigma}_{{\rm H} }$ &  ${\bf 0.697}(5)$ &  $0.196, \; 0.192$ \\
                 &          &                &$-5\bar{\sigma}_{{\rm H} }$ &  ${\bf 0.695}(5)$ &  $0.197, \; 0.192$ \\
\hline
\hline
\end{tabular}
\caption{ Coverage, mean, and median of the $1\sigma$ confidence interval for $\mu=\mu^{\prime}_{1}$ for the toy model with $N=2\times 10^6$, $n=100$, $k=1$, and $\epsilon=0.03$, obtained from an ensemble of $10^4$ identically repeated pseudo-experiments. Cases where the observed coverage agrees with the expectation of $68.3\%$ within a relative $\pm5\%$ tolerance are highlighted with a bold font. When present, the number within parenthesis refers to the statistical error on the last digit.}
    \label{tab:nominal_bins100}
\end{table}

\begin{table}
\scriptsize
    \centering
\begin{tabular}{cccccc}
Likelihood     & Minimim. & CI method      & $\mu_{\rm t}$ & Coverage    &  $\hat{\sigma}$ (mean, median)          \\
\hline
\hline
Gauss (asympt.)  & Analytic & Hessian        & 0                     &  $0.683$    &  $0.094, \; 0.094$ \\
                 &          &                &$+5\bar{\sigma}_{{\rm H} }$   &  $0.683$    &  $0.094, \; 0.094$ \\
                 &          &                &$-5\bar{\sigma}_{{\rm H} }$   &  $0.683$    &  $0.094, \; 0.094$ \\
\hline
Poisson          & Numeric  & Hessian        & 0                     &  $0.514(5)$ &  $0.088, \; 0.088$ \\
                 &          &                &$+5\bar{\sigma}_{{\rm H} }$   &  $0.450(5)$ &  $0.088, \; 0.087$ \\
                 &          &                &$-5\bar{\sigma}_{{\rm H} }$   &  $0.442(5)$ &  $0.088, \; 0.088$ \\
\hline
Gauss            & Numeric  & Hessian        & 0                     &  $0.514(5)$ &  $0.088, \; 0.088$ \\
                 &          &                &$+5\bar{\sigma}_{{\rm H} }$   &  $0.450(5)$ &  $0.088, \; 0.087$ \\
                 &          &                &$-5\bar{\sigma}_{{\rm H} }$   &  $0.441(5)$ &  $0.088, \; 0.088$ \\
\hline
Gauss            & Analytic & Hessian        & 0                     &  $0.514(5)$ &  $0.088, \; 0.088$ \\
                 &          &                &$+5\bar{\sigma}_{{\rm H} }$   &  $0.450(5)$ &  $0.088, \; 0.087$ \\
                 &          &                &$-5\bar{\sigma}_{{\rm H} }$   &  $0.441(5)$ &  $0.088, \; 0.088$ \\
\hline
Gauss + MC stat. & Analytic & Hessian        & 0                     &  ${\bf 0.676}(5)$ &  $0.125, \; 0.124$ \\
                 &          &                &$+5\bar{\sigma}_{{\rm H} }$   &  $0.604(5)$ &  $0.125, \; 0.124$ \\
                 &          &                &$-5\bar{\sigma}_{{\rm H} }$   &  $0.593(5)$ &  $0.125, \; 0.124$ \\
\hline
Gauss + BB-lite  & Numeric  & Hessian        & 0                     &  ${\bf 0.676}(5)$ &  $0.125, \; 0.124$ \\
                 &          &                &$+5\bar{\sigma}_{{\rm H} }$   &  $0.603(5)$ &  $0.124, \; 0.124$ \\
                 &          &                &$-5\bar{\sigma}_{{\rm H} }$   &  $0.594(5)$ &  $0.125, \; 0.124$ \\
\hline
Gauss + BB       & Numeric  & Hessian        & 0                     &  $0.641(5)$ &  $0.136, \; 0.134$ \\
                 &          &                &$+5\bar{\sigma}_{{\rm H} }$   &  ${\bf 0.653}(5)$ &  $0.143, \; 0.141$ \\
                 &          &                &$-5\bar{\sigma}_{{\rm H} }$   &  ${\bf 0.653}(5)$ &  $0.144, \; 0.141$ \\
\hline
Gauss + BB       & Numeric  & PLR       & 0                     &  $0.649(5)$ &  $0.136, \; 0.135$ \\
                 &          &                &$+5\bar{\sigma}_{{\rm H} }$   &  ${\bf 0.660}(5)$ &  $0.148, \; 0.145$ \\
                 &          &                &$-5\bar{\sigma}_{{\rm H} }$   &  ${\bf 0.660}(5)$ &  $0.140, \; 0.138$ \\
\hline
Gauss + BB       & Numeric  & sPFC& 0                     &  $0.511(5)$ &  $0.101, \; 0.100$ \\
                 &          &                &$+5\bar{\sigma}_{{\rm H} }$   &  $0.501(5)$ &  $0.104, \; 0.102$ \\
                 &          &                &$-5\bar{\sigma}_{{\rm H} }$   &  $0.495(5)$ &  $0.100, \; 0.099$ \\
\hline
Gauss + BB       & Numeric  & PFC & 0                     &  $0.569(5)$ &  $0.127, \; 0.116$ \\
                 &          &                &$+5\bar{\sigma}_{{\rm H} }$   &  $0.588(5)$ &  $0.138, \; 0.120$ \\
                 &          &                &$-5\bar{\sigma}_{{\rm H} }$   &  $0.592(5)$ &  $0.130, \; 0.115$ \\
\hline
Gauss + BB       & Numeric  & Barlett    & 0                     &  $0.559(5)$ &  $0.123, \; 0.115$ \\
                 &          &                &$+5\bar{\sigma}_{{\rm H} }$   &  $0.568(5)$ &  $0.133, \; 0.118$ \\
                 &          &                &$-5\bar{\sigma}_{{\rm H} }$   &  $0.572(5)$ &  $0.125, \; 0.114$ \\
\hline
Gauss + BB       & Numeric  & CH     & 0                   &  ${\bf 0.683}(5)$ &  $0.149, \; 0.140$ \\
                 &          &                           &$+5\bar{\sigma}_{{\rm H} }$ &  $0.724(4)$ &  $0.166, \; 0.146$ \\
                 &          &                           &$-5\bar{\sigma}_{{\rm H} }$ &  $0.723(4)$ &  $0.155, \; 0.139$ \\
\hline
Gauss + BB       & Numeric  & FC Cheat       & 0                     &  ${\bf 0.688}(5)$ &  $0.148, \; 0.130$ \\
                 &          &                &$+5\bar{\sigma}_{{\rm H} }$   &  ${\bf 0.681}(5)$ &  $0.160, \; 0.137$ \\
                 &          &                &$-5\bar{\sigma}_{{\rm H} }$   &  ${\bf 0.685}(5)$ &  $0.149, \; 0.129$ \\
\hline
Gauss + BB       & Numeric  & Heuristic   & 0                   &  ${\bf 0.675}(5)$ &  $0.145, \; 0.144$ \\
                 &          &                &$+5\bar{\sigma}_{{\rm H} }$ &  ${\bf 0.687}(5)$ &  $0.152, \; 0.151$ \\
                 &          &                &$-5\bar{\sigma}_{{\rm H} }$ &  ${\bf 0.688}(5)$ &  $0.153, \; 0.152$ \\
\hline
\hline
\end{tabular}
\caption{ Coverage, mean, and median of the $1\sigma$ confidence interval for $\mu=\mu^{\prime}_{1}$ for the toy model with $N=2\times 10^6$, $n=20$, $k=1$, and $\epsilon=0.03$, obtained from an ensemble of $10^4$ identically repeated pseudo-experiments. Cases where the observed coverage agrees with the expectation of $68.3\%$ within a relative $\pm5\%$ tolerance are highlighted with a bold font. When present, the number within parenthesis refers to the statistical error on the last digit.}
    \label{tab:nominal_bins20}
\end{table}

\begin{table}
\scriptsize
    \centering
    \begin{tabular}{cccccc}
Likelihood     & Minimim. & CI method      & $\mu_{\rm t}$ & Coverage    &  $\hat{\sigma}$ (mean, median)          \\
\hline
\hline
Gauss (asympt.)  & Analytic & Hessian        & 0                     &  $0.683$    &  $0.030, \; 0.030$ \\
                 &          &                &$+5\bar{\sigma}_{\rm H }$   &  $0.683$    &  $0.030, \; 0.030$ \\
                 &          &                &$-5\bar{\sigma}_{\rm H }$   &  $0.683$    &  $0.030, \; 0.030$ \\
\hline
Poisson          & Numeric  & Hessian        & 0                     &  $0.518(5)$ &  $0.028, \; 0.028$ \\
                 &          &                &$+5\bar{\sigma}_{\rm H }$   &  $0.443(5)$ &  $0.028, \; 0.027$ \\
                 &          &                &$-5\bar{\sigma}_{\rm H }$   &  $0.452(5)$ &  $0.028, \; 0.028$ \\
\hline
Gauss            & Numeric  & Hessian        & 0                     &  $0.518(5)$ &  $0.028, \; 0.028$ \\
                 &          &                &$+5\bar{\sigma}_{\rm H }$   &  $0.443(5)$ &  $0.028, \; 0.027$ \\
                 &          &                &$-5\bar{\sigma}_{\rm H }$   &  $0.451(5)$ &  $0.028, \; 0.028$ \\
\hline
Gauss            & Analytic & Hessian        & 0                     &  $0.518(5)$ &  $0.028, \; 0.028$ \\
                 &          &                &$+5\bar{\sigma}_{\rm H }$   &  $0.443(5)$ &  $0.028, \; 0.027$ \\
                 &          &                &$-5\bar{\sigma}_{\rm H }$   &  $0.451(5)$ &  $0.028, \; 0.028$ \\
\hline
Gauss + MC stat. & Analytic & Hessian        & 0                     &  ${\bf 0.690}(5)$ &  $0.039, \; 0.039$ \\
                 &          &                &$+5\bar{\sigma}_{\rm H }$   &  $0.604(5)$ &  $0.039, \; 0.039$ \\
                 &          &                &$-5\bar{\sigma}_{\rm H }$   &  $0.608(5)$ &  $0.039, \; 0.039$ \\
\hline
Gauss + BB-lite  & Numeric  & Hessian        & 0                     &  ${\bf 0.690}(5)$ &  $0.039, \; 0.039$ \\
                 &          &                &$+5\bar{\sigma}_{\rm H }$   &  $0.604(5)$ &  $0.039, \; 0.039$ \\
                 &          &                &$-5\bar{\sigma}_{\rm H }$   &  $0.608(5)$ &  $0.039, \; 0.039$ \\
\hline
Gauss + BB       & Numeric  & Hessian        & 0                     &  $0.646(5)$ &  $0.042, \; 0.042$ \\
                 &          &                &$+5\bar{\sigma}_{\rm H }$   &  $0.642(5)$ &  $0.042, \; 0.042$ \\
                 &          &                &$-5\bar{\sigma}_{\rm H }$   &  $0.644(5)$ &  $0.042, \; 0.042$ \\
\hline
Gauss + BB       & Numeric  & PLR       & 0                     &  $0.647(5)$ &  $0.042, \; 0.042$ \\
                 &          &                &$+5\bar{\sigma}_{\rm H }$   &  $0.643(5)$ &  $0.043, \; 0.043$ \\
                 &          &                &$-5\bar{\sigma}_{\rm H }$   &  $0.646(5)$ &  $0.042, \; 0.042$ \\
\hline
Gauss + BB       & Numeric  & sPFC& 0                     &  $0.518(5)$ &  $0.032, \; 0.032$ \\
                 &          &                &$+5\bar{\sigma}_{\rm H }$   &  $0.521(5)$ &  $0.032, \; 0.032$ \\
                 &          &                &$-5\bar{\sigma}_{\rm H }$   &  $0.520(5)$ &  $0.032, \; 0.032$ \\
\hline
Gauss + BB       & Numeric  & PFC & 0                     &  $0.579(5)$ &  $0.039, \; 0.036$ \\
                 &          &                &$+5\bar{\sigma}_{\rm H }$   &  $0.599(5)$ &  $0.040, \; 0.036$ \\
                 &          &                &$-5\bar{\sigma}_{\rm H }$   &  $0.594(5)$ &  $0.040, \; 0.036$ \\
\hline
Gauss + BB       & Numeric  & Barlett    & 0                     &  $0.571(5)$ &  $0.038, \; 0.035$ \\
                 &          &                &$+5\bar{\sigma}_{\rm H }$   &  $0.582(5)$ &  $0.039, \; 0.036$ \\
                 &          &                &$-5\bar{\sigma}_{\rm H }$   &  $0.576(5)$ &  $0.039, \; 0.036$ \\
\hline
Gauss + BB       & Numeric  & CH     & 0                   &  ${\bf 0.693}(5)$ &  $0.046, \; 0.043$ \\
                 &          &                           &$+5\bar{\sigma}_{\rm H }$ &  $0.737(4)$ &  $0.048, \; 0.044$ \\
                 &          &                           &$-5\bar{\sigma}_{\rm H }$ &  $0.729(4)$ &  $0.048, \; 0.044$ \\
\hline
Gauss + BB       & Numeric  & FC Cheat       & 0                     &  ${\bf 0.687}(5)$ &  $0.046, \; 0.040$ \\
                 &          &                &$+5\bar{\sigma}_{\rm H }$   &  ${\bf 0.677}(5)$ &  $0.046, \; 0.040$ \\
                 &          &                &$-5\bar{\sigma}_{\rm H }$   &  ${\bf 0.685}(5)$ &  $0.046, \; 0.040$ \\
\hline
Gauss + BB       & Numeric  & {Heuristic}   & 0                   &  ${\bf 0.690}(5)$ &  $0.046, \; 0.046$ \\
                 &          &                &$+5\bar{\sigma}_{\rm H }$ &  ${\bf 0.682}(5)$ &  $0.046, \; 0.046$ \\
                 &          &                &$-5\bar{\sigma}_{\rm H }$ &  ${\bf 0.681}(5)$ &  $0.046, \; 0.046$ \\
\hline
\hline
\end{tabular}
\caption{ Coverage, mean, and median of the $1\sigma$ confidence interval for $\mu=\mu^{\prime}_{1}$ for the toy model with $N=2\times 10^7$, $n=200$, $k=1$, and $\epsilon=0.03$, obtained from an ensemble of $10^4$ identically repeated pseudo-experiments. Cases where the observed coverage agrees with the expectation of $68.3\%$ within a relative $\pm5\%$ tolerance are highlighted with a bold font. When present, the number within parenthesis refers to the statistical error on the last digit.}
    \label{tab:nominal_x10}
\end{table}

\begin{table}
\scriptsize
    \centering
\begin{tabular}{cccccc}
Likelihood     & Minimim. & CI method      & $\mu_{\rm t}$ & Coverage    &  $\hat{\sigma}$ (mean, median)          \\
\hline
\hline
Gauss (asympt.)  & Analytic & Hessian        & 0                     &  $0.683$    &  $0.009, \; 0.009$ \\
                 &          &                &$+5\bar{\sigma}_{\rm H }$   &  $0.683$    &  $0.009, \; 0.009$ \\
                 &          &                &$-5\bar{\sigma}_{\rm H }$   &  $0.683$    &  $0.009, \; 0.009$ \\
\hline
Poisson          & Numeric  & Hessian        & 0                     &  $0.529(5)$ &  $0.009, \; 0.010$ \\
                 &          &                &$+5\bar{\sigma}_{\rm H }$   &  $0.516(5)$ &  $0.009, \; 0.010$ \\
                 &          &                &$-5\bar{\sigma}_{\rm H }$   &  $0.514(5)$ &  $0.009, \; 0.010$ \\
\hline
Gauss            & Numeric  & Hessian        & 0                     &  $0.529(5)$ &  $0.009, \; 0.010$ \\
                 &          &                &$+5\bar{\sigma}_{\rm H }$   &  $0.516(5)$ &  $0.009, \; 0.010$ \\
                 &          &                &$-5\bar{\sigma}_{\rm H }$   &  $0.514(5)$ &  $0.009, \; 0.010$ \\
\hline
Gauss            & Analytic & Hessian        & 0                     &  $0.529(5)$ &  $0.009, \; 0.010$ \\
                 &          &                &$+5\bar{\sigma}_{\rm H }$   &  $0.516(5)$ &  $0.009, \; 0.010$ \\
                 &          &                &$-5\bar{\sigma}_{\rm H }$   &  $0.514(5)$ &  $0.009, \; 0.010$ \\
\hline
Gauss + MC stat. & Analytic & Hessian        & 0                     &  ${\bf 0.693}(5)$ &  $0.013, \; 0.013$ \\
                 &          &                &$+5\bar{\sigma}_{\rm H }$   &  ${\bf 0.678}(5)$ &  $0.013, \; 0.013$ \\
                 &          &                &$-5\bar{\sigma}_{\rm H }$   &  ${\bf 0.679}(5)$ &  $0.013, \; 0.013$ \\
\hline
Gauss + BB-lite  & Numeric  & Hessian        & 0                     &  ${\bf 0.693}(5)$ &  $0.013, \; 0.013$ \\
                 &          &                &$+5\bar{\sigma}_{\rm H }$   &  ${\bf 0.678}(5)$ &  $0.013, \; 0.013$ \\
                 &          &                &$-5\bar{\sigma}_{\rm H }$   &  ${\bf 0.678}(5)$ &  $0.013, \; 0.013$ \\
\hline
Gauss + BB       & Numeric  & Hessian        & 0                     &  ${\bf 0.688}(5)$ &  $0.013, \; 0.013$ \\
                 &          &                &$+5\bar{\sigma}_{\rm H }$   &  ${\bf 0.675}(5)$ &  $0.013, \; 0.013$ \\
                 &          &                &$-5\bar{\sigma}_{\rm H }$   &  ${\bf 0.674}(5)$ &  $0.013, \; 0.013$ \\
\hline
Gauss + BB       & Numeric  & PLR       & 0                     &  ${\bf 0.688}(5)$ &  $0.013, \; 0.013$ \\
                 &          &                &$+5\bar{\sigma}_{\rm H }$   &  ${\bf 0.675}(5)$ &  $0.013, \; 0.013$ \\
                 &          &                &$-5\bar{\sigma}_{\rm H }$   &  ${\bf 0.674}(5)$ &  $0.013, \; 0.013$ \\
\hline
Gauss + BB       & Numeric  & sPFC& 0                     &  $0.518(5)$ &  $0.009, \; 0.009$ \\
                 &          &                &$+5\bar{\sigma}_{\rm H }$   &  $0.522(5)$ &  $0.010, \; 0.010$ \\
                 &          &                &$-5\bar{\sigma}_{\rm H }$   &  $0.518(5)$ &  $0.010, \; 0.010$ \\
\hline
Gauss + BB       & Numeric  & PFC & 0                     &  $0.590(5)$ &  $0.012, \; 0.011$ \\
                 &          &                &$+5\bar{\sigma}_{\rm H }$   &  $0.591(5)$ &  $0.012, \; 0.011$ \\
                 &          &                &$-5\bar{\sigma}_{\rm H }$   &  $0.602(5)$ &  $0.012, \; 0.011$ \\
\hline
Gauss + BB       & Numeric  & Barlett    & 0                     &  $0.583(5)$ &  $0.011, \; 0.011$ \\
                 &          &                &$+5\bar{\sigma}_{\rm H }$   &  $0.582(5)$ &  $0.011, \; 0.011$ \\
                 &          &                &$-5\bar{\sigma}_{\rm H }$   &  $0.593(5)$ &  $0.011, \; 0.011$ \\
\hline
Gauss + BB       & Numeric  & CH     & 0                   &  ${\bf 0.686}(5)$ &  $0.014, \; 0.013$ \\
                 &          &                           &$+5\bar{\sigma}_{\rm H }$ &  ${\bf 0.694}(5)$ &  $0.014, \; 0.013$ \\
                 &          &                           &$-5\bar{\sigma}_{\rm H }$ &  ${\bf 0.701}(5)$ &  $0.014, \; 0.013$ \\
\hline
Gauss + BB       & Numeric  & FC Cheat       & 0                     &  ${\bf 0.683}(5)$ &  $0.013, \; 0.012$ \\
                 &          &                &$+5\bar{\sigma}_{\rm H }$   &  ${\bf 0.671}(5)$ &  $0.014, \; 0.012$ \\
                 &          &                &$-5\bar{\sigma}_{\rm H }$   &  ${\bf 0.680}(5)$ &  $0.013, \; 0.012$ \\
\hline
Gauss + BB       & Numeric  & {Heuristic}   & 0                   &  ${\bf 0.692}(5)$ &  $0.013, \; 0.014$ \\
                 &          &                &$+5\bar{\sigma}_{\rm H }$ &  ${\bf 0.677}(5)$ &  $0.013, \; 0.014$ \\
                 &          &                &$-5\bar{\sigma}_{\rm H }$ &  ${\bf 0.679}(5)$ &  $0.013, \; 0.014$ \\
\hline
\hline
\end{tabular}
\caption{ Coverage, mean, and median of the $1\sigma$ confidence interval for $\mu=\mu^{\prime}_{1}$ for the toy model with $N=2\times 10^8$, $n=200$, $k=1$, and $\epsilon=0.03$, obtained from an ensemble of $10^4$ identically repeated pseudo-experiments. Cases where the observed coverage agrees with the expectation of $68.3\%$ within a relative $\pm5\%$ tolerance are highlighted with a bold font. When present, the number within parenthesis refers to the statistical error on the last digit.}
    \label{tab:nominal_x100}
\end{table}

\newpage

\section{Adding external constraints to the likelihood}\label{app:additional_constr}

In section~\ref{sec:method}, all parameters of the likelihood were assumed to be unconstrained, though this is certainly not the most general case of interest. Indeed, it often happens that at least some of the nuisance parameters can be constrained by independent auxiliary measurements.

Let us then assume that the likelihood function includes both constrained and unconstrained nuisance parameters. If the set of $q$ constrained nuisance parameters is denoted by $\boldsymbol{\tau}$, the right-hand side of eq.~\eqref{eq:lin} gets modified to    
\begin{equation}\label{eq:linconstr}
{\bf f}(\mu, \boldsymbol{\theta}, \boldsymbol{\tau}) = 
{\bf f}_{0} + {\bf j}(\mu -\mu_0) + {\bf J} (\boldsymbol{\theta} - \boldsymbol{\theta}_{0}) +  {\bf K}( \boldsymbol{\tau} - \boldsymbol{\tau}_0) + \cdots,
\end{equation}
where ${\bf K}$ is the $n\times q$ Jacobian matrix associated with the NP subject to a constraint. In some cases, the effect of a constrained nuisance parameter relative to the nominal prediction is known exactly: for example, a luminosity uncertainty on the total MC normalization can be modeled as a multiplicative correction, i.e. ${\bf K} \propto {\bf f}_0$. In other cases, however, ${\bf K}$ needs to be estimated numerically, giving rise to an additional source of systematic uncertainty, if MC samples of finite size are used to determine its elements. The negative log-likelihood function in eq.~\eqref{eq:chi2} should be then extended accordingly to
\begin{align}\label{eq:chi2constr}
-\ln { L}(\mu, \boldsymbol{\theta}, \boldsymbol{\tau}) & \approx \frac{1}{2} \left({\bf y} - {\bf f}(\mu,\boldsymbol{\theta},\boldsymbol{\tau}) \right)^T {\bf V}^{-1} \left({\bf y} - {\bf f}(\mu,\boldsymbol{\theta},\boldsymbol{\tau}) \right)  \nonumber \\
& + \frac{1}{2}(\boldsymbol{\tau} - \boldsymbol{\tau}_0)^T(\boldsymbol{\tau} - \boldsymbol{\tau}_0) + {\rm const.}
\end{align}
where the estimators from the auxiliary measurements have been assumed as normally-distributed and independent from each other~\cite{Pinto:2023yob}.
The additional parameters $\boldsymbol{\tau}$ can be profiled at a fixed value of $(\mu,\boldsymbol{\theta})$ yielding the well-known result~\cite{Pinto:2023yob,Behnke}:
\begin{equation}\label{eq:chi2constr2}
-2  \ln \frac{L(\mu, \boldsymbol{\theta}, \hat{\boldsymbol{\tau}}_{(\mu,\boldsymbol{\theta}}) )}{{ L}(\hat{\mu}, \hat{\boldsymbol{\theta}}, \hat{\boldsymbol{\tau}} )}  = \left({\bf y} - {\bf f}(\mu,\boldsymbol{\theta},\boldsymbol{\tau}_0) \right)^T \left( {\bf V} + {\bf K}{\bf K}^T \right)^{-1} \left({\bf y} - {\bf f}(\mu,\boldsymbol{\theta},\boldsymbol{\tau}_0) \right), 
\end{equation}
that is, the Poisson covariance matrix of the data ${\bf V}$ is augmented by the non-diagonal matrix ${\bf K}{\bf K}^T$, with their sum still symmetric and positive-definite. By means of Woodbury's matrix identity, we can expand
\begin{align}\label{eq:chi2constr3}
\left( {\bf V} + {\bf K}{\bf K}^T\right)^{-1} & = {\bf V}^{-\frac12}  \left( {\bf 1} - {\bf H} \left( 1 + {\bf H}^T{\bf H} \right)^{-1}{\bf H}^T \right) {\bf V}^{-\frac12}  \nonumber \\
& = {\bf V}^{-\frac12} {\bf Z}^{-1} {\bf V}^{-\frac12}, 
\end{align}
with ${\bf H}={\bf V}^{-\frac12}{\bf K}$. The newly introduced matrix ${\bf Z}$ is a symmetric and positive-definite matrix, so it is invertible and has a unique square-root matrix ${\bf Z}^{\frac12}$. By means of eq.~\eqref{eq:chi2constr3}, eq.~\eqref{eq:chi2constr2} can be ultimately recast in the same form of eq.~\eqref{eq:rearrange}, namely 
 \begin{equation}\label{eq:rearrangeconstr}
 \lVert {\bf r}_{\mu}^\prime - {\bf A}^\prime (\boldsymbol{\theta} - \boldsymbol{\theta}_{0}) \rVert^2,
\end{equation}
where ${\bf r}_{\mu}^\prime$ and ${\bf A}^\prime$ are defined as in eq.~\eqref{eq:rearrange2} with the replacement ${\bf V}^{-1/2} \to \left({\bf V} {\bf Z}\right)^{-1/2}$. In this framework, statistical fluctuations affecting the entries of the Jacobian matrix ${\bf K}$ are propagated into the NLL function by the matrix ${\bf Z}$ which mixes them with fluctuations of ${\bf J}$ in a possibly intricate way. However, due to the formal identity between eq.~\eqref{eq:rearrange} and~\eqref{eq:rearrangeconstr}, the same analysis pursued in section~\ref{sec:hessuncMCfinite} still holds. Notice, however, that in the case of eq.~\eqref{eq:chi2constr3} some of the scaling laws of eq.~\eqref{eq:scaling} are broken due to a non-trivial combination of statistical and systematic uncertainties. Also, the inclusion of external constraints in the likelihood function is likely to reduce $\rho_\mu$ and hence improve the validity of the asymptotic approximation.

Constrained nuisance parameters naturally arise in problems where the nominal model prediction is affected by sources of systematic uncertainty for which some {\it a priori} knowledge is available from independent measurements or theoretical prejudice. In this respect, $K_{ij}$ can be seen as the ``error'' on the expectation value in the $i^{\rm th}$ bin as due to a $\pm 1\sigma$ ``error'' of the $j^{\rm th}$ source of systematic uncertainty. Thus, when ${\bf K}$ is modeled via MC simulation, statistical fluctuations of the MC sample induce ``errors on the errors''. This subject has a connection with ref.~\cite{Cowan:2018lhq}, where breaking of asymptotic properties of the MLE has been also observed. 

\newpage 

\section{Supplementary material for section~\ref{sec:general}}\label{app:additional}

This appendix contains intermediate steps and supplementary material to support the main results of section~\ref{sec:general}.

\subsection{Additional results for the profile-likelihood ratio}\label{app:additional_hessian}

The profile likelihood estimator for the NP is given by:
\begin{align}\label{eq:minimum}
\hat{ \boldsymbol{\theta} }_{\mu} = \boldsymbol{\theta}_0 + ({\bf A}^T{\bf A})^{-1} {\bf A}^T {\bf r}_{\mu}.
\end{align}
where ${\bf r}_{\mu}={\bf d} - {\bf b}(\mu - \mu_0)$.
Notice that the matrix ${\bf A}^T{\bf A}$ is positive-definite, hence invertible, because ${\bf A}$ was assumed to have full rank. By inserting the right-hand side of eq.~\eqref{eq:minimum} into eq.~\eqref{eq:rearrange}, the log-likelihood function becomes a quadratic function of $\mu$, which can be again minimized analytically yielding the estimator:
\begin{equation}\label{eq:chi2minhat}
\hat{\mu} = \mu_0 + \left( {\bf b}^T {\bf U} {\bf b} \right)^{-1} {\bf b}^T {\bf U} {\bf d}.
\end{equation}

An alternative way to prove the relation between the $S$ variable defined in eq.~\eqref{eq:sigmaH} and the variance of the MLE estimator $\hat{\mu}$ is to derive directly the full set of MLE estimators, that is the point in the parameter space that minimizes eq.~\eqref{eq:chi2} globally, obtaining:
\begin{equation}\label{eq:jacvar}
\begin{pmatrix}
\hat{\boldsymbol{\theta}} \\
\hat{\mu}
\end{pmatrix} = 
\begin{pmatrix}
{\boldsymbol{\theta}}_0 \\
{\mu}_0
\end{pmatrix} + \left( {\bf B}^T {\bf B} \right)^{-1}  {\bf B}^T {\bf d},
\end{equation}
where ${\bf B}$ is a $n\times (p+1)$ matrix obtained by adding the column ${\bf j}$ on the right of ${\bf J}$, and ${\bf C} = \left( {\bf B}^T {\bf B} \right)^{-1}$ is the covariance matrix of $(\hat{\boldsymbol{\theta}},\hat{\mu})$. By using the formula for the inverse block matrix, one can easily verify that
\begin{align}\label{eq:invB}
{\bf C} =
%\begin{pmatrix}
%{\bf A}^T{\bf A}  & {\bf A}^T{\bf b} \\
%{\bf b}^T{\bf A} & {\bf b}^T{\bf b} 
%\end{pmatrix}^{-1} \equiv 
\begin{pmatrix}
{\bf W}  & {\bf A}^T{\bf b} \\
{\bf b}^T{\bf A} & {\bf b}^T{\bf b} 
\end{pmatrix}^{-1} = 
\begin{pmatrix}
{\bf W}^{-1}  + S^{-1}{\bf W}^{-1} {\bf A}^T{\bf b}  {\bf b}^T{\bf A}  {\bf W}^{-1}  & -S^{-1}{\bf W}^{-1}{\bf A}^T{\bf b}  \\
-S^{-1} {\bf b}^T{\bf A} {\bf W}^{-1} & S^{-1}
\end{pmatrix}.
\end{align}
where ${\bf W}={\bf A}^T{\bf A}$. Equation~\eqref{eq:invB} implies that
\begin{equation}
{\sigma}^2_{\rm H} = \left[ {\bf C}\right]_{p+1,\,p+1} =S^{-1},
\end{equation}
thus showing that, in the Gaussian approximation, the standard deviation of $\hat{\mu}$ obtained from the Hessian matrix and from the concavity of the profile-likelihood ratio function coincide\footnote{Formally, eq.~\eqref{eq:invB} implies that $S$ is the Schur complement of the upper-left block ${\bf W}$ of the Hessian matrix ${\bf C}^{-1}$.}. 

An obvious consequence of eq.~\eqref{eq:sigmaH} is that $\mu$ cannot be measured (i.e. $\sigma_{\rm H}$ would be virtually infinite) if ${\bf b}$ can be expressed as a linear combination of the columns of ${\bf A}$. In fact,
%\footnote{Physically, this would imply that a point in the NP space can be always found such that the right-hand side of eq.~\eqref{eq:rearrange} is invariant under a variation of $\mu$.}: 
in this case one could write ${\bf b}={\bf A}{\bf w}$ for some vector of weights ${\bf w}$, so that
\begin{align}
    {\bf b}^T{\bf U}{\bf b}  & = {\bf b}^T({\bf 1}  - {\bf A}({\bf A}^T{\bf A})^{-1}{\bf A}^T){\bf A}{\bf w} \nonumber \\
    & = {\bf b}^T{\bf A}{\bf w} - {\bf b}^T{\bf A}({\bf A}^T{\bf A})^{-1}({\bf A}^T {\bf A}) {\bf w} \nonumber  \\  
    & = {\bf b}^T({\bf A}{\bf w} - {\bf A}{\bf w}) = {\bf 0}.
\end{align}
which would then imply $\sigma_{\rm H}\to\infty$.

By means of eq.~\eqref{eq:invB}, it can be easily proved that $\rho^2_\mu$ is equivalent to the global correlation coefficient
\begin{align}\label{eq:globalrho2}
 1 - \left( \left[{ \bf C}\right]_{p+1 \,p+1} \cdot \left[{\bf C}^{-1}\right]_{p+1 \, p+1} \right)^{-1}.
\end{align}
%where ${\bf C}$ is again defined as the covariance matrix of $(\boldsymbol{\theta},\mu)$.

Finally, we can use the results obtained so far to show that $\rho^2_\mu {\sigma}^2_{\rm H}$ is also equal to the difference in quadrature between ${\sigma}^2_{\rm H}$ and the variance of $\mu$ when $\boldsymbol{\theta}$ are fixed to their post-fit values. Hence, $\rho_\mu$ can be in principle computed also from the {\it group impact} ($I_{\boldsymbol{\theta}}$) on $\mu$ from {\it all} nuisance parameters~\cite{Pinto:2023yob,CMS:2024lrd}, namely:
\begin{align}\label{eq:globalrho3}
\rho_\mu = \frac{I_{\boldsymbol{\theta}}}{{\sigma}_{\rm H}}, \;\;\; {\rm with} \;\;\; I_{\boldsymbol{\theta}}=\sqrt{ \left[ {\bf C} \right]_{\boldsymbol{\theta}\mu}^T \left[ {\bf C} \right]_{\boldsymbol{\theta}\boldsymbol{\theta}}^{-1} \left[ {\bf C} \right]_{\boldsymbol{\theta}\mu} },
\end{align}
where $\left[{\bf C}\right]_{\boldsymbol{\theta}\boldsymbol{\theta}}$ and $\left[{\bf C}\right]_{\boldsymbol{\theta}\mu}$ stand for the upper-left $p\times p$ and upper-right $p\times 1$ blocks of the covariance matrix ${\bf C}$\footnote{For a single parameter $\theta_j$, the impact can be more simply written as $I_{\theta_j}=\frac{\left[{\bf C}\right]_{\mu\theta_j}}{\sqrt{\left[{\bf C}\right]_{\theta_j\theta_j} }}=|\rho_{\mu\theta_j}| \cdot \sigma_{\rm H}$, with $\rho_{\mu\theta_j}$ equal to the linear correlation between $\mu$ and $\theta_j$.}. 

\subsection{Derivation of eq.~\eqref{eq:U3terAvg}}\label{app:additional_Avg}

By expanding $S$ around its true value, we obtain an {\it estimator} $\hat{S}$ given by
%$\tilde{\bf u}_j = {\bf u}_j+ \boldsymbol{\nu}_j$, with $\boldsymbol{\nu}_j$ being ``small'' perturbations around the true eigenvectors, and write eq.~\eqref{eq:chi2min} as
\begin{align}\label{eq:U3ter}
\hat{S} & = \hat{\bf b}^T\left( 1 - \sum_{j=1}^p \hat{\bf u}_j\hat{\bf u}_j^T \right)\hat{\bf b}  ={S} + 2\boldsymbol{\beta}^T{\bf U}{\bf b}  + \boldsymbol{\beta}^T{\bf U}\boldsymbol{\beta}  \nonumber \\
&  - 2\sum_{j=1}^p \left({\bf b}^T {\bf u}_j \right) \left( {\bf b}^T\boldsymbol{\nu}_j\right) - {\bf b}^T\left( \sum_{j=1}^p\boldsymbol{\nu}_j\boldsymbol{\nu}_j^T \right) {\bf b} \nonumber \\
& +2\boldsymbol{\beta}^T\left( -\sum_{j=1}^p {\bf u}_j\boldsymbol{\nu}_j^T - \sum_{j=1}^p \boldsymbol{\nu}_j{\bf u}_j^T \right){\bf b} + \hdots,
\end{align}
where the dots stand for terms that are of higher-order in the perturbations. Equation~\ref{eq:U3terAvg} follows by taking the expectation value of the right-hand side of eq.~\eqref{eq:U3ter}.

\subsection{Contribution from the matrix ${\bf A}$ for generic column vectors}\label{app:additional_general}

We consider the more general case for the columns vectors ${\bf a}_j$. In this case, we can always find a set of $p$ orthonormal vectors such that 
\begin{equation}
    {\bf u}_j = \sum_{k=1}^p w_{jk}{\bf a}_k
\end{equation}
where $w_{jk}$ are the elements of a non-singular square matrix. After some straightforward algebra, the expectation value of the two last terms on the second line of eq.~\eqref{eq:U3terAvg} can be written as:
\begin{align} \label{eq:extrachi7}
    \sum_{j,k,l=1}^p w_{jk}w_{jl}\left( {\bf b}^T{\bf a}_k \right)\left( {\bf b}^T{\bf a}_l \right)\langle \boldsymbol{\alpha}_l^2\rangle\left( 1 + \frac{\delta_{kl}}{n-1}\right)  - \frac{{\bf b}^T{\bf b}}{n-1}\sum_{j,k=1}^p w^2_{jk} \langle \boldsymbol{\alpha}_k^2\rangle
\end{align}
Again, we first focus on the case of large overlap between ${\bf b}$ and the linear space ${\cal V}_{\bf A}$, which allows us to consider only the first term on the right-hand side of eq.~\eqref{eq:extrachi7} since the remaining terms are relatively suppressed by a factor of $1/n$. If we denote by ${\bf v}$ a vector with $k^{\rm th}$-element equal to ${\bf b}^T{\bf a}_k$, we see that the triple sum of eq.~\eqref{eq:extrachi7} can be written as
\begin{equation} \label{eq:extrachi8}
    {\bf v}^T \left( \sum_{j=1}^p {\bf w}_j {\bf w}_j^T  
    {\rm diag}(  \langle \boldsymbol{\alpha}_1^2 \rangle, \hdots, \langle \boldsymbol{\alpha}_p^2 \rangle ) \right) {\bf v}
\end{equation}
The matrix defined between parentheses in eq.~\eqref{eq:extrachi8} is positive (semi-)definite if $\langle \boldsymbol{\alpha}_k^2 \rangle > 0$ for all (at least one) index $k$, so that the quadratic form cannot be negative. In the limiting case of small $\rho_\mu$, the rightmost term in eq.~\eqref{eq:extrachi7} dominates, and the same conclusions derived for the case of orthogonal columns holds.

\subsection{The large correlation case for arbitrary $n$}\label{app:additional_cheb}

By expanding ${\bf b}$ as a linear combination of the vectors ${\bf a}_j$, and neglecting the ${\bf b}_{\perp}$ term, we have
\begin{align}
& \frac{n}{n-1}\sum_{j=1}^p
\left( {\bf b}^T {\bf a}_j \right)^2\langle \boldsymbol{\alpha}_i^2 \rangle - \frac{{\bf b}^T{\bf b}}{n-1}\left( \sum_{j=1}^p \langle \boldsymbol{\alpha}_i^2 \rangle \right)  > \nonumber \\
& \frac{1}{n-1} \left[p \sum_{j=1}^p
\left( {\bf b}^T {\bf a}_j \right)^2\langle \boldsymbol{\alpha}_i^2 \rangle -  \left( \sum_{j=1}^p
\left( {\bf b}^T {\bf a}_j \right)^2\right)\left( \sum_{j=1}^p \langle \boldsymbol{\alpha}_i^2 \rangle \right) \right]
\end{align}
where we have used the fact that $n > p$. The quantity within square brackets is positive-definite for all $n$ as for Chebyshev's sum inequality, provided that ${\bf b}^T {\bf a}_1 \geq \hdots \geq  {\bf b}^T {\bf a}_p$ and $\langle \boldsymbol{\alpha}_1^2 \rangle \geq \hdots \geq \langle \boldsymbol{\alpha}_p^2 \rangle$, which includes both the case of uniform uncertainties on the columns, as well as the case of increasingly larger uncertainty for column vectors that have larger overlap with ${\bf b}$, which is also an intuitive result.

\subsection{The two-dimensional problem}\label{app:additional_2D}

\begin{figure}
    \centering
    \includegraphics[width=0.55\linewidth]{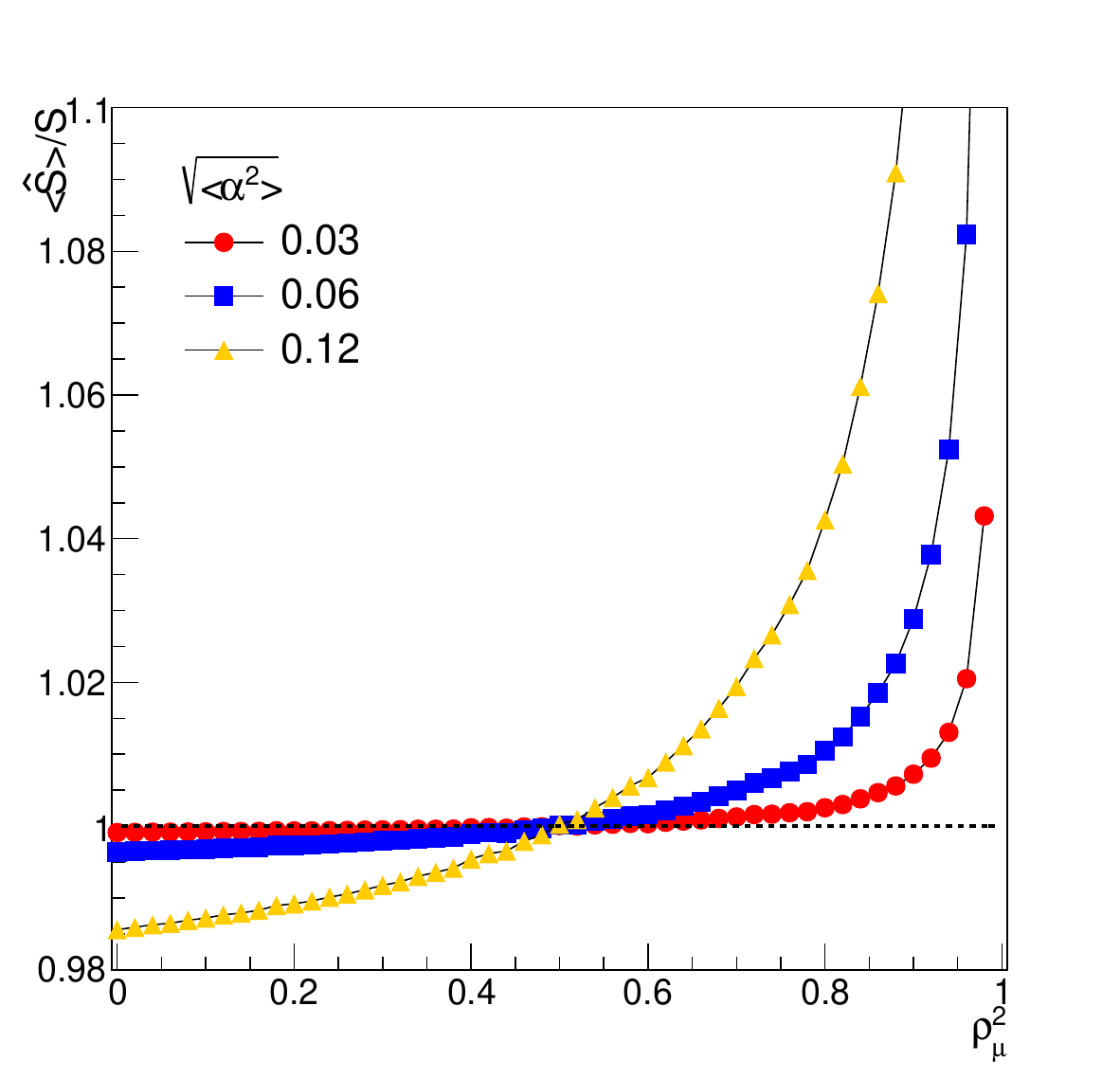}
    \caption{The ratio between the expectation value $\langle \hat{S} \rangle$ and the true value $S$ for the two-dimensional problem of eq.~\eqref{eq:simple} as a function of the squared correlation coefficient $\rho^2_\mu$, and for three representative values of $\langle\alpha^2\rangle$.}
    \label{fig:simple2}
\end{figure}

Equations~\eqref{eq:extrachi5} and~\eqref{eq:extrachi6} can be verified, for example, in a simple two-dimensional setup, namely for $n=2$ and $p=1$. Modulo an overall normalization factor to the vector ${\bf b}$ and to the single-column matrix ${\bf A}$, we can define
\begin{equation}\label{eq:simple}
    {\bf b} = \left(
    \begin{matrix}
         1\\
        0
    \end{matrix}
    \right)
    \;\;\;{\rm and}\;\;\;
    {\bf A} = 
    \begin{pmatrix}
    s_\gamma \\
    c_\gamma
    \end{pmatrix}
    \Rightarrow
    {\bf U} = 
    \begin{pmatrix}
    c^2_\gamma & 1-s_\gamma c_\gamma\\
    1-s_\gamma c_\gamma & s^2_\gamma
    \end{pmatrix},
\end{equation}
where $s_\gamma\equiv \sin\gamma$, $c_\gamma\equiv \cos\gamma$, with $\gamma\in[0,\pi]$ an angle that parametrizes the degree of correlation between $\mu$ and the nuisance parameter $\theta$. From eq.~\eqref{eq:simple} we have that $S=U_{11}=c^2_\gamma$ and $\rho^2_\mu=s^2_\gamma=1-S$. We treat statistical fluctuations on ${\bf A}$ as a pair of independent and Gaussian-distributed random variables $(\alpha_1,\alpha_2)$, which, for simplicity, are assumed to have the same variance $\langle\alpha^2\rangle$. Under fluctuations, the matrix ${\bf U}$ is then changed to
\begin{equation}\label{eq:simple2}
    {\bf U} \to \hat{ \bf U} = 
    \begin{pmatrix}
    \frac{(c_\gamma + \alpha_2)^2}{(s_\gamma + \alpha_1)^2  + (c_\gamma + \alpha_2)^2}  &  \frac{(s_\gamma+\alpha_1-c_\gamma-\alpha_2)^2 + (s_\gamma+\alpha_1)(c_\gamma+\alpha_2)}{(s_\gamma + \alpha_1)^2  + (c_\gamma + \alpha_2)^2}  \\
     - &  \frac{(s_\gamma + \alpha_1)^2}{(s_\gamma + \alpha_1)^2  + (c_\gamma + \alpha_2)^2}
    \end{pmatrix}  
\end{equation}
Using eq.~\eqref{eq:simple2} evaluated at randomly generated points $(\alpha_1,\alpha_2)$, the estimator $\hat{S}=\hat{U}_{11}$ can be calculated and its mean value $\langle \hat{S} \rangle$ extracted from a large number of random samplings. The results are shown in figure~\ref{fig:simple2} as a function of $\rho^2_\mu$ and for three representative values of $\langle\alpha^2\rangle$. For small values of $\rho^2_\mu$, the ratio $\langle \hat{S} \rangle/S$ is slightly less than one, i.e. the bias on $\hat{S}$ is negative, in agreement with eq.~\eqref{eq:extrachi6}. Conversely, when $\rho^2_\mu$ increases, the ratio becomes larger than one (hence, the bias gets positive), eventually growing as $\langle\alpha^2\rangle/(1 - \rho^2_\mu)$ for $\rho^2_\mu\to 1$, in agreement with eq.~\eqref{eq:extrachi5}.

\subsection{Scaling laws for the Barlow-Beeston lite likelihood}\label{app:additional_scaling}

Equation~\eqref{eq:scaling4} assumes that the likelihood does not contain any explicit term to account for the statistical uncertainties on the MC templates. It can be easily extended to the case where the MC statistical uncertainties are treated {\it \`a la} Barlow-Beeston lite by exploiting the Gauss + MC stat. approximation (see sec.~\ref{sec:gaus}), thus modifying the scaling of ${\bf b}$ via an additional multiplicative factor $\left( 1 + \frac{1}{k} \right)^{-{1}/{2}}$,
%\begin{align}\label{eq:scalingBB}
%{\bf b} & \sim (\mu-\mu_0)\frac{L}{\sqrt{L + \frac{L^2}{N}}}, 
%\end{align}
yielding:
\begin{align}\label{eq:scaling5}
\bar{\sigma}_{\rm H} =  \left(\frac{r_\xi \sqrt{\xi -1} }{\sqrt{\alpha(1+\frac{1}{k})r^2_\xi - (1+\frac{\xi}{k}) } } \right) \hat{\sigma}_{{\rm H}}.
\end{align}

\subsection{Gradient of $\hat{\mu}$ with respect to ${\bf B}$ and ${\bf d}$}\label{app:additional_dermu}

By applying the chain rule for differentiation, and using the known formula for the derivative of an inverse matrix, the Jacobian of $(\hat{\boldsymbol{\theta}},\hat{\mu})$ with respect to $B_{ik}$ can be computed as
\begin{align}\label{eq:jacvar2}
& \frac{\partial  }{\partial B_{ik}} \begin{pmatrix}
\hat{\boldsymbol{\theta}} \\
\hat{\mu}
\end{pmatrix}   = \left(\frac{\partial  \left( {\bf B}^T{\bf B} \right)^{-1} }{\partial B_{ik}} \right) {\bf B}^T {\bf d} +  \left( {\bf B}^T{\bf B} \right)^{-1}  \left(\frac{\partial {\bf B}^T}{\partial B_{ik}} \right) {\bf d}  \nonumber \\
& = - \left( {\bf B}^T{\bf B} \right)^{-1} \frac{\partial \left( {\bf B}^T{\bf B} \right) }{\partial B_{ik}} \left( {\bf B}^T{\bf B} \right)^{-1} {\bf B}^T {\bf d} + \left( {\bf B}^T{\bf B} \right)^{-1} \left(\frac{\partial {\bf B}^T}{\partial B_{ik}} \right) {\bf d} \nonumber \\
& = - \left( {\bf B}^T{\bf B} \right)^{-1} \left( \boldsymbol{\Pi}^T_{ik} {\bf B} + {\bf B}^T \boldsymbol{\Pi}_{ik} \right) \begin{pmatrix}
\hat{\boldsymbol{\theta}} \\
\hat{\mu}
\end{pmatrix}  + \left( {\bf B}^T{\bf B} \right)^{-1}{\boldsymbol{\Pi}^T_{ik}}{\bf d} \nonumber  \\ 
& = {\bf C} \left( \boldsymbol{\Pi}^T_{ik} \left( {\bf d} - {\bf B} \begin{pmatrix}
\hat{\boldsymbol{\theta}} \\
\hat{\mu}
\end{pmatrix} \right) - {\bf B}^T \boldsymbol{\Pi}_{ik} \begin{pmatrix}
\hat{\boldsymbol{\theta}} \\
\hat{\mu}
\end{pmatrix}  \right),
\end{align}
where $\boldsymbol{\Pi}_{ik} = {\bf e}_i {\bf e}_k^T$, with ${\bf e}_l$ the $l^{\rm th}$ vector of the canonical basis. The gradient $\partial\hat{\mu}/\partial B_{ik}$ is provided by the last element of eq.~\eqref{eq:jacvar2}.

The gradient of ${\hat \mu}$ with respect to ${\bf d}$ can be readily obtained, giving:
\begin{equation}\label{eq:jacvar3}
    \frac{\partial \hat{\mu}}{\partial d_{i}}= \sum_{k=1}^{p+1} { C}_{p+1 \, k}  B_{i k}.
\end{equation}

\section*{Acknowledgements}
C.A., L.B., and D.B. acknowledge financial support from the
European Research Council (ERC) under the European Union’s Horizon 2020 research
and innovation program (Grant agreement N. 101001205).

\bibliographystyle{elsarticle-num-names} 
\bibliography{biblio.bib}

\end{document}